\documentclass[twocolumn]{aastex62}

\usepackage{graphicx}	
\usepackage{amsmath}	
{\LARGE 
\usepackage{hyperref}
}
\usepackage{wrapfig}
\usepackage{float}
\usepackage{subfigure}

\newcommand{\eazy}{{\tt{EAZY}}}
\newcommand{\bagpipes}{{\tt{Bagpipes}}}
\newcommand{\qsogen}{{\tt{qsogen}}}
\newcommand{\sextractor}{{\tt{SExtractor}}}

\newcommand{\webbpsf}{{\tt{WebbPSF}}}
\newcommand{\cigale}{{\tt{CIGALE}}}
\newcommand{\galfind}{{\tt{galfind}}}

\newcommand{\sfhdelayed}{{\tt{sfhdelayed}}}
\newcommand{\bc}{{\tt{bc03}}}
\newcommand{\mara}{{\tt{m2005}}}
\newcommand{\nebular}{{\tt{nebular}}}
\newcommand{\dustattmodifiedstarburst}{{\tt{dustatt\_modified\_starburst}}}
\newcommand{\dl}{{\tt{dl2014}}}
\newcommand{\skirtor}{{\tt{skirtor2016}}}
\newcommand{\scipy}{{\tt{Scipy}}}
\newcommand{\astropy}{{\tt{Astropy}}}
\newcommand{\matplotlib}{{\tt{Matplotlib}}}
\newcommand{\mast}{{\tt{MAST}}}
\newcommand{\MAST}{{\tt{Mikulski\ Archive\ for\ Space\ Telescopes}}}

\received{June 4, 2025}
\revised{July 25, 2025}
\accepted{July 30, 2025}

\submitjournal{ApJL}

\shorttitle{The Nature of Little Red Dots through Light Emission and Clustering}
\shortauthors{.}

\def\deg{$^{\circ}\,$}

\def\deg{$^{\circ}\,$}

\def\casgm20{CAS-G-M$_{20}\,$}
\def\m20{M$_{20}\,$}

 \def\HST{\textit{Hubble Space Telescope}}
 \def\ACS{\textit{Advanced Camera for Surveys}}
 \def\JWST{\textit{James Webb Space Telescope}}
 \def\ALMA{\textit{Atacama Large Millimeter Array}}
 
\begin{document}  

\title{Lonely Little Red Dots: Challenges to the AGN-nature of little red dots through their clustering and spectral energy distributions}

\correspondingauthor{María Carranza-Escudero}
\email{maria.carranzaescudero@physics.ox.ac.uk}

\author[0009-0002-0274-2676]{María Carranza-Escudero}
\affiliation{Jodrell Bank Centre for Astrophysics, University of Manchester, Oxford Road, Manchester M13 9PL, UK}

\author[0000-0003-1949-7638]{Christopher J. Conselice}
\affiliation{Jodrell Bank Centre for Astrophysics, University of Manchester, Oxford Road, Manchester M13 9PL, UK}

\author[0000-0003-4875-6272]{Nathan Adams}
\affil{Jodrell Bank Centre for Astrophysics, University of Manchester, Oxford Road, Manchester M13 9PL, UK}

\author[0000-0002-4130-636X]{Thomas Harvey}
\affil{Jodrell Bank Centre for Astrophysics, University of Manchester, Oxford Road, Manchester M13 9PL, UK}

\author[0000-0003-0519-9445]{Duncan Austin}
\affil{Jodrell Bank Centre for Astrophysics, University of Manchester, Oxford Road, Manchester M13 9PL, UK}

\author[0000-0002-2517-6446]{Peter Behroozi}
\affiliation{Steward Observatory and Department of Astronomy, University of Arizona, Tuscon, AZ 85721, USA}

\author[0000-0002-8919-079X]{Leonardo Ferreira}
\affil{Department of Physics \& Astronomy, University of Victoria, Finnerty Road, Victoria, British Columbia, V8P 1A1, Canada}

\author[0000-0003-2000-3420]{Katherine Ormerod}
\affil{Jodrell Bank Centre for Astrophysics, University of Manchester, Oxford Road, Manchester M13 9PL, UK}
\affil{Astrophysics Research Institute, Liverpool John Moores University, 146 Brownlow Hill, Liverpool, L3 5RF}

\author[0009-0009-8105-4564]{Qiao Duan}
\affiliation{Jodrell Bank Centre for Astrophysics, University of Manchester, Oxford Road, Manchester M13 9PL, UK}

\author[0000-0002-9081-2111]{James Trussler}
\affil{Center for Astrophysics $|$ Harvard \& Smithsonian, 60 Garden St., Cambridge MA 02138, USA}

\author[0000-0002-3119-9003]{Qiong Li}
\affil{Jodrell Bank Centre for Astrophysics, University of Manchester, Oxford Road, Manchester M13 9PL, UK}


\author[0009-0008-8642-5275]{Lewi Westcott}
\affiliation{Jodrell Bank Centre for Astrophysics, University of Manchester, Oxford Road, Manchester M13 9PL, UK}




\author[0000-0001-8156-6281]{Rogier A. Windhorst}
\affiliation{School of Earth and Space Exploration, Arizona State University, Tempe, AZ 85287-1404}



\author[0000-0001-7410-7669]{Dan Coe} 
\affiliation{AURA for the European Space Agency (ESA), Space Telescope Science
Institute, 3700 San Martin Drive, Baltimore, MD 21218, USA}

\author[0000-0003-3329-1337]{Seth H. Cohen} 
\affiliation{School of Earth and Space Exploration, Arizona State University,
Tempe, AZ 85287-1404}

\author[0000-0003-0202-0534]{Cheng Cheng}
\affiliation{Chinese Academy of Sciences South America Center for Astronomy, National Astronomical Observatories, CAS, Beijing 100101, People's Republic of China}
\affiliation{2 CAS Key Laboratory of Optical Astronomy, National Astronomical Observatories, Chinese Academy of Sciences, Beijing 100101, People's Republic of China}

\author[0000-0001-9491-7327]{Simon P. Driver} 
\affiliation{International Centre for Radio Astronomy Research (ICRAR) and the
International Space Centre (ISC), The University of Western Australia, M468,
35 Stirling Highway, Crawley, WA 6009, Australia}

\author[0000-0003-1625-8009]{Brenda Frye} 
\affiliation{University of Arizona, Department of Astronomy/Steward
Observatory, 933 N Cherry Ave, Tucson, AZ85721}

\author[0000-0001-6278-032X]{Lukas J. Furtak}
\affiliation{Department of Physics, Ben-Gurion University of the Negev, P.O. Box 653, Be'er-Sheva 84105, Israel}

\author[0000-0001-9440-8872]{Norman A. Grogin} 
\affiliation{Space Telescope Science Institute, 3700 San Martin Drive, Baltimore, MD 21218, USA}

\author[0000-0001-6145-5090]{Nimish P. Hathi}
\affiliation{Space Telescope Science Institute, 3700 San Martin Drive, Baltimore, MD 21218, USA}


\author[0000-0003-1268-5230]{Rolf A. Jansen} 
\affiliation{School of Earth and Space Exploration, Arizona State University,
Tempe, AZ 85287-1404}

\author[0000-0002-6610-2048]{Anton M. Koekemoer} 
\affiliation{Space Telescope Science Institute, 3700 San Martin Drive, Baltimore, MD 21218, USA}

\author[0000-0001-6434-7845]{Madeline A. Marshall} 
\affiliation{National Research Council of Canada, Herzberg Astronomy \&
Astrophysics Research Centre, 5071 West Saanich Road, Victoria, BC V9E 2E7,
Canada; \& ARC Centre of Excellence for All Sky Astrophysics in 3
Dimensions (ASTRO 3D), Australia}

\author[0000-0003-3351-0878]{Rosalia O'Brien}
\affiliation{School of Earth and Space Exploration, Arizona State University,
Tempe, AZ 85287-1404}

\author[0000-0003-3382-5941]{Norbert Pirzkal}
\affiliation{Space Telescope Science Institute, 3700 San Martin Drive, Baltimore, MD 21218, USA}

\author[0000-0001-7411-5386]{Maria Polletta}
\affiliation{INAF – Istituto di Astrofisica Spaziale e Fisica Cosmica Milano, Via
A. Corti 12, I-20133 Milano, Italy}

\author[0000-0003-0429-3579]{Aaron Robotham} 
\affiliation{International Centre for Radio Astronomy Research (ICRAR) and the
International Space Centre (ISC), The University of Western Australia, M468,
35 Stirling Highway, Crawley, WA 6009, Australia}

\author[0000-0001-7016-5220]{Michael~J.~Rutkowski}
\affiliation{Minnesota State University-Mankato, Department of Physics \& Astronomy, Trafton Science Center North 141, Mankato, MN, 56001 USA}

\author[0000-0002-7265-7920]{Jake Summers}
\affiliation{School of Earth and Space Exploration, Arizona State University,
Tempe, AZ 85287-1404}

\author[0000-0003-3903-6935]{Stephen M. Wilkins}
\affiliation{Astronomy Centre, Department of Physics and Astronomy, University of Sussex, Brighton, BN1 9QH, UK}

\author[0000-0001-9262-9997]{Christopher N. A. Willmer} 
\affiliation{Steward Observatory, University of Arizona, 933 N Cherry Ave, Tucson, AZ, 85721-0009}

\author[0000-0001-7592-7714]{Haojing Yan} 
\affiliation{Department of Physics and Astronomy, University of Missouri,
Columbia, MO 65211}

\author[0000-0002-0350-4488]{Adi Zitrin}
\affiliation{Department of Physics, Ben-Gurion University of the Negev, P.O. Box 653, Be'er-Sheva 84105, Israel}




\begin{abstract}
Observations with the $\JWST$ (JWST) reveal a previously unseen population of compact red objects, known as ``little red dots`` (LRDs). We study a new photometrically selected sample of 124 LRDs in the redshift range $z$ $\sim$ 3 - 10 selected from NIRCam coverage of the CEERS, NEP-TDF, JADES and JEMS surveys. For JADES, the NEP-TDF and CEERS, we compare SED models with and without AGN components and analyse the impact of an AGN component on the goodness of fit using the Bayesian information criterion (BIC). We find that whilst the $\chi^{2}$ of the majority of models containing AGN components is improved compared to models without AGN components, we show that the BIC suggests models without AGN are a more appropriate fit to LRD SEDs, especially when MIRI data is available. We also measure LRD clustering in the CEERS field, JADES field, and NEP-TDF, where we compare the spatial distribution of LRDs and galaxies with Kolmogorov-Smirnov tests of equality of distribution. We find that the neighbourhood of LRDs tends to be less dense compared to galaxies at all selections and masses and at similar redshifts.  We further measure upper limit estimates for the halo masses of LRDs using abundance matching.  Whilst the population of LRDs could be a mixture of several different inherent populations, as a whole it does appear that these systems are mostly hosting compact galaxies or star clusters in formation.

\end{abstract}


\vspace{1em}
\section{Introduction} \label{sec:intro}

The discovery of a mysterious set of red objects exhibiting a v-shaped continuum and point-source morphology has received a large amount of attention due to their puzzling nature \citep[e.g.][]{labbe2023uca, matthee2024,kocevski2023, Furtak_2023, Furtak2024}.  These so-called Little Red Dots (LRDs) prove challenging to interpret and understand due to the similarity of their spectral energy distributions to both stellar populations and dust-reddened active galactic nuclei (AGN).  LRDs are typically found around redshift $z$ $\sim$ 5 \citep{labbe2023uca}.  These LRDs are a unique high redshift population whose nature is still very much uncertain.  

The spectral energy distributions (SEDs) of LRDs do not resemble any typical SEDs, and seem to be composed of a blue component starting at roughly 3600\AA, and a red component at longer wavelengths \citep{2024setton}. Investigations into the blue rest-ultraviolet (UV) of LRD SEDs \citep{labbe2023uca, akins2023, pérezgonzález2024} suggest, for example, that their origin could be unobscured star formation (SF) or, alternatively, light scattered from an active galactic nucleus' (AGN) accretion disk.  Attempts have been made to constrain the origin of the red rest-frame optical SEDs, some making use of Mid-Infrared Instrument (MIRI) data, in which few LRDs are detected.  The red component is often attributed to warm dust heated by some of the following scenarios.  Using MIRI data, \cite{pérezgonzález2024} find that the red optical and near infrared (NIR) data fit an obscured accretion disk, but stellar-dominated (often starburst) models provide an even better fit.  \cite{williams2023} find that data using the $\HST$ (HST) and Near Infrared Camera (NIRCam), but no MIRI and $\ALMA$ (ALMA) data, results in larger masses and star formation rates (SFRs) than when MIRI+ALMA data are used.  \cite{li2024} find no need for stellar emission and scattered AGN light and instead fit LRD SEDs with AGN embedded in extended dusty medium and a relatively grey extinction curve.

Employing spectroscopic observations of LRDs, \cite{Greene_2024} find that $\sim$ 75 $\%$ of their LRD sample exhibit broad-line H$\alpha$ emission, suggesting that these are dust-reddened AGN.  Supermassive black hole (SMBH) mass estimates from the broadness of LRD H$\alpha$ and $H\beta$ lines give rise to a range of masses \citep[10$^5$ - 10$^9$ M$_\odot$;][]{Furtak2024, kocevski2023,Greene_2024}.  The abundance of LRDs would suggest a much higher density of AGN \citep{akins2024, Kocevski_2025, Kokorev_2023} than predicted by ground-based surveys \citep{he2023,Matsuoka_2018,Niida_2020}, with implications for Lyman continuum radiation and reionisation \citep[e.g.][]{madau2024cosmicreionizationjwstera, Grazian_2024}. Theories discussing "black hole stars" \citep{2025Naidu, 2025deGraaff} or hyperdense gas cocoons surrounding smaller black holes also exist \citep{2025ApJ...980L..27I}, with early ALMA results providing upper mass limits that are in line with this \citep{2025Casey}.

Investigations into the local Universe ($z <$ 0.1) demonstrate evidence for a strong evolutionary relationship between SMBHs and their host galaxies \citep{gultekin2009, hu2008, kormendyho2013}.  As most LRDs lack detectable extended components, it is possible to place upper limits on stellar mass based on their maximum physical size.  Using the empirical relationship between the SMBH to host galaxy mass ratio presented in \cite{kormendyho2013}, and then comparing it to LRD SMBH mass estimates from spectroscopy \citep{Kokorev_2023, Greene_2024} reveals a SMBH to host galaxy mass ratio that is startlingly higher for LRDs, and in the case of a lensed LRD is perhaps unphysically high with a very high broad line width of $\sim 2000$ km s$^{-1}$ \citep{Furtak2024}.   
 
Other challenges to the AGN interpretation of LRDs exist.  For example, \cite{kokubo2024} report that LRDs do not exhibit the typical photometric variability associated with standard AGNs.  However, \cite{kokubo2024} suggest that this could be the result of intrinsically non-variable AGN or the dominance of AGN emission through scattering.  \cite{zhang2024analysismultiepochjwstimages} also find that LRDs do not usually display strong photometric variability.  Another challenge is that unlike what is expected of type-I AGN, the majority of LRDs are non-detected in X-rays \citep{yue2024, maiolino2024}.  A sample of LRDs exhibiting broad-line H$\alpha$ emission studied by \cite{Ananna_2024} show a stacked signal of only $\sim$ $2.6\sigma$.  Theories on the lack of X-ray detection of AGN are discussed in \cite{maiolino2024}, who suggest that the lack of X-ray detections are due to a mixture of scenarios including X-ray absorption, intrinsic X-ray weakness, and non-AGN sources. \cite{Baggen_2024} find an alternative explanation to the broad lines found in LRDs, suggesting that they are indicative of a brief phase in which galaxies have high central densities.

In this work, we select and present a sample of 124 LRDs using our own photometric selection based on previous LRD selection criteria. Our sample spans the redshift range $z$ $\sim$ 3 - 10 in the fields covered by the Cosmic Evolution Early Release Science Survey \citep[CEERS;][]{Bagley_2023}, the North Ecliptic Pole Time Domain Field \citep[NEP-TDF;][]{Windhorst_2023} survey and the JWST Advanced Deep Extragalactic Survey \citep[JADES;][]{2023ApJS..269...16R}.  Most LRDs appear around $z$ $\sim$ 5 \citep{labbe2023uca}, although some can reach photometric redshifts of $z$ $>$ 9 \citep{leung2024exploringnaturelittlered}.  In fact, the highest redshift LRD presented in our sample has $z$ = $10.4_{-1.3}^{+0.6}$.  

In this paper, we investigate SED modelling with models containing an AGN and those with no AGN component, and compare the best fit $\chi^{2}$ statistics and Bayesian Information Criterion (BIC) of these to determine which models are most suitable.  We briefly discuss the differences in stellar mass and dust for AGN and non-AGN models.  We also carry out a study of the local environment of LRDs and compare it to the local environment of galaxies using Kolmogorov-Smirnov tests.  Finally, we create estimates for the upper limits of halo masses for our LRD sample using abundance matching.

Section \ref{sec:reduc} describes the imaging and data reduction of the fields used in this work.  The selection criteria for LRDs are described in \S \ref{sec:LRDsample}.  In \S \ref{sec:LRD_cluster}, we analyse the local environment of LRDs and compare this to the local environment of galaxies at similar redshifts.  We investigate SED modelling and how the inclusion of AGN components affects fitting in \S \ref{sec:SEDmodel}.

Throughout this work, unless stated otherwise, we assume a standard cosmology with $H_0=70$\,km\,s$^{-1}$\,Mpc$^{-1}$, $\Omega_{\rm M}=0.3$ and $\Omega_{\Lambda} = 0.7$. All magnitudes listed follow the AB magnitude system \citep{oke1974, Oke1983}.

\section{Imaging and Data Reduction} \label{sec:reduc}

For this study we use JWST NIRCam imaging of CEERS, NEP-TDF survey, and JADES.
To cover objects at lower redshifts ($z$ $\sim$ 4.5 - 6) we make use of datasets from the $\HST$ (HST).  For CEERS we use the Cosmic Assembly Near-infrared Deep Extragalactic Legacy Survey \citep[CANDELS;][]{Grogin_2011, Koekemoer_2011} imaging, specifically the Extended Groth Strip \citep[EGS;][]{davis2007}.  For NEP-TDF we incorporate HST $\ACS$ Wide Field Channel (ACS/WFC) imaging from programs GO-15278 (PI: R. Jansen) and GO-16252/16793 (PIs: R. Jansen $\&$ N. Grogin).  To cover JADES, which lies on the Great Observatories Origins Deep Survey South (GOODS-South) footprint, we make use of HST data from the most recent mosaic (v2.5) from the Hubble Legacy Fields team \citep{illingworth2017hubblelegacyfieldshlfgoodss, Whitaker_2019}.  This section describes the details of the observations and data reduction used, as well as source identification and extraction.  A more detailed overview of this can be found in \cite{conselice2024epochsidiscoverystar}.

\subsection{CEERS JWST NIRCam and HST Imaging}
The JWST/NIRCam observations of CEERS (ID:1345, PI:S.Finkelstein; \citep[ID: 1345, PI: S. Finkelstein, see also][]{Bagley_2023} consist of 10 pointings covering  66.40 arcmin$^{2}$ in the EGS field.  The observations cover 7 photometric bands: F115W, F150W, F200W, F277W, F356W, F410M, and F444W.  We process CEERS data independently as described in \autoref{sec:reduction}.  Further details on the calibration process and data products can be found in \cite{adams2024epochspaperiiultraviolet} and \cite{conselice2024epochsidiscoverystar}.

Due to the lack of F090W imaging, we include HST CANDELS imaging \citep{Grogin_2011, Koekemoer_2011} of the F606W and F814W filters to cover a bluer wavelength range.  This imaging was reduced by the CANDELS team and is aligned using the Gaia EDR3 \citep{brown2021}.  The average depth is 28.5 mag and 28.3 mag for the F606W and F814W filters respectively.

\subsection{NEP-TDF JWST NIRCam and HST Imaging}
NEP-TDF is part of the JWST Prime Extragalactic Areas for Reionization and Lensing Science (PEARLS) project \citep{Frye_2023,diego2023,Windhorst_2023}.  NEP-TDF observations have 8 pointings covered by 8 filters: F090W, F115W, F150W, F200W, F277W, F356W, F410M, F444W.  We process the data as described in \autoref{sec:reduction}.  This is described in more depth in \cite{adams2024epochspaperiiultraviolet}.  The total area is 57.32 arcmin$^{2}$, with a resolution of 0.03 arcsec/pixel.

To cover bluer wavelengths, we incorporate HST observations that utilise the F606W filter \citep{obrien2024treasurehunttransientsvariabilitydiscovered}.  These observations were obtained through the GO-15278 (PI: R. Jansen) and GO-16252/16793 (PIs: R. Jansen $\&$ N. Grogin) programs from October 1 2017 through October 31 2022.

\subsection{JADES JWST NIRCam and HST Imaging}
In this paper we also use JADES DR1 \citep{2023ApJS..269...16R}, covering 22.98 arcmin$^{2}$ in the GOODS-S field footprint \citep[PID:1180, PI: D. Eisenstein;][]{eisenstein2023overviewjwstadvanceddeep}.  JADES consists of 6 overlapping pointings of 9 filters: F090W, F115W, F150W, F200W, F277W, F335M, F356W, F410M, and F444W. We reduce these data using our own pipeline once again to ensure consistency with other fields as described in \autoref{sec:reduction} and in greater detail in \cite{adams2024epochspaperiiultraviolet}.

We include HST/ACS data of the F435W, F606W, F775W and F814W filters.  This imaging is derived from the v2.5 GOODS-S mosaic from the Hubble Legacy Fields team \citep{illingworth2017hubblelegacyfieldshlfgoodss, Whitaker_2019}.

\subsection{MIRI imaging}
We make use of the Systematic Mid-infrared Instrument Legacy Extragalactic Survey (SMILES) \citep[PID 1207; PI: G. Rieke;][]{smiles} coverage of JADES, which has 15 pointings in the F560W, F770W, F1000W, F1280W, F1500W, F1800W, F2100W, and F2550W filters.  The total area covered by SMILES is $\sim$ 34 arcmin$^{2}$.  The public data release can be found on the MAST website\footnote{\url{https://archive.stsci.edu/hlsp/smiles}}.  Details on the reduction process and alignment can be found in \cite{smiles}.

\subsection{Reduction Process} \label{sec:reduction}

We process all uncalibrated lower-level JWST NIRCam data products with a modified version of the STScI JWST Pipeline v1.8.2 \citep{bushouse_2022} and use Calibration Reference Data System (CRDS) v1084 for the most up-to-date NIRCam calibration files at the time of writing.  After running stage 1 of the JWST pipeline, we subtract templates of 'wisps', large-scale artefacts affecting F150W and F200W imaging \citep{adams2024epochspaperiiultraviolet}.  After stage 2 of the pipeline we apply a 1/f noise correction derived by Chris Willott\footnote{\url{https://github.com/chriswillott/jwst}}.  We then perform background subtraction on each NIRCam frame before continuing to stage 3, after which we align the final images.  The final resolution of the drizzled images is 0.03 arcsec/pixel.

\subsection{Source Extraction}
To carry out source identification and extraction we make use of $\sextractor$  \citep{bertinarnouts1996}.  We take the inverse variance weighted stack of the F277W, F356W, and F444W bands and run this in dual-image mode to detect and select objects.
We carry out forced aperture photometry for multi-band measurements.  Photometry is calculated within circular apertures with 0.32 arcsecond diameters, chosen to enclose the central and brightest 70 - 80$\%$ of flux of a point source, and yet small enough to avoid contamination.  We include an aperture correction derived from simulated $\webbpsf$ point spread functions for each band used \citep{perrin2012, Perrin2014}. 

To avoid underestimating photometric errors we use the 5$\sigma$ local depth as the error.  Local depth is calculated using \galfind \footnote{\url{https://github.com/duncanaustin98/galfind}} by placing apertures with an approximately uniform number density of around $\sim$ 2500 arcmin$^{-2}$ in the 'empty' regions of the images, where 'empty' refers to an aperture where no pre-existing source is in the image. Overlapping apertures are not permitted to ensure that these provide independent measurements of the background.  To calculate the photometric error for each source we use the normalised mean absolute deviation \citep{hoaglin2000understanding} of the nearest 200 apertures, making the depth resistant to potential outliers.

For each field we carefully mask areas of the images affected by defects such as diffraction spikes and snowballs. The apertures are placed at least 30 pix from the mask.  More details of this process can be found in our EPOCHS paper I \citep{conselice2024epochsidiscoverystar} and paper II \citep{adams2024epochspaperiiultraviolet}.

\subsection{Photometric redshifts} \label{sec:red_qual}
We use the $\eazy$  photometric redshift code \citep{brammer2008} to calculate both the photometric redshift probability distribution and the most likely photometric redshift.  We include templates from \cite{larson2023}, which expand the default template sets that use 12 templates generated with the Flexible Stellar Population Synthesis code \citep{conroyjames2010}.  We use and test the SED templates used in \cite{hainline2024}, but find that the \cite{larson2023} templates match current spectroscopic results somewhat more closely. 

To determine the quality of our photometric redshifts, we employ the outlier fraction $\eta$, and the Normalized Median Absolute Deviation (NMAD).  The outlier fraction is defined as the fraction of photometric redshifts that differ from spectroscopic redshifts by more than 15$\%$, and is given by
\begin{equation} \label{eq1}
\begin{split}
\eta = \frac{N_{115}+N_{85}}{N_{total}}
\end{split}
\end{equation}
where $N_{115}$ represents the number of points above the line $z_{phot} = 1.15 \times z_{spec}$ and $N_{85}$ represents the number of points below the line $z_{phot} = 0.85 \times z_{spec}$.

The NMAD quantifies the dispersion in the redshift variances and is normalised. It is defined as:
\begin{equation} \label{eqNMAD}
\begin{split}
\sigma_{\mathrm{NMAD}} = 1.48 \times \mathrm{median}  \bigg | \frac{z_{spec} - z_{phot}}{1 + z_{spec}}\bigg | 
\end{split}
\end{equation}
Note that the factor of 1.48 in \autoref{eqNMAD} normalises the expectation value of the NMAD to be equivalent to the standard deviation of a normal distribution.

We match objects to published spectroscopic redshifts, including those from: the JADES DR3 \citep{deugenio2024jadesdatarelease3} release, spectra and redshifts from the EGS region from CEERS \citep{ArrabalHaro_2023}, including a followup DDT programme \citep[PID 2750]{Arrabal_Haro_2023} and PID 2565 \citep{glazebrook}.  To increase our sample for the calibration of photometric redshifts we also include spectroscopic redshifts for the GLASS-z12 object \citep{castellano2024jwstnirspecspectroscopyremarkable}, and results from the MACS-0416 field \citep{ma2024jwstviewinfantgalaxies} and the SMACS-0723 ERO programme \citep{Pontoppidan_2022}.  We compare these spectroscopic redshifts to our calculated photometric redshifts and find that for redshifts $z >$ 6.5 the NMAD value is 0.021. We find our outlier fraction to be $\eta$ = 9/86, or $\sim10\%$.  These measures indicate a high quality photometric redshift sample.  The lack of F115W for the SMACS-0723 cluster makes some redshifts uncertain at 7.5 $ < z < $ 9.5.  When SMACS-0723 is omitted from the redshift sample, the fraction of outliers drops to $\eta$ = 6/86, or $\sim7\%$. Further details can be found in \cite{adams2024epochspaperiiultraviolet} and \cite{conselice2024epochsidiscoverystar}.

\subsection{LRD NIRSpec spectra}
We use the spectra of 26 LRDs (see $\S$\ref{sec:broad}) found in the Cosmic Dawn Center (DAWN) JWST Archive \footnote{https://dawn-cph.github.io/dja/} (DJA) \citep{2024heintz, brammer_2023_8319596}. The majority of these spectra are taken as part of the RUBIES program \citep[GO-4233; PI: A. de Graaff;][]{degraaff2024rubiescompletecensusbright}.
The remaining spectra are from the NIRSpec WIDE GTO Survey \citep[GTO-1211 to 1215; PI: M. Maseda;][]{Maseda_2024}, CEERS, and from JADES DR1 \citep[][]{Bunker_2024}. The reduction process of these spectra is described in \cite{2024heintz, degraaff2024rubiescompletecensusbright}.

\section{Methodology}
\subsection{LRD sample selection} \label{sec:LRDsample}

LRDs were selected following the general previous criteria used by \cite{kokorev2024}, which aims to identify compact sources with a red rest-frame optical continuum and blue rest-frame UV light.  We require objects to be strongly detected in the F444W band by applying a F444W $<$ 27.7 AB mag cut and a $>$14$\sigma$ detection. 
Using the same redness and significance criteria as \cite{kokorev2024}, we select red objects under the criteria red1 or red2, where
\begin{equation} \label{eq2}
\begin{split}
\mbox{red1} & = F115W - F150W < 0.8 \text{ \&}\\
 & F200W - F277W > 0.7  \text{ \&}\\
 & F200W - F356W > 1.0,
\end{split}
\end{equation}
\begin{equation} \label{eq3}
\begin{split}
\mbox{red2} & = F150W - F200W < 0.8 \text{ \&}\\
 & F277W - F356W > 0.6  \text{ \&}\\
 & F277W - F444W > 0.7.
\end{split}
\end{equation}
but only where one band of each colour cut is $>$ 2$\sigma$ detected and the other $>$ 3$\sigma$ detected.

These two sets of redness criteria target two redshift bins.  We find that LRDs that meet the red1 criteria typically have $z$ $\lesssim$ 6, whilst those that meet the red2 criteria have $z$ $\gtrsim$ 5.  We find $\sim80\%$ of our LRDs in red1 and $\sim40\%$ LRDs in red2.  Around 10$\%$ of the LRDs are present in both red1 and red2.

To ensure the compactness of objects in the LRD sample we require the condition
\begin{equation} \label{eq4}
\begin{split}
\mbox{compact} & = f_{F444W}(0\farcs5)/f_{F444W}(0\farcs32) < 1.4.
\end{split}
\end{equation}
This is different to other works on LRDs as we use aperture sizes 0\farcs5 and 0\farcs32, as opposed to the more commonly used 0\farcs4 and 0\farcs2, which requires a ratio $< 1.7$ \citep[e.g][]{kokorev2024, Greene_2024}.

To reduce the amount of contaminants with strong emission lines rather than a red continuum, we add the colour cut criteria:
\begin{equation} \label{eq5}
\begin{split}
F200W - F410M > 0.9,
\end{split}
\end{equation}
when F410M photometry is available.  To choose this colour cut, we compare objects with emission lines in the F200W or F410M bands to those without emission lines in either band.  This is similar to \cite{Kocevski_2025}, who define a limit of F277W $-$ F410M $>$ 0.42 instead.  We find a surplus of objects below this colour cut in the latter category.  This removed a further 23 objects ($\sim14\%$) from the sample.  Finally, we also remove objects that appear to be diffuse or hot pixels by eye. 

To increase the size of our clustering samples we attempted to select a sample of LRDs from the COSMOS-Web field.  However, we find that the relatively small number of bands means that criteria could not be adjusted to avoid selecting a very different population.  We also find that less than 20$\%$ of our sample in CEERS, NEP-TDF, and JADES overlaps with the colour cut criteria in \cite{akins2024}.  We remark that this criterion is meant to select only the subset of red objects known as Extremely Red Objects (EROs).  For the reasons outlined above we decide not to include COSMOS-Web in our LRD sample.

\subsection{Brown dwarfs}
It is possible that some compact red sources are in fact brown dwarfs rather than LRDs \citep{langeroodi}.  In fact, the fraction of brown dwarfs found in some samples is between up to $\sim25\%$ \citep{langeroodi} and $\sim5\%$ \citep{Kocevski_2025}. It may not be possible to rule out brown dwarfs in JWST data based on their sizes \citep{2024Holwerda}. To remove brown dwarfs, we fit the SEDs of the sample using brown dwarf templates from the Sonora Bobcat \citep{marley2021} and Sonora Cholla \citep{hainline2024bd} models.  We consider an object in the sample to be a brown dwarf if $\chi^2 < 20$ for the best fitting brown dwarf model.  Following the suggestion of \cite{Greene_2024}, we investigate using an additional colour cut to remove potential brown dwarfs from LRD samples. The additional colour cut is given by
\begin{equation} \label{eq6}
\begin{split}
\mbox(bd\_removal) & = F115W - F200W > -0.5.
\end{split}
\end{equation}
We compare this cut to the cuts made using our Sonora Bobcat and Sonora Cholla fits.  In comparison to the objects selected as brown dwarfs by our $\chi^2 < 20$ cut, we find that the cut in \autoref{eq6} removes 8 contaminant brown dwarfs and 6 LRDs, but misses one brown dwarf.  We conclude that this colour cut is effective in producing a mostly clean sample.  Ultimately, we remove 9 brown dwarfs using our own $\chi^2 < 20$ brown dwarf cut to remove this form of contaminant whilst maximising our sample size. 

\subsection{Final LRD sample}

The number of objects in our final sample is 63 LRDs in the CEERS field, 42 in the NEP-TDF, and 19 in the JADES field, totalling 124 LRDs.  One LRD in JADES is outside the SMILES footprint and thus has no MIRI data.  We match 33 out of the 44 LRDs in CEERS found by \cite{kokorev2024}.  We also match 30 of the 64 LRDs in CEERS and 12 of 46 LRDs in JADES found by \cite{Kocevski_2025}. Our sample of LRDs in the NEP-TDF are the first to be published.

\subsection{Robust 4 $< z <$ 9 galaxy sample} 
To investigate the local environment of LRDs, we compare the clustering of LRDs and galaxies. To create a sample of comparison galaxies to use in our clustering studies, we follow the method of \cite{li2024epochspaperxenvironmental}.  Limited to the range 4 $< z <$ 9 due to the available photometric bands, we use the following criteria \citep{adams2024epochspaperiiultraviolet, conselice2024epochsidiscoverystar} to identify robust 4 $< z <$ 9 galaxies:

\begin{enumerate}
    \item The first and second bands redward of the break are $\geq$ 5$\sigma$ detected and any other bands redward of the break are $\geq$ 2$\sigma$ detected to ensure a strong Lyman-break detection. 
    \item We also require any bands blueward of the Lyman-break to not be $\geq$ 3$\sigma$ detected.
    \item The majority of the redshift probability density function (PDF) $P(z)$ must be located inside the primary peak, achieved with the criteria $\int_{0.9\times z_{phot}}^{1.10\times z_{phot}} P(z) dz \geq 0.6$.
    \item If a secondary peak exists we require it to be less than 50$\%$ of the higher probability z solution, so $P(z_{sec}) < 0.5 \times P(z_{phot})$. 
    \item For a best-fitting SED we require $\chi_{red}^{2} < 3$ for the SED fit to be considered robust.
\end{enumerate}

The number of objects in the final galaxy sample is 3685.  We later put galaxies into redshift bins for clustering analysis and restrict our galaxy sample to 4.75 $< z <$ 8.25 as described in $\S$\ref{sec:KStest}. We note that the majority of LRDs selected by the criteria in $\S$\ref{sec:LRDsample} do not meet the criteria described above.

\subsection{Redshift and number distribution of LRDs} \label{sec:redshift}

\begin{figure}
\centering
\setkeys{Gin}{draft=False}
\includegraphics[width=1\columnwidth]{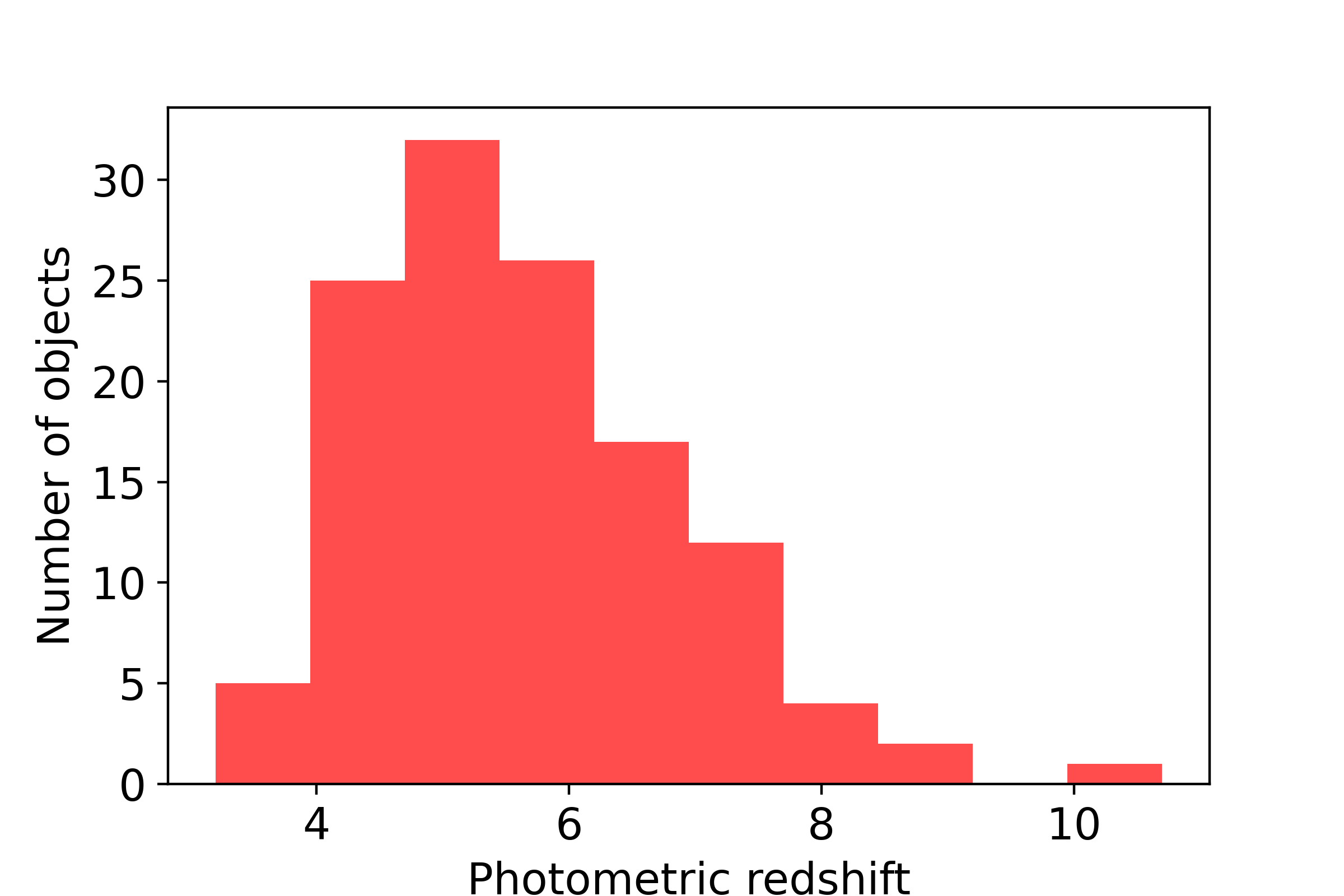}
\caption{The redshift distribution for our LRD sample.  The distribution peaks around $z \sim$ 5 and contains most LRDs in the range 4 $\lesssim z \lesssim$ 6, similar to \cite{labbe2023uca}, \cite{kokorev2024} and \cite{kocevski2023}.}
\label{fig:lrd_redshifts}
\end{figure}

\begin{figure}
\centering
\setkeys{Gin}{draft=False}
\includegraphics[width=1\columnwidth]{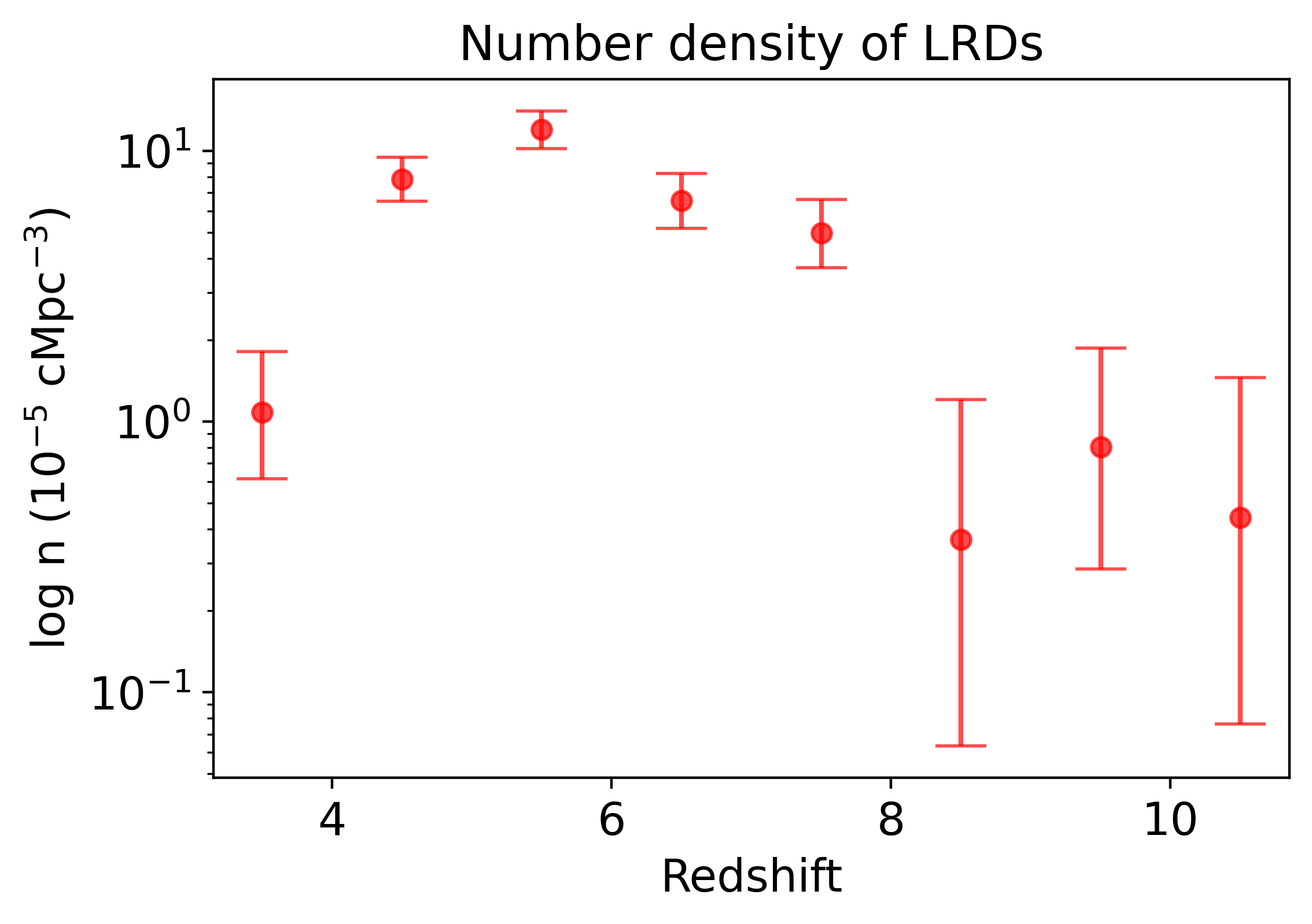}
\caption{The number density evolution of our LRD sample.  The errors on the number density are calculated assuming a Poissonian error on the count of LRDs in each bin.}
\label{fig:lrd_density}
\end{figure}

To investigate the quality of our photometric redshifts for LRDs, we look for matches of our LRDs in the DJA \citep{2024heintz, brammer_2023_8319596} for spectra and find matches for 26 out of 124 in total.  We investigate the quality of the photometric redshifts of our sample of LRDs using the same method as in $\S$ \ref{sec:red_qual}, focusing on the grade 3 redshift estimates, where grade 3 refers to spectra whose fits have been visually checked.  Of the 26 matched LRDs, 24 have grade 3 redshift estimates.  The outlier fraction of redshifts for our LRD sample is $\eta = 9/24$, or $\sim40\%$.  The NMAD for this sample is 0.112.  This is noticeably poorer than for our parent sample ($\S$\ref{sec:red_qual}). This difference in quality is likely tied to our current lack of understanding of the physics involved in LRDs. Parameter estimates produced by various SED modelling codes may make non-representative assumptions about LRDs, meaning that photometric redshift estimates are less accurate.

Most LRDs in our sample have redshift 4 $\lesssim z \lesssim$ 6, similar to \cite{labbe2023uca}, \cite{kokorev2024} and \cite{kocevski2023}, spanning a total range of 3.5 $< z <$ 10.4 as shown in \autoref{fig:lrd_redshifts}.  To calculate the number density of LRDs, we split our sample into redshift bins of size $\Delta z$ = 
1 from $z = 3$ to $z = 11$ as shown in \autoref{fig:lrd_density} and assuming a Poissonian error on the count of LRDs per bin.  We find that the number density of our LRD sample in most of these bins is $\sim10^{-5}$ cMpc$^{-3}$, in agreement with \cite{pizzati2024littlereddotsreside}.  For the bins 8 $< z <$ 9 and 10 $< z <$ 11, which contain only 1 LRD each, the number density drops to $\sim10^{-6}$ cMpc$^{-3}$.

\subsection{Broad-lines in LRDs} \label{sec:broad}
\begin{figure}
\centering
\setkeys{Gin}{draft=False}
\includegraphics[width=1\columnwidth]{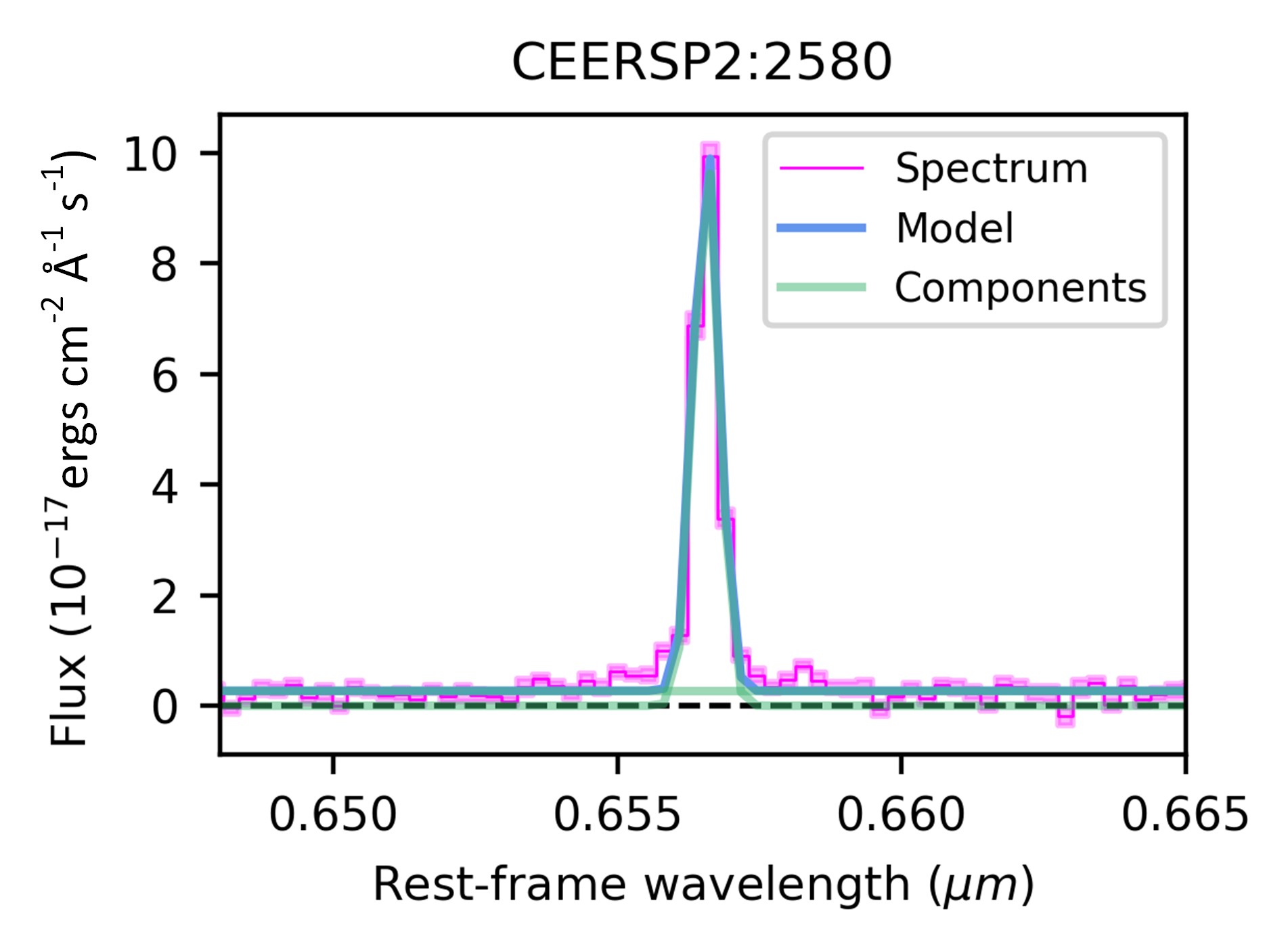}
\caption{Spectrum and model of H$\alpha$ line for CEERSP2:2580.  The best model selected for this LRD is a single component Gaussian with a full width at half maximum (FWHM) of $\sim$ 240 km/s.}
\label{fig:lrd_2580}
\end{figure}
\begin{figure}
\centering
\setkeys{Gin}{draft=False}
\includegraphics[width=1\columnwidth]{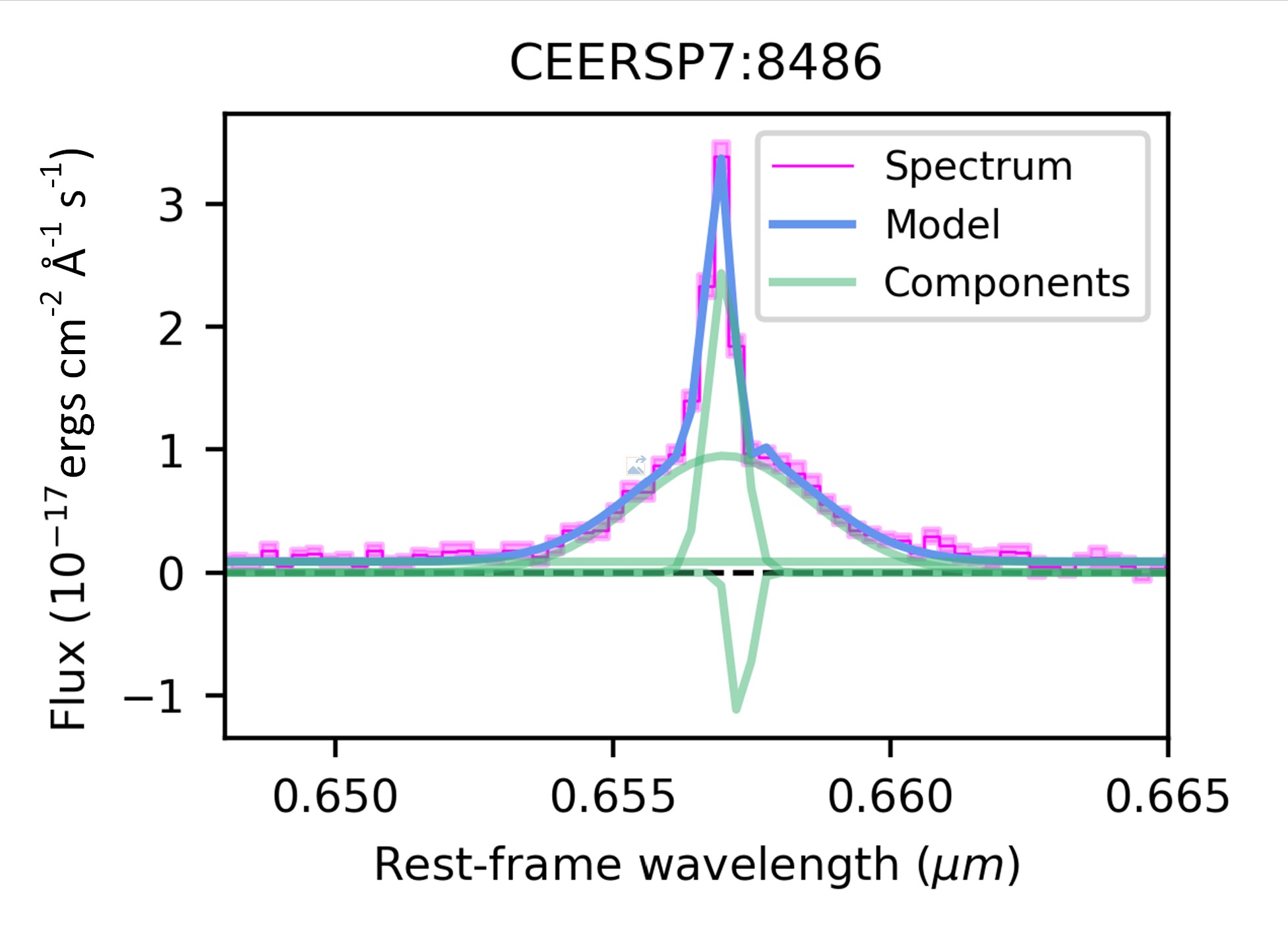}
\caption{Spectrum and model of H$\alpha$ line for CEERSP7:8486.  The best model selected is a double component Gaussian with an absorption feature.  The FWHM of the components is $\sim$ 1980 km/s and $\sim$ 560 km/s for the broad and narrow component respectively.  We note that the supposed detection of an absorption feature may instead be due to the presence of [\textrm{N}~\textsc{ii}] emission creating a small secondary peak. This was the only spectrum that displayed a possible hint of [\textrm{N}~\textsc{ii}] emission.}
\label{fig:lrd_8486}
\end{figure}
\begin{figure}
\centering
\setkeys{Gin}{draft=False}
\includegraphics[width=1\columnwidth]{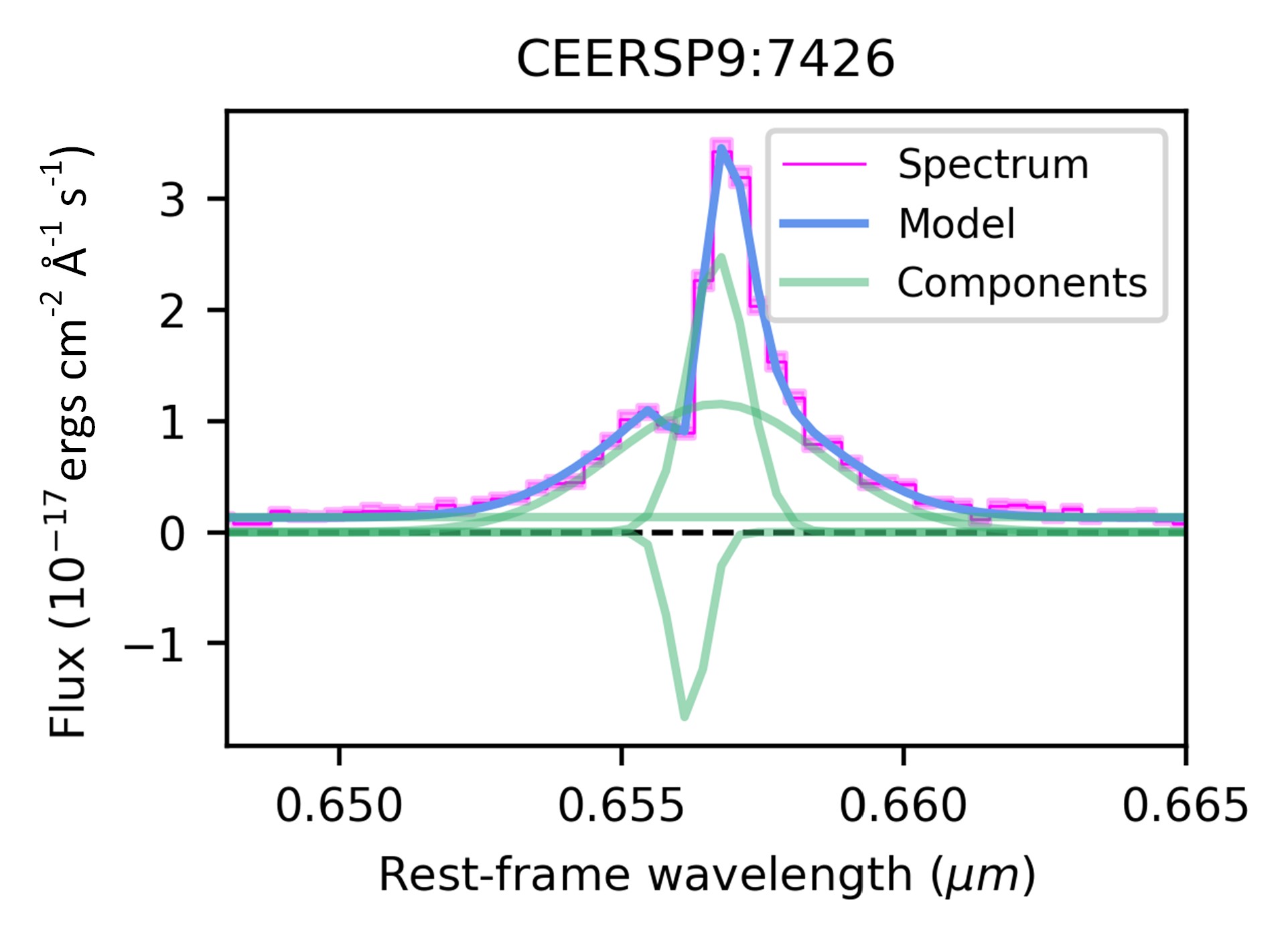}
\caption{Spectrum and model of H$\alpha$ line for CEERSP9:7426.  The best model selected is a double component Gaussian with an absorption feature.  The FWHM of the components is $\sim$ 1700 km/s and $\sim$ 320 km/s for the broad and narrow component respectively.}
\label{fig:lrd_7426}
\end{figure}
\begin{figure}
\centering
\setkeys{Gin}{draft=False}
\includegraphics[width=1\columnwidth]{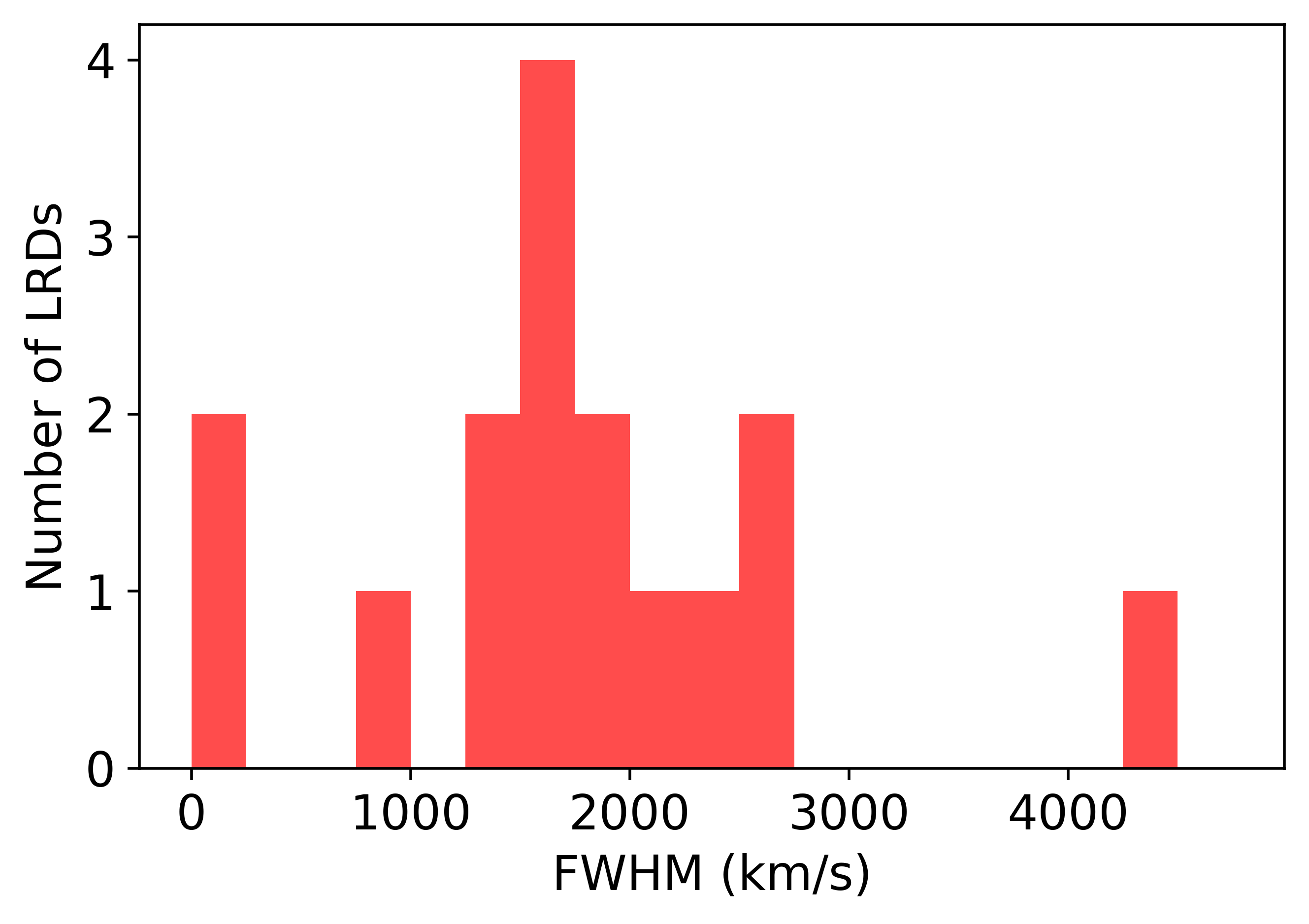}
\caption{Distribution of FWHM in km/s of the broadest component of each H-alpha line for LRDs.}
\label{fig:FWHM_dist}
\end{figure}

Whilst the presence of somewhat broad H$\alpha$ lines (2000 km s$^{-1}$ $>$ FWHM $>$ 1000 km s$^{-1}$) could be due to other processes, broader lines (2000 km s$^{-1}$ $>$ FWHM) are indicative of broad-line AGN \citep{2025habouzit}. For this reason, we investigate the broad line fraction of our sample. We use the spectra of our 26 LRDs found in the DJA to search for broad H$\alpha$ lines (FWHM $>$ 1000 km s$^{-1}$).  If there are multiple spectra of one object taken with different resolution gratings, we select the spectrum with the highest resolution grating and reject objects that only have PRISM spectra covering the wavelength range for H-alpha, as the PRISM resolution is generally not high enough to resolve broad lines.  This leaves 19 spectra. To determine the presence of broad H-alpha lines, we fit the spectra using a model with a single component Gaussian curve and a model with a double component Gaussian curve. We fix the centre of the previously mentioned Gaussian curves for simplicity.  To ensure that absorption features do not alter our results, we also use a version of the double component model that models an absorption feature as a negative Gaussian curve of variable centre. Whilst it is possible that some spectra could contain [\textrm{N}~\textsc{ii}] emission, this was not obvious upon inspection and thus was not included in the models.

\section{Results}

\begin{table*}
\centering
\caption{List of variables used in $\cigale$ modelling.  Any variables that are not listed were kept as the defaults given by $\cigale$.}
\label{tab:cigale}
\setlength{\tabcolsep}{3pt}
\begin{tabular}{|l|l|l|}
\hline \hline
Module         & Parameter  & Values  \\ \hline \hline
Star formation history          & Main population e-folding time & 100, 400, 800, 1000, 2000, 3000, 4000, 5000 Myrs \\
$\sfhdelayed$ & Main population age & 100, 200, 300, 400, 500, 750, 1000, 1250, 1500, 1750 Myrs \\ 
 & Burst population e-folding time & 10, 50, 100 Myrs \\
 & Burst population age & 10, 30, 50 Myrs \\
 & Burst fraction & 0, 0.001, 0.01, 0.1, 0.2, 0.3, 0.4, 0.5\\ \hline
Stellar population & Initial mass function & Chabrier \\
$\bc$ & & \\ \hline
AGN emission & AGN contribution & 0.1, 0.2, 0.3, 0.4, 0.5, 0.6, 0.7, 0.8, 0.9, 0.99 \\
$\skirtor$& Viewing angle & 30\deg \\
&Extinction in polar direction & 0.1, 0.5, 1, 2, 3, 4, 5, 6 [E(B-V)]\\ \hline\hline

\end{tabular}
\end{table*}

To determine whether or not an H-alpha line is broad, we compare the fits for the three models.  To select the best model, we use a $\chi^{2}$ difference test \citep{werner}, where we take the difference of the values of $\chi^{2}$ and compare this to the critical $\chi^{2}$ for the corresponding number of degrees of freedom at the 95th percentile.  If the double component Gaussian model (with or without absorption) is selected, we check the standard deviation of the two components to ensure these are not too similar and are in line with the expected values for broad and narrow lines.  We take the expected values for the full width at half maximum (FWHM) $\sim5000$ km/s to $\sim1000$ km/s and $\sim800$ km/s to $\sim100$ km/s respectively \citep{Greene_2024, 2025Ji, Baggen_2024}.  Out of 19 LRD spectra, 3 have H-alpha lines that are redshifted out of the range of both gratings and the PRISM, leaving 16 spectra that can be fit.  Of these 16, 14 are better fit by a double component Gaussian with or without an absorption feature and 2 are better fit by a single narrow Gaussian component.  One of the two single narrow lines found is shown in \autoref{fig:lrd_2580}.  We provide two examples of the fits of the double component Gaussian models with absorption features in \autoref{fig:lrd_8486} and \autoref{fig:lrd_7426}.  The distribution of FWHM of the broad line fits is shown in \autoref{fig:FWHM_dist}.  In summary, we find $\sim80\% \pm10\%$ of LRDs with grating spectra have broad-lines, in approximate agreement with \cite{Greene_2024} ($\sim75\%$).

\subsection{SED modelling} \label{sec:SEDmodel}
We investigate the composition of LRDs through SED modelling.  We compare the results of models with and without AGN components for LRDs selected from CEERS, NEP-TDF, and JADES.

\subsubsection{Fitting code}

We analyse SEDs using $\cigale$ v2022.1 \citep{burgarella2005, noll2009, boquien2019}, which models the spectrum of galaxies between the far-UV (FUV) and radio.  $\cigale$ builds composite models using templates that describe stellar populations with a flexible star formation history (SFH), emission from ionised gas, AGN emission, dust emission attenuation, and nebular emission.  To search for redshifts, $\cigale$ also has a photometric redshift mode. To determine whether CIGALE gives us different photometric redshift estimates, we run a subset of LRDs with and without MIRI data with CIGALE in photometric redshift mode, and then compare these to an identical run with fixed photometric redshifts produced by EAZY. The resulting models are not significantly affected by the redshift setting used. Due to our large sample and the resulting computation time, we keep our photometric redshifts as a fixed variable. This is also consistent with how we have carried out similar analyses with the EPOCHS sample \citep[e.g.,][]{conselice2024epochsidiscoverystar}.

The $\sfhdelayed$ SED module models a standard delayed $\tau$ SF model.  We choose this module as it includes an optional exponential burst.  We use a similar method to \cite{durodola2024} to select the range of main stellar ages for the module, where we use the range of redshifts of our LRD sample to determine the range of possible main stellar ages.  We make use of the $\bc$ module \citep{bruzualcharlot} to model a simple stellar population and set the IMF to \cite{Chabrier_2003}.  Unlike the $\mara$ module \citep{2005Maraston}, the alternative stellar population model available in $\cigale$, the $\bc$ module can be combined with the $\nebular$ module.  The choice of IMF is known to have a strong impact on stellar mass estimates.  We select a \cite{Chabrier_2003} IMF as it includes fewer low-mass stars than a Salpeter IMF, reducing stellar mass, and seems to be somewhat more in line with available data \citep{van_Dokkum_2008, cappell}.

To model nebular emission we employ $\nebular$, whose nebular templates are based on \cite{inoue_nebular}.  To model dust attenuation we use $\dustattmodifiedstarburst$, which follows the \cite{calzetti} starburst attenuation curve.  We select this dust attenuation module as it allows the E(B - V) of the continuum to be varied with one factor.  We briefly investigated dust attenuation with higher E(B - V) values but found that this significantly worsens the quality of the fits.  For this reason, we do not change the standard $\cigale$ values for $\dustattmodifiedstarburst$.  To model dust emission we apply the $\dl$ module that is based on the \cite{draine2007} models and updated in \cite{draine2014}.  This is one of the more up-to-date dust emission modules and focuses on dust in the galaxies.  We select this module as it does not contain an AGN component, which allows an AGN to be modelled as a separate module.  

Finally, to model an AGN component we use the clumpy AGN $\skirtor$ model \citep{2012skirtor, 2016skirtor} SED module, which has controls for the gradient of dust density with both angle and radius, and the fraction of the total dust mass contained in clumps. The $\skirtor$ module also includes parameters for the opening angle of the torus, the edge-on optical depth at 9.7 $\mu$m, polar dust extinction, and allows for both Type I and II AGN. However, we find that allowing Type II AGN causes $\cigale$ to select the lowest AGN fraction and the highest polar dust extinction that we allow. For this reason, as well as because of the high broad line fraction of our LRD sample, we model AGN of Type I only.  Note that these models are for those AGN with central tori, and thus other physical models of AGN, including those which are not understood or fully modelled, may give different SED shapes and forms. 

To create fits we combine the $\sfhdelayed$, $\bc$, $\nebular$, $\dustattmodifiedstarburst$, and $\dl$ SED modules.  We run the fitting procedure twice, once with $\skirtor$ to model AGN, and once without.  Following the example of \cite{durodola2024}, we only allow a small set of variables to vary.  Our choice of variables we vary is listed in \autoref{tab:cigale}.  Any other variables are kept the same as the default single value given by $\cigale$. 

Due to the shallow nature of MIRI, many objects that are $ > 5 \sigma$ detected in NIRCam data are not detected in MIRI data.  HST data also contain non-detections due to the Lyman break of objects in our sample.  To create our $\cigale$ fits, we treat any bands with $ < 5\sigma$ detections as upper limits.

\begin{figure*}
    \centering
    \includegraphics[width=\textwidth]{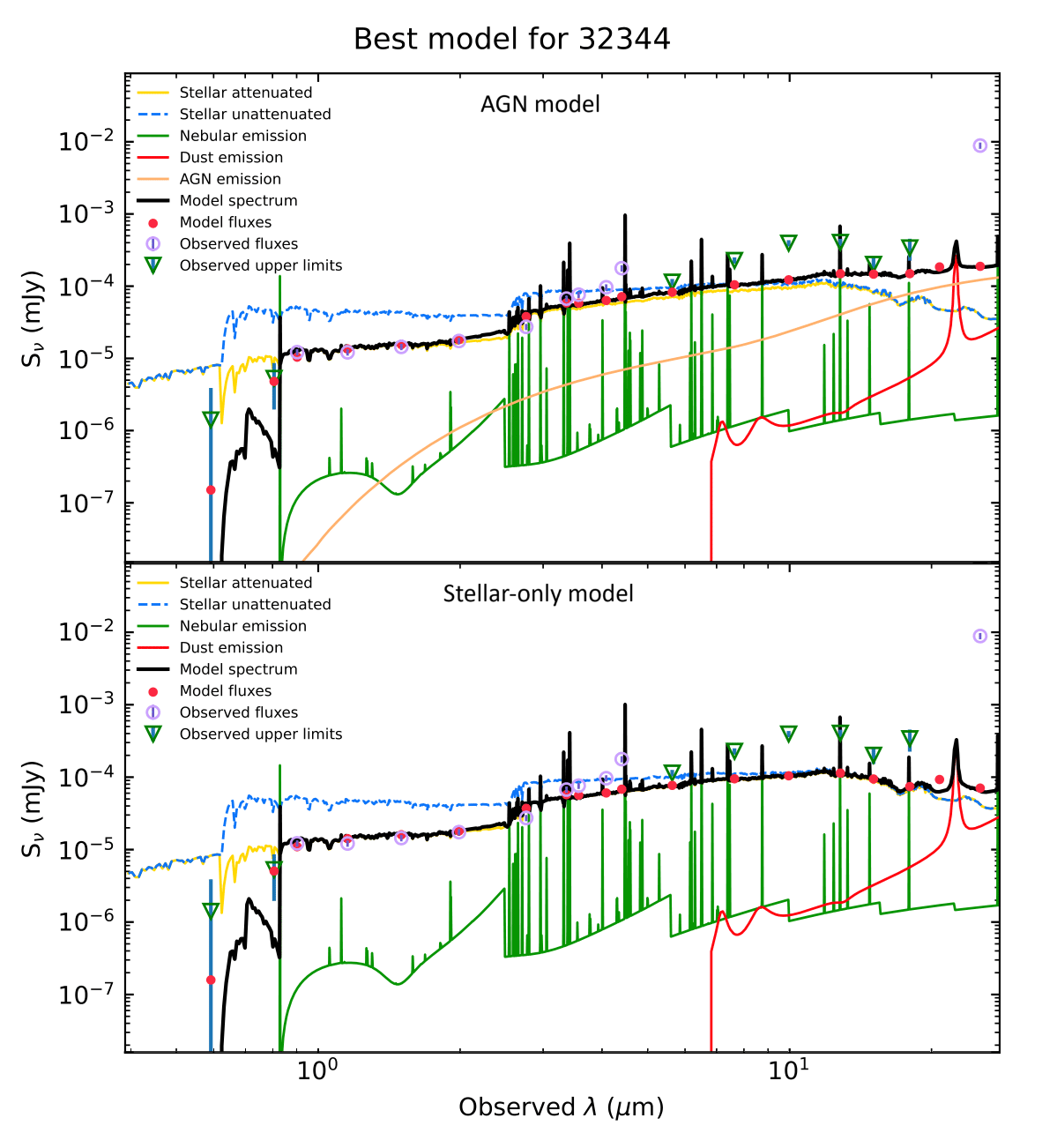}
    \caption{AGN (top) and non-AGN (bottom) $\cigale$ models for JADES:32344 ($z=5.82$), which has MIRI coverage.  Note that six of the longest wavelength model flux points corresponding to MIRI data are treated as upper limits, whilst the F2550W band is treated as an observed flux.  Both models produce a stellar mass of $\sim 10^{9.3} M_{\odot}$.  The AGN IR luminosity fraction for this LRD is f$_{AGN} = $ 0.9.}
    \label{fig:SED_JADES}
\end{figure*}

\begin{figure*}
\centering
\includegraphics[width=\textwidth]{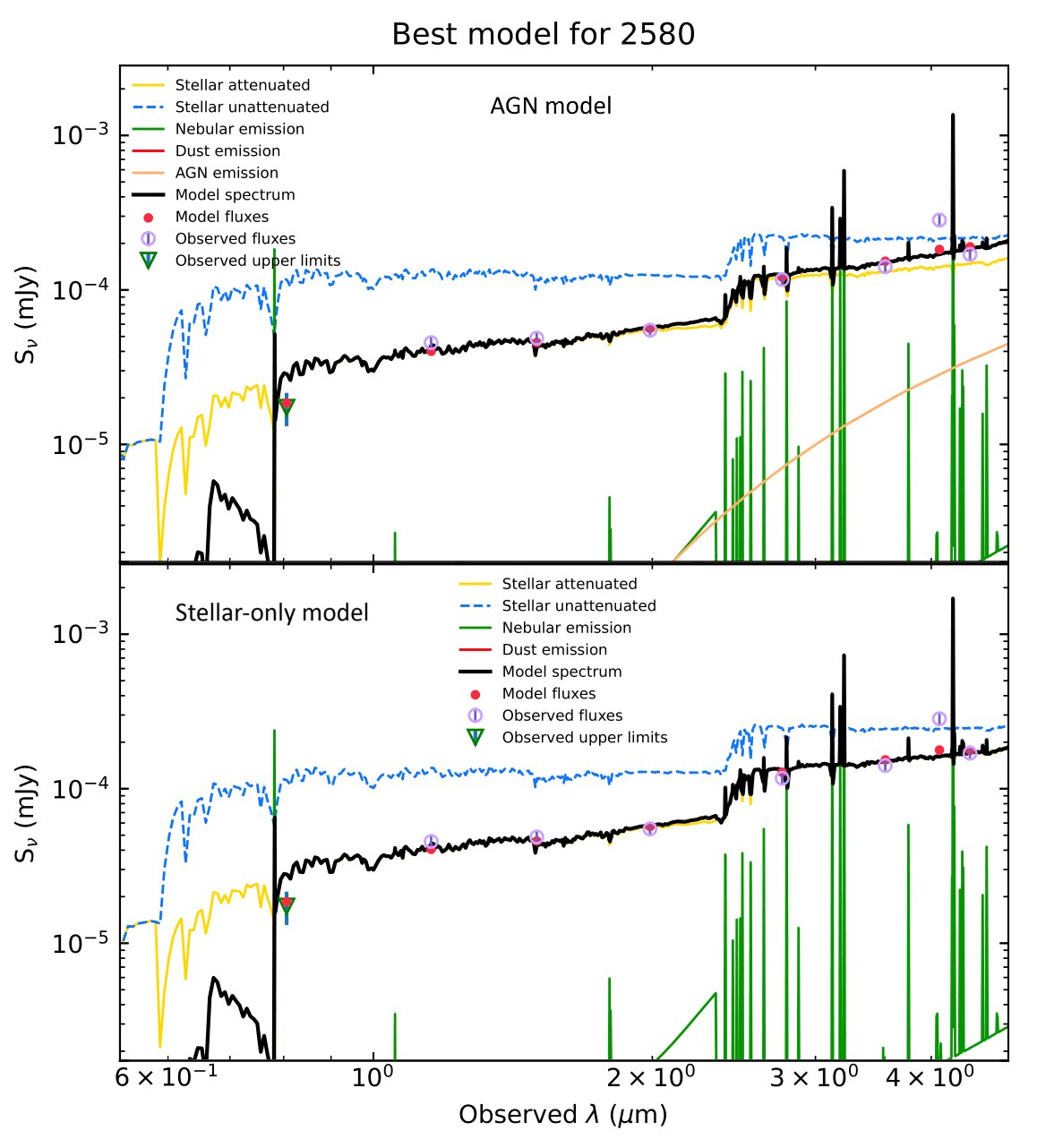}
\caption{AGN (top) and non-AGN (bottom) $\cigale$ models for CEERS:2580, which has redshift $z = 5.43$.}
\label{fig:SED_CEERS}
\end{figure*}

\subsubsection{AGN vs. non-AGN model composition}

\begin{figure}
\centering
\includegraphics[width=1\columnwidth]{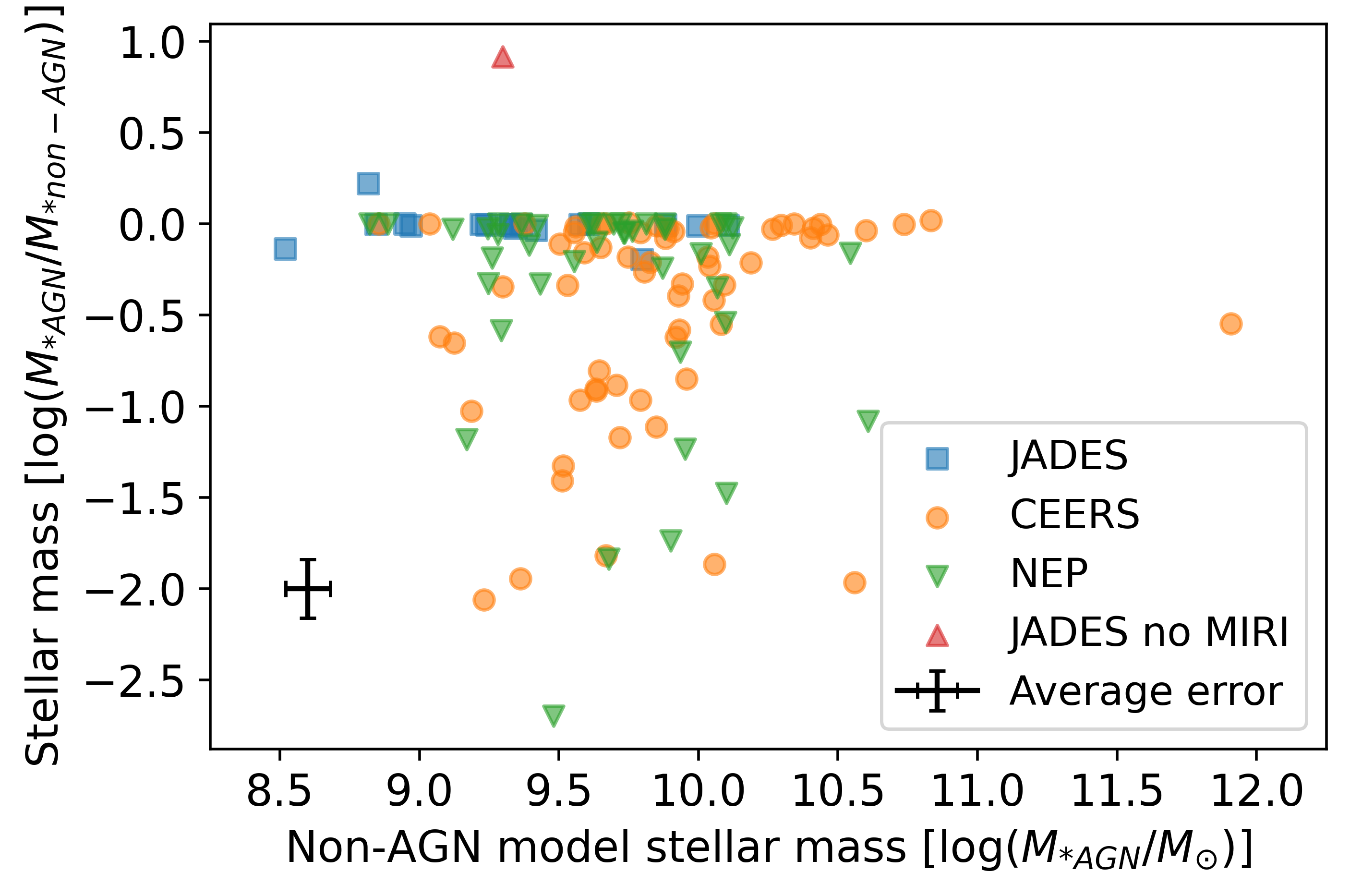}
\caption{The stellar mass calculated by $\cigale$ for non-AGN models compared to the difference in stellar mass for AGN and non-AGN models, expressed logarithmically.  The average error is included in the bottom left corner.   The stellar mass of the non-AGN model is higher for all but two LRDs in JADES and two LRDs in CEERS.}
\label{fig:stellarmass}
\end{figure}

\begin{figure}
\centering
\includegraphics[width=1\columnwidth]{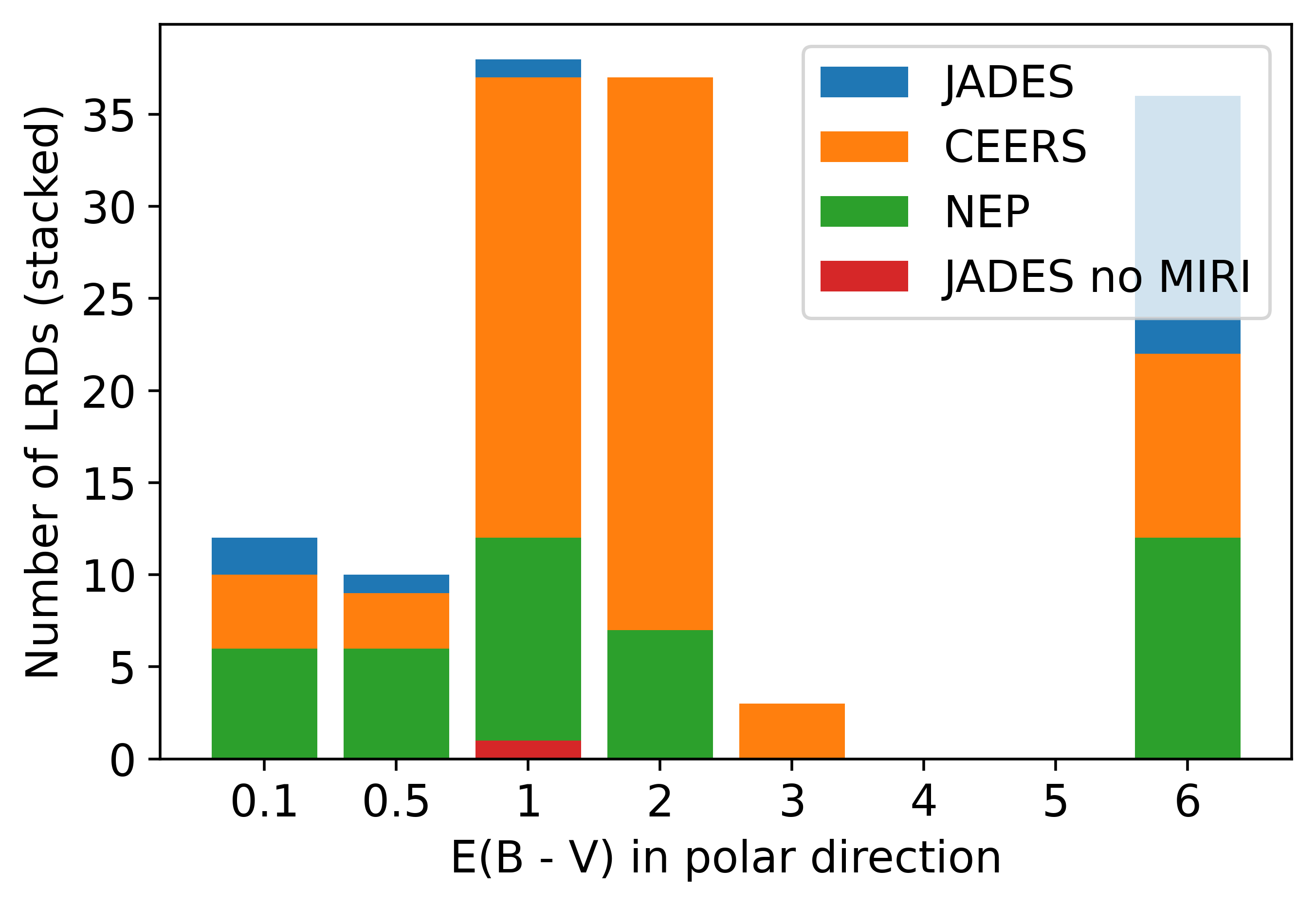}
\caption{Stack plot of the dust extinction calculated by $\cigale$ for AGN models.  Most AGN models are fit by a dust extinction close to extremes (0.1 and 6) of the allowed values.}
\label{fig:ebv}
\end{figure}

We briefly investigate the composition of the models created by $\cigale$, examples of which are shown in \autoref{fig:SED_JADES} and \autoref{fig:SED_CEERS}.  A comparison of the stellar mass of the AGN and non-AGN models is shown in \autoref{fig:stellarmass}.  All of our LRDs, save two in the JADES and two in the CEERS fields, produce a higher stellar mass for non-AGN models.  A total of 61 LRDs ($\sim 50 \%$) have a mass difference less than $\log_{10}(M_{* non-AGN}/M_{\odot})-\log_{10}(M_{* AGN}/M_{\odot}) = $ 0.1.  This includes all except two JADES LRDs with MIRI data. The single LRD in JADES without MIRI data has a higher stellar mass when using an AGN component. This could be partly due to the greater depth of NIRCam data of JADES, but the reason for this outlier is not clear.

The dust extinction of AGN models is shown in \autoref{fig:ebv}.  Most AGN models have a dust extinction at extremes of the allowed values with E(B - V) = 0.1, 0.5, 1, 2, or 6, with only three LRD in CEERS having E(B - V) = 3.  No LRDs have E(B - V) = 4 or 5. We note that allowing even higher values of E(B - V) typically results in LRDs with E(B - V) moving to these higher values. For some of the LRDs, the unphysically high dust masses might discount tori surrounded AGN.

\begin{figure}
\centering
\includegraphics[width=1\columnwidth]{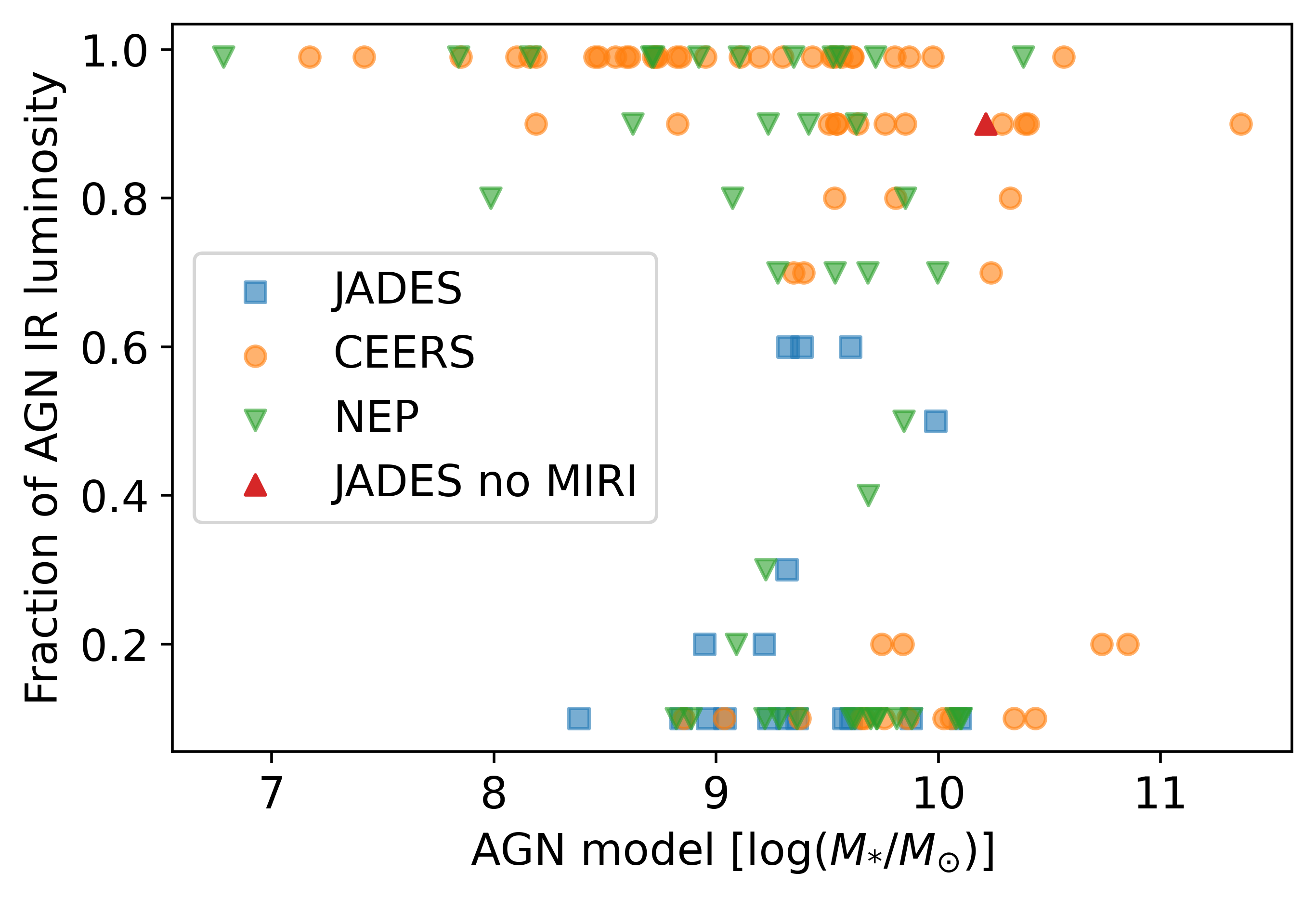}
\caption{Comparison of the stellar mass with the fraction of AGN IR luminosity to total IR luminosity f$_{AGN}$. The fraction f$_{AGN}$ is quantised because of the choice of values allowed.  We find no correlation between the fraction of AGN and stellar mass.}
\label{fig:stellarmass_AGN}
\end{figure}

We compare the stellar mass in LRDs with the fraction of AGN IR luminosity to total IR luminosity in \autoref{fig:stellarmass_AGN}.  We find no correlation between the fraction of AGN IR luminosity and the stellar mass.  Most fractions of AGN produce similar ranges of stellar masses for LRDs. However, the highest fraction of AGN IR luminosity ( f$_{AGN} = $ 0.99) produces a larger range of stellar masses than other fractions.

We note the $\chi_{r}^2$ given by $\cigale$ is not necessarily the most suitable value of $\chi_{r}^2$, as it is calculated using the number of photometric bands available to determine the number of degrees of freedom.  There is also some difficulty associated with determining the number of orthogonal parameters.  For this reason, we utilise the Bayesian information criterion (BIC) to determine whether AGN or non-AGN models are more suitable, rather than comparing the $\chi^2_r$ of models.

\subsubsection{Bayesian Information Criterion}

\begin{figure}
\centering
\includegraphics[width=1\columnwidth]{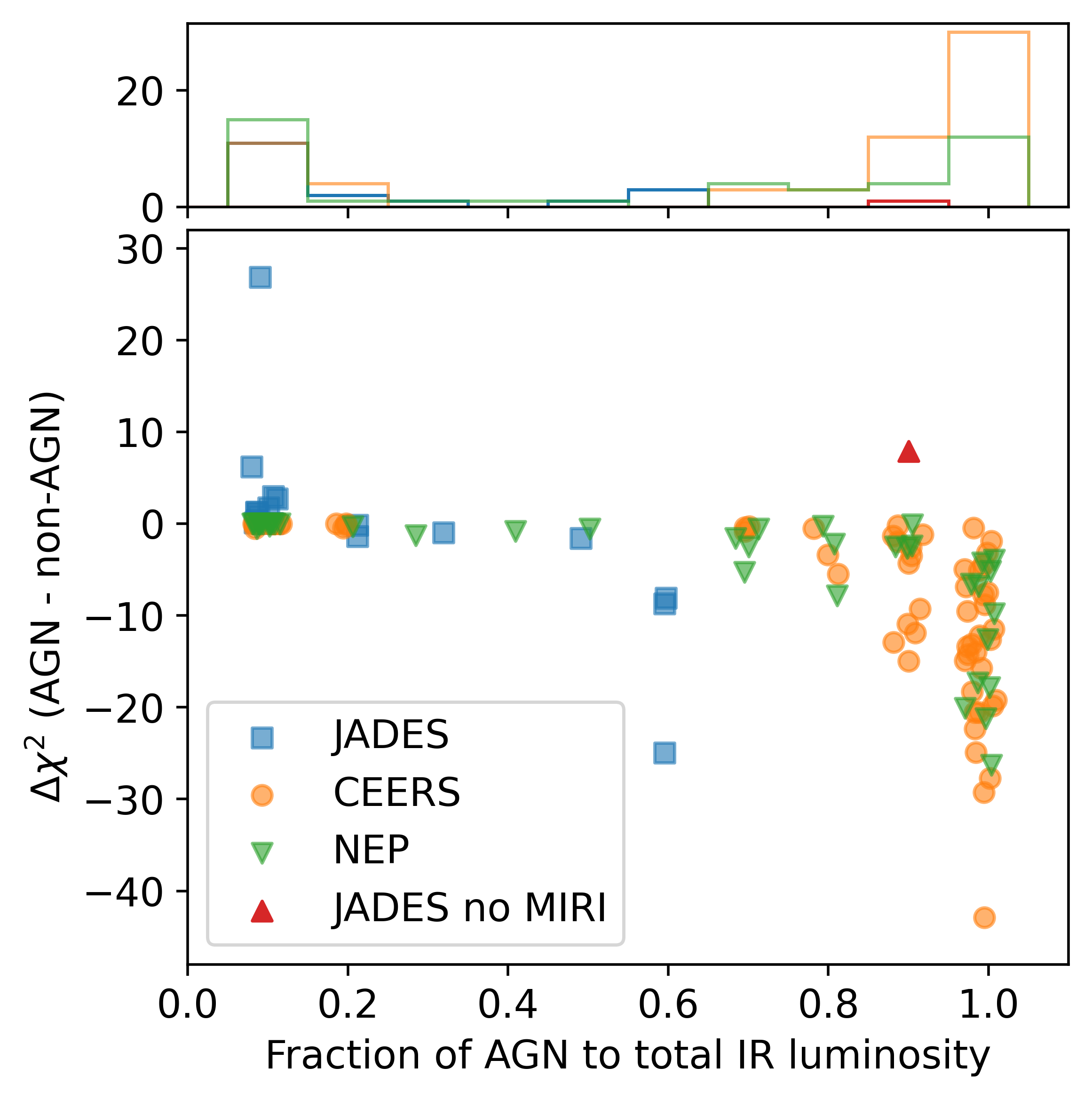}
\caption{Comparison of $\Delta\chi^{2}$ across CEERS, NEP-TDF and JADES with the fraction of AGN IR luminosity to total IR luminosity.  Note that a jitter has been added to distinguish data points.  The red dot is the LRD in JADES without MIRI data.  Note that some values of $\chi^{2}$ remain effectively the same. The LRDs with the highest AGN fractions tend to have a considerably improved $\chi^{2}$ compared to non-AGN models.}
\label{fig:chi}
\end{figure}

\begin{figure}
\centering
\includegraphics[width=1\columnwidth]{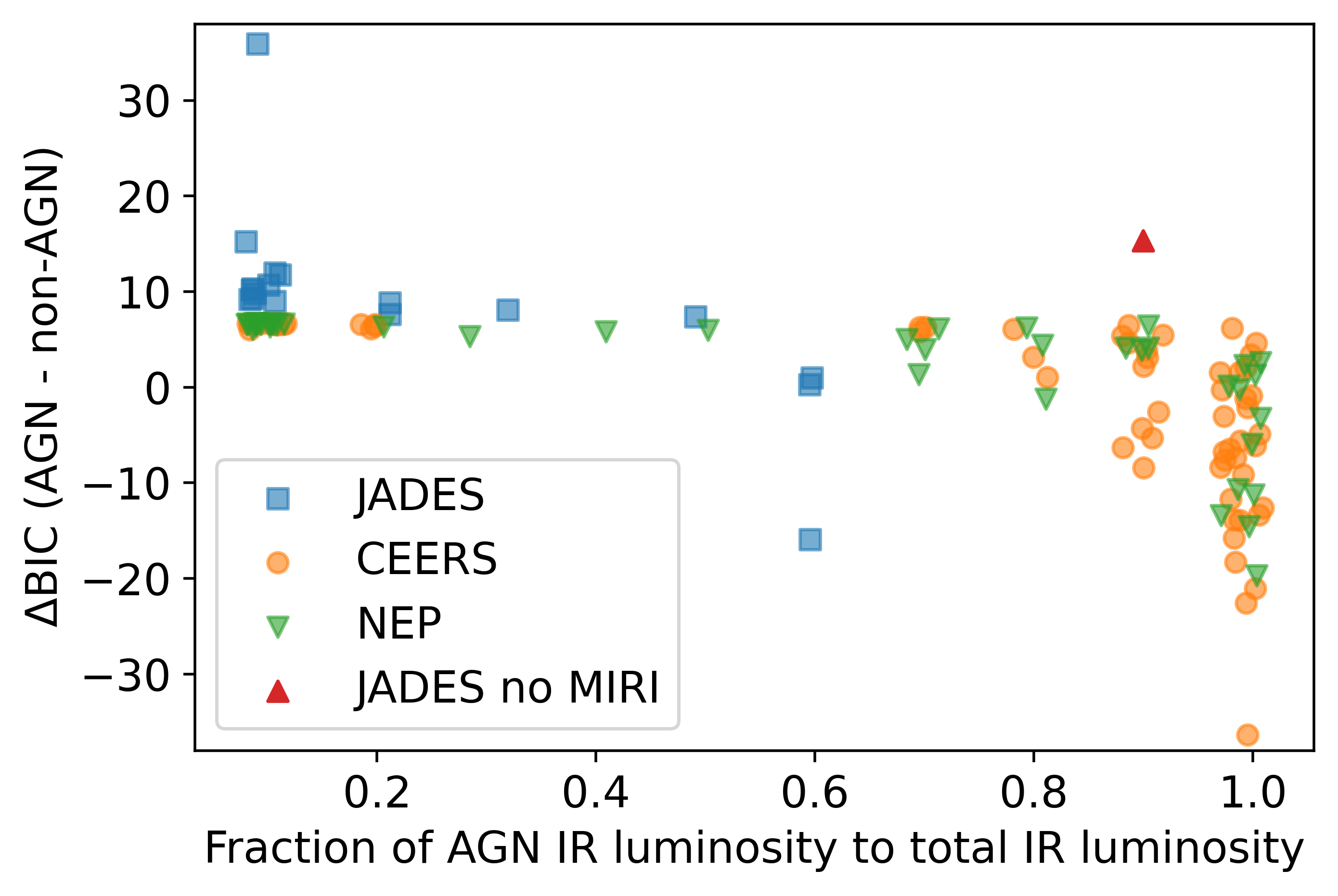}
\caption{Comparison of the BIC with the fraction of AGN IR luminosity to total IR luminosity.  Note that a jitter has been added to distinguish data points.  Around $\sim$ 70 $\%$ of pairs of AGN and non-AGN models have a positive $\Delta$BIC value.  For LRDs with MIRI data, this is $\sim 9$.  In comparison, CEERS and NEP-TDF LRDs tend to have a lower $\Delta$BIC than JADES LRDs.  Most of the values for CEERS and NEP-TDF LRDs with lower AGN fractions are  $\sim$ 6.}
\label{fig:bic}
\end{figure}

Due to the difficulty in determining the number of orthogonal degrees of freedom of the $\cigale$ models, we make use of the Bayesian information criterion (BIC) to determine whether adding the $\skirtor$ SED module results in the overfitting of LRD SEDs.  The BIC is used in model selection and penalises models based on the number of free parameters used. When comparing BICs between models, a lower BIC is generally preferred. For two given models, a difference $>$ 3.2 between the BICs indicates a substantial preference for the model with the lower BIC, and a difference $>$ 10 indicates a very strong preference \citep{Raftery1995}.
The BIC is given by
\begin{equation} \label{eq11}
    BIC = k\log N + \chi^{2}
\end{equation}
where $k$ is the number of free parameters, $N$ is the number of photometric bands, and $\chi^{2}$ is defined as usual.  This particular expression for the BIC relies on the assumption that the errors are independent of each other and distributed as identical Gaussian distributions.  The AGN model we use has eight free parameters, whilst the non-AGN model has five free parameters.  Out of our sample of 124 LRDs, we find 93 LRDs have a lower $\chi^{2}$ for their AGN models (\autoref{fig:chi}).  However, 84 out of 124 LRDs (68$\%$) have a lower BIC for the non-AGN model than for the corresponding AGN model (\autoref{fig:bic}).

Most $\chi^{2}$ values are higher for the non-AGN models, and when $\chi^{2}$ is higher for the AGN model, it is only slightly higher than the $\chi^{2}$ value for the corresponding non-AGN model (\autoref{fig:chi}).  There is also a clear preference for models with higher fractions of AGN IR luminosity, where $\Delta\chi^{2}$ tends to a larger range of more negative values.  

Around $\sim$ 70 $\%$ or 84 out of 124 pairs of AGN and non-AGN models have a positive $\Delta$BIC value, with most of the values for fields with MIRI data around $\sim 9$ (\autoref{fig:bic}). As BIC penalises models with more parameters, positive values of $\Delta$BIC suggest that most LRDs may be overfit by AGN components. Of these 84, 72 of them show substantial evidence of overfitting with $\Delta$BIC $>$ 3.2, and 8 of these show very strong evidence with $\Delta$BIC $>$ 10. Notably, these 8 are all in JADES. Generally, CEERS and NEP-TDF LRDs tend to have a lower $\Delta$BIC than JADES LRDs. This is to be expected, as an AGN component is expected to have a larger impact when fitting MIRI data than NIRCam data, thus broadening the gap between BIC values when MIRI data is used. Most of the values for CEERS and NEP-TDF LRDs with lower AGN fractions are $\Delta$BIC $\simeq$ 6.

We briefly investigate if there exists a relation between the broadness of the H-alpha line in LRDs and the BICs of our AGN fits. We may expect the very broadest of lines to be a sign of AGN, and so expect that these would be better fit by SED models with an AGN component. However, no relationship or cut-off is found between the FWHMs of the broadest line component and either the $\Delta$BICs or BICs of the AGN fits.

\subsection{Clustering of LRDs} \label{sec:LRD_cluster}

\subsubsection{Local environment} \label{sec:loc_env}
To study the local environment of galaxies and LRDs we follow \cite{li2024epochspaperxenvironmental} and use the nearest neighbour method.  We make use of the k-dimensional tree (KDTree) data structure to search for the nearest neighbours of objects and determine the separation between them.  We use the term 'objects' to refer to both LRDs and galaxies.  We use the term 'galaxy' to exclude all LRDs to avoid confusion.  For any given object we search for a nearest neighbour in the set of all objects, rather than searching for the nearest neighbour in the subset of like objects.  We constrict our nearest neighbour search to a specific maximum redshift offset $\Delta z$ relative to the object in question.  We define a redshift offset mask as:
\begin{equation} \label{eq9}
\mathcal{M}_{redshift}  = \begin{cases}
            1 & \text{if }  \Delta z < \text{0.2} \\
      0 & \text{otherwise.}
         \end{cases}
\end{equation}
The value 0.2 is chosen to ensure that $\Delta z$ is larger than the NMAD of the photometric redshifts for LRDs, which is found to be 0.112 (see \S \ref{sec:redshift}). Any objects falling outside of this mask relative to the object in question are not included in the nearest neighbour search. 

\subsubsection{Impact of Image Depth and Borders}
As the nearest neighbour(s) of objects close to the edges of an image may fall outside the image itself, the local density of these objects is usually underestimated.  Whilst this should have little impact on the comparison between LRDs and galaxies, accounting for image edges will give a somewhat more complete measure of the local environment around LRDs.
To reduce the impact of image borders we simply exclude objects that are expected to be significantly affected.  We calculate the shortest distance to the boundary for each object, $d_{edge}$.  We define a mask for objects that are affected by the boundary:
\begin{equation} \label{eq12}
\mathcal{M}_{edge}  = \begin{cases}
            0 & \text{if }  d_{edge} < \text{1cMpc} \\
      1 & \text{otherwise.}
         \end{cases}
\end{equation}
Objects falling within this mask are not included in the sample.

The local galaxy density in units of galaxies per Mpc$^{2}$, $\Sigma_{n}$, is then given by
\begin{equation} \label{eq10}
\begin{split}
\Sigma_{n} = \frac{n}{\pi d_{n}^{2}}
\end{split}
\end{equation}
where $d_{n}$ is the projected distance to the $n^{th}$ nearest neighbour in Mpc.  Whilst $\Sigma_{5}$ is typically used to study galaxy clusters \citep{lopes2016}, we opt for $n = 5$ following the method given by \cite{li2024epochspaperxenvironmental} and use it to compare the local density surrounding LRDs to the local density of galaxies, and do not require $n$ to be less than the number of objects in a cluster. 

\begin{figure}
\centering
\setkeys{Gin}{draft=False}
\includegraphics[width=1\columnwidth]{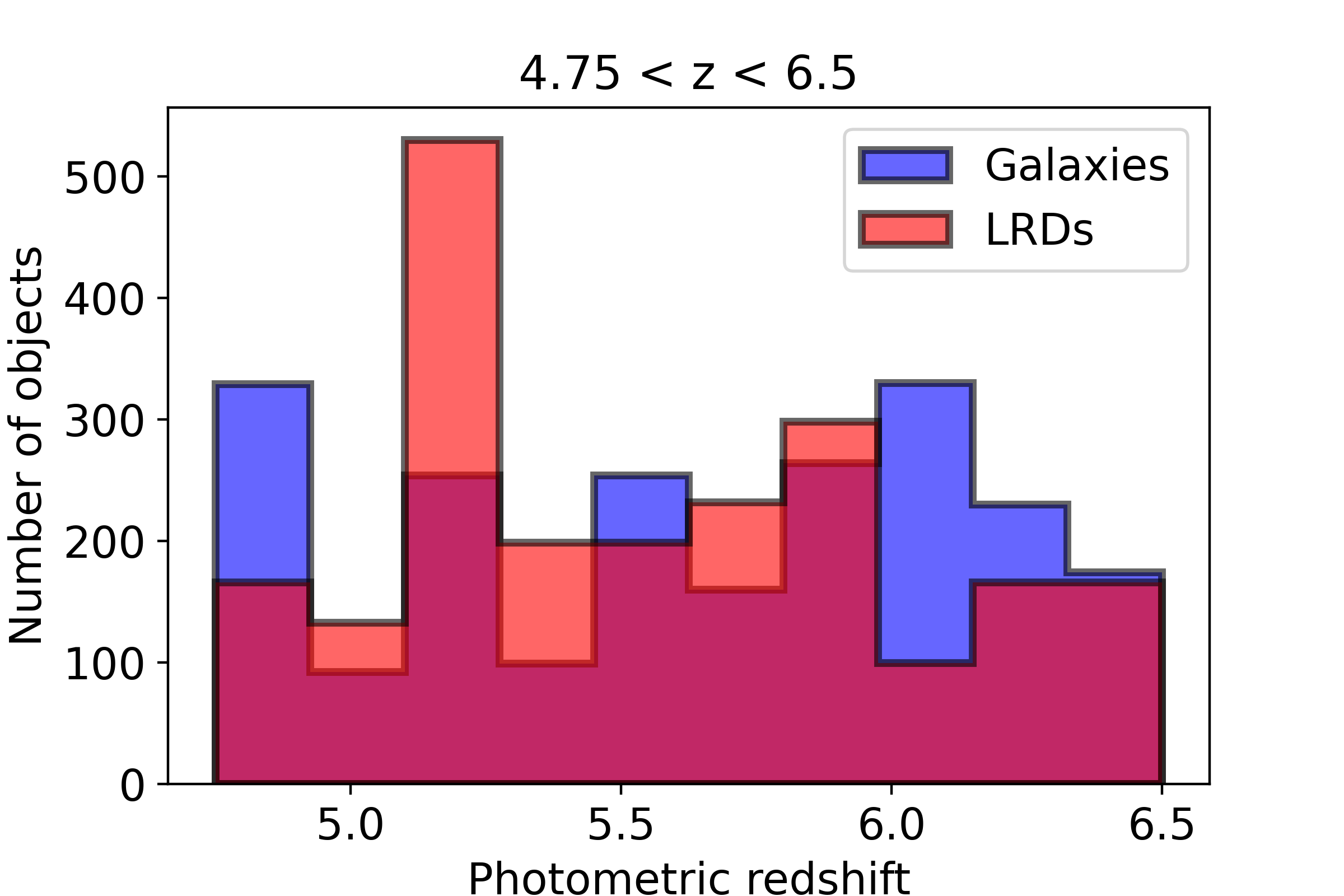}\\
\vspace{0.5cm}
\includegraphics[width=1\columnwidth]{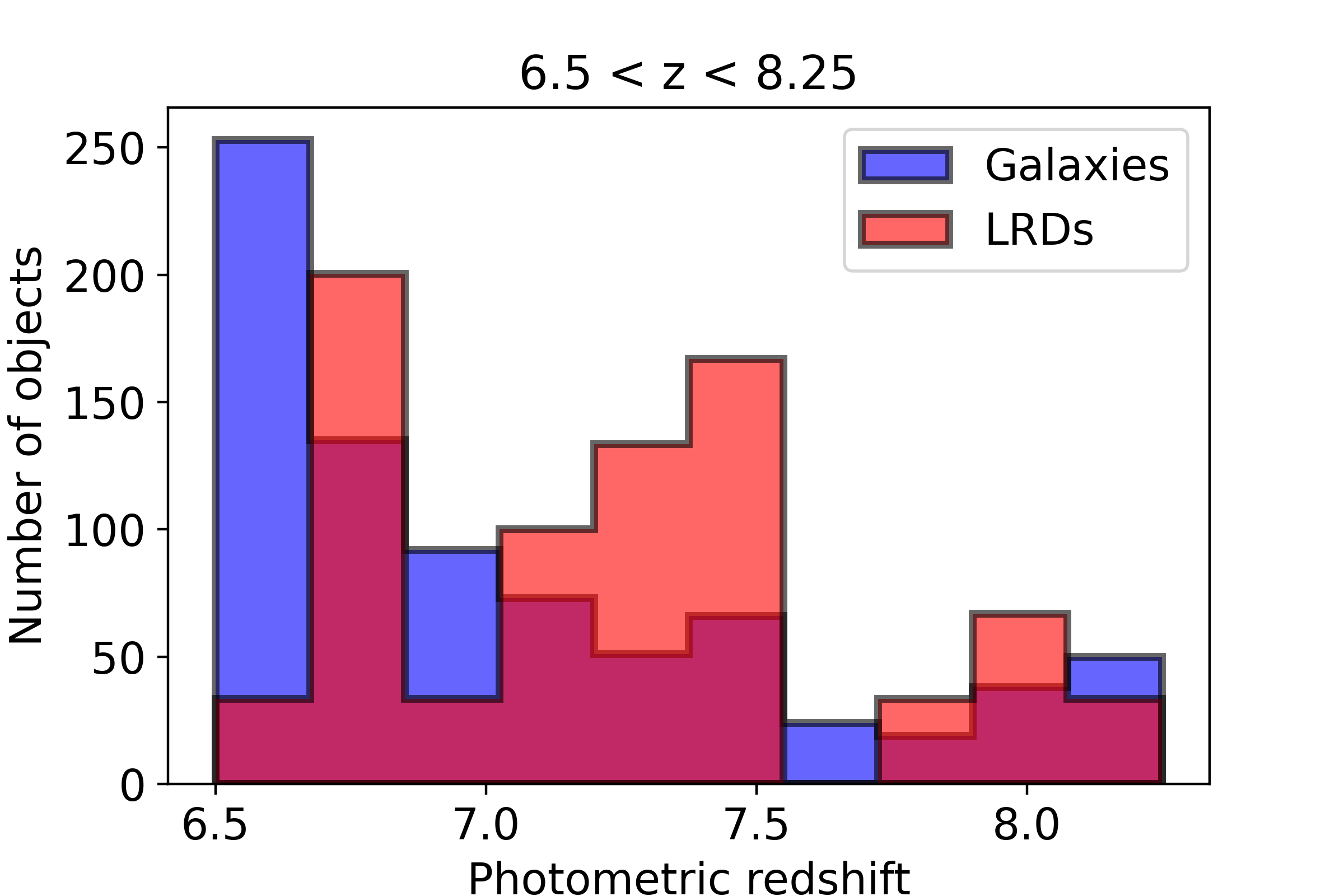}
\caption{Distributions of LRD and galaxy redshifts for the 4.75 $ < z <$ 6.5 and 6.5 $ < z < $ 8.25 bins.  Note that the LRD histograms are weighted to be normalised compared to the the galaxy histograms for ease of visual comparison.  The redshift distribution of LRDs and galaxies is roughly similar, especially for the 4.75 $< z <$ 6.5 bin.}
\label{fig:redshifts}
\end{figure}

We combine the data from the three fields and divide the samples of galaxies and LRDs into the redshift bins 4.75 $< z <$ 6.5 and 6.5 $< z <$ 8.25, described in \autoref{tab:redshift}.  We choose these bins to include the largest possible sample of LRDs and to ensure that the distributions of LRDs and galaxies are similar as seen in \autoref{fig:redshifts}, thus minimising biases to our nearest neighbour search. Ideally, stellar masses should be matched to avoid further biases \citep{matthee2024environmentalevidenceoverlymassive}, but due to the difficulty in determining the stellar masses of LRDs through SED fitting and the necessity of choosing between AGN and non-AGN models, this is not possible. We find the $\langle\Sigma_{5}\rangle$ of non-LRD galaxies and LRDs to be $14.91_{-0.65}^{+0.79}$ cMpc$^{-2}$ and $9.56_{-1.38}^{+1.51}$ cMpc$^{-2}$ respectively, at 4.75 $< z <$ 6.5, and for 6.5 $< z <$ 8.25, $7.80_{-4.56}^{+4.40}$ cMpc$^{-2}$ and $4.65_{-1.83}^{+1.63}$ cMpc$^{-2}$ respectively.  We show the $\Sigma_{5}$ distribution of both categories in \autoref{fig:cluster}. We also calculate and plot $\langle\Sigma_{5}\rangle$ for random points with the same redshift distribution as LRDs and the same number density as galaxies and LRDs combined. The $\langle\Sigma_{5}\rangle$ for random points are 10.51 cMpc$^{-2}$ and 3.48 cMpc$^{-2}$ at redshift 4.75 $< z <$ 6.5 and 6.5 $< z <$ 8.25 respectively. For LRDs, this distribution is high-end deficient and significantly more common than the galaxy distribution at the very lowest $\Sigma_{5}$ values.

The fields used in this study are not completely homogeneous in depth, which in theory could reduce the $\langle\Sigma_{5}\rangle$ by a small amount compared to completely homogeneous fields. However, the magnitude limit for our galaxy selection is significantly brighter than the brightest variation of the depths across each of the three fields which makes this variation unimportant for our analysis.

\subsubsection{Distribution testing} \label{sec:KStest}
\begin{table*}
\centering
\caption{List of redshift bins and number of objects for galaxies and LRDs per field and in total.  The bins were chosen to maximise the sample sizes whilst ensuring similar redshift distributions so as to avoid skewing $\Sigma_{5}$ measurements and distributions.}
\label{tab:redshift}
\begin{tabular}{|l|llll|llll|}
\hline \hline
Redshift              & \multicolumn{4}{l}{Number of LRDs} & \multicolumn{4}{|l|}{Number of galaxies}   \\ \hline \hline
                          & CEERS     & NEP-TDF      & JADES & Total     & CEERS        & NEP-TDF       & JADES   & Total  \\ \hline
4.75 $< z <$ 6.5 & 33      & 24       & 10      & 67   & 840            & 869         & 498    & 2207     \\ \hline
6.5 $< z <$ 8.25 & 10         & 10       & 4      & 24    & 283            & 273         & 253  & 809 \\ \hline   \hline
\end{tabular}
\end{table*}

\begin{figure}
\centering
\setkeys{Gin}{draft=False}
\includegraphics[width=1\columnwidth]{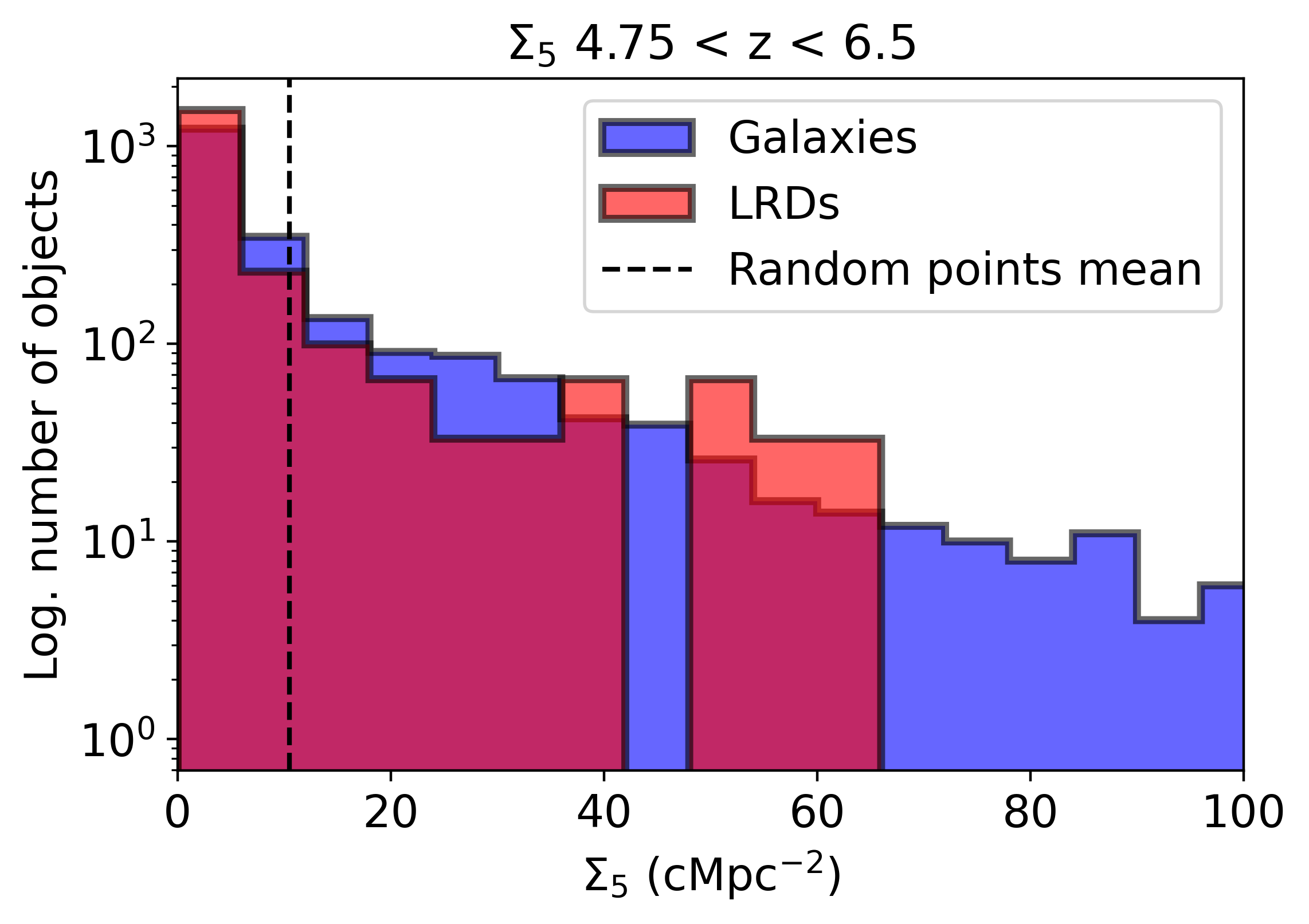}
\includegraphics[width=1\columnwidth]{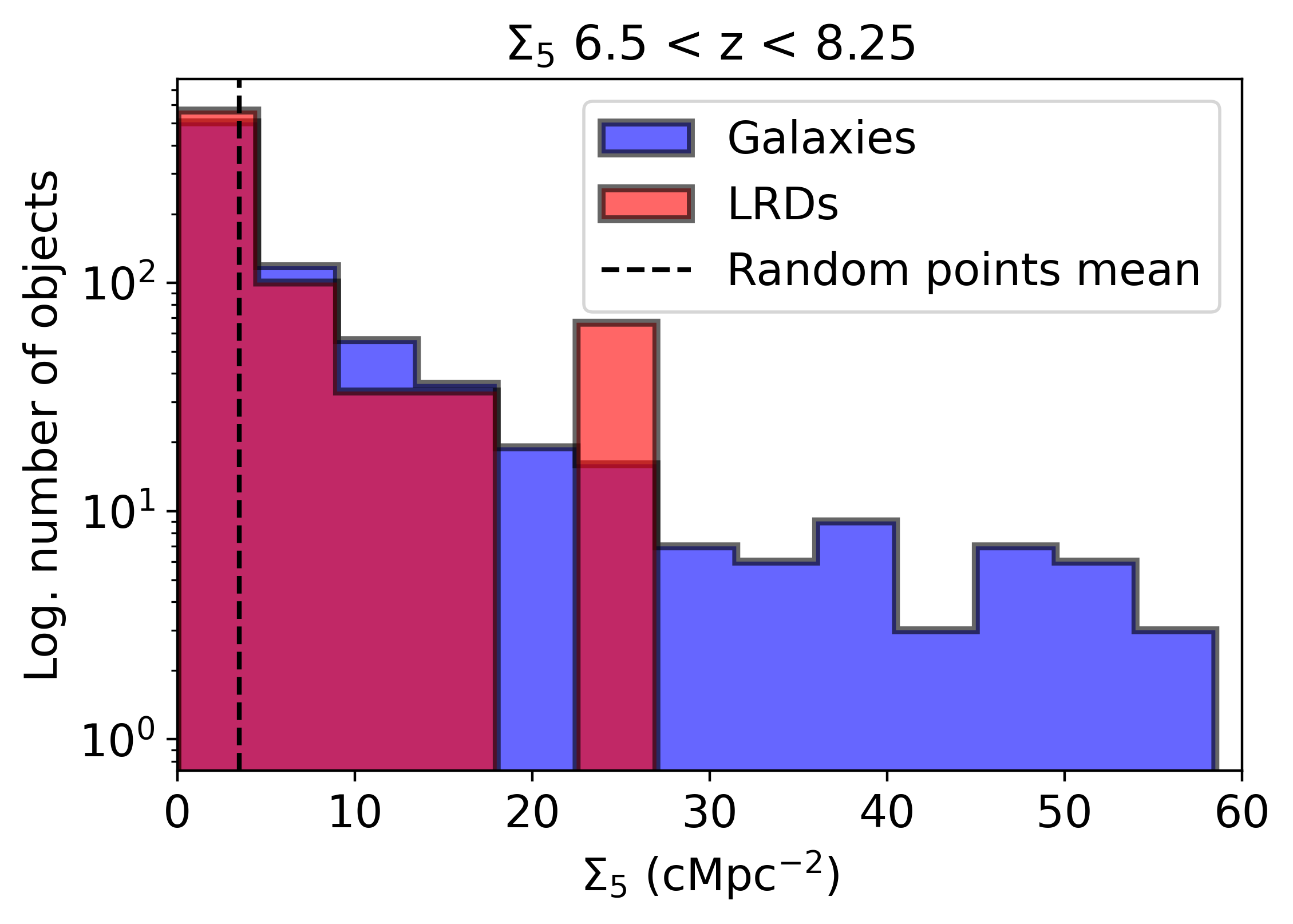}
\caption{Comparison of $\Sigma_{5}$ for galaxies and LRDs for redshift bin 4.75 $ < z < $ 6.5 (top) and 6.5 $ < z < $ 8.25 (bottom). The black dashed line represents the $\langle\Sigma_{5}\rangle$ of random points. Note that the LRD histograms are weighted to be normalised compared to the galaxy histograms and have a logarithmic y-axis.  LRDs occupy a smaller range of lower densities than galaxies do and are not found in dense environments.}
\label{fig:cluster}
\end{figure}

We apply a Kolmogorov-Smirnov (K-S) test using $\scipy$ to test and compare the distributions of $\Sigma_{5}$ of LRDs and non-LRD galaxies.  K-S tests can be used to determine whether one sample came from a given probability distribution, or in the case of a two sample K-S test, whether two samples came from the same parent distribution. Our choice of redshift bins 4.75 $< z <$ 6.5 and 6.5 $< z <$ 8.25 show similarity in redshift distributions between LRDs and galaxies. This should ensure that the K-S tests are not biased by differing redshift distributions.

For each redshift bin we run a two sample K-S test on the $\Sigma_{5}$ distribution of LRDs and of galaxies.  If a p-value smaller than the default value of 0.003 is calculated, corresponding to a 3$\sigma$ detection, then we reject the null hypothesis that the distribution of $\Sigma_{5}$ of LRDs and galaxies originate from the same parent distribution. The K-S tests give a p-value of 0.044 for 4.75 $< z <$ 6.5 and 0.014 for 6.5 $< z <$ 8.25, thus we cannot reject the null hypothesis at a 3$\sigma$ level. Regardless, this result is a tentative indication at the 2$\sigma$ level that LRDs and galaxies do not have the same distribution.

There is a considerably larger proportion of LRDs with the lowest $\Sigma_{5}$ values ($\lesssim 5$ cMpc$^{-2}$) when compared to galaxies. To investigate the impact of this on the strength of K-S test results, we randomly remove galaxies in all but the lowest $\Sigma_{5}$ bin shown in \autoref{fig:cluster} until the proportion of galaxies matches that of LRDs in the lowest $\Sigma_{5}$ bin and re-run the K-S tests. The resulting p-values are 0.84 for 4.75 $< z <$ 6.5 and 0.03 for 6.5 $< z <$ 8.25. In the lowest redshift bin, the p-value is far higher than its original value. This means the difference in distribution, at least at lower redshift, is largely due to the large proportion of LRDs in the very lowest density environments.

To further investigate this difference, we run additional K-sample Darling-Anderson (D-A) tests, which determine whether a set of samples originate from a given population. The results are somewhat strong, with the lower redshift and higher redshift bins producing p-values of 0.024 and 0.018, and statistic of 2.79 and 3.11 respectively, pointing to a low likelihood that LRDs and galaxies have the same distribution. This, alongside the K-S tests, indicates a possibly significant difference in distribution of LRDs and galaxies.

\begin{figure}
\centering
\setkeys{Gin}{draft=False}
\includegraphics[width=1\columnwidth]{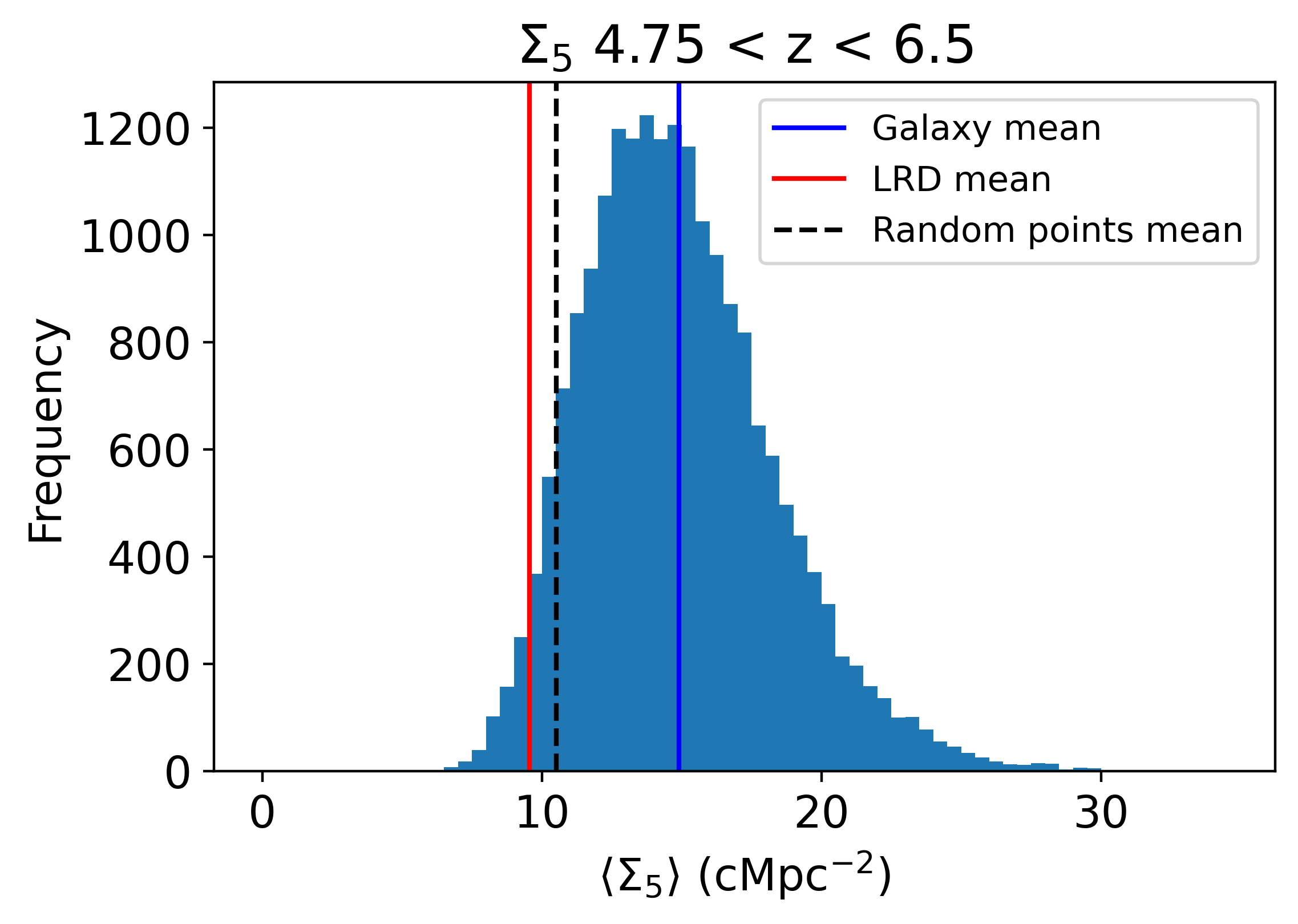}
\includegraphics[width=1\columnwidth]{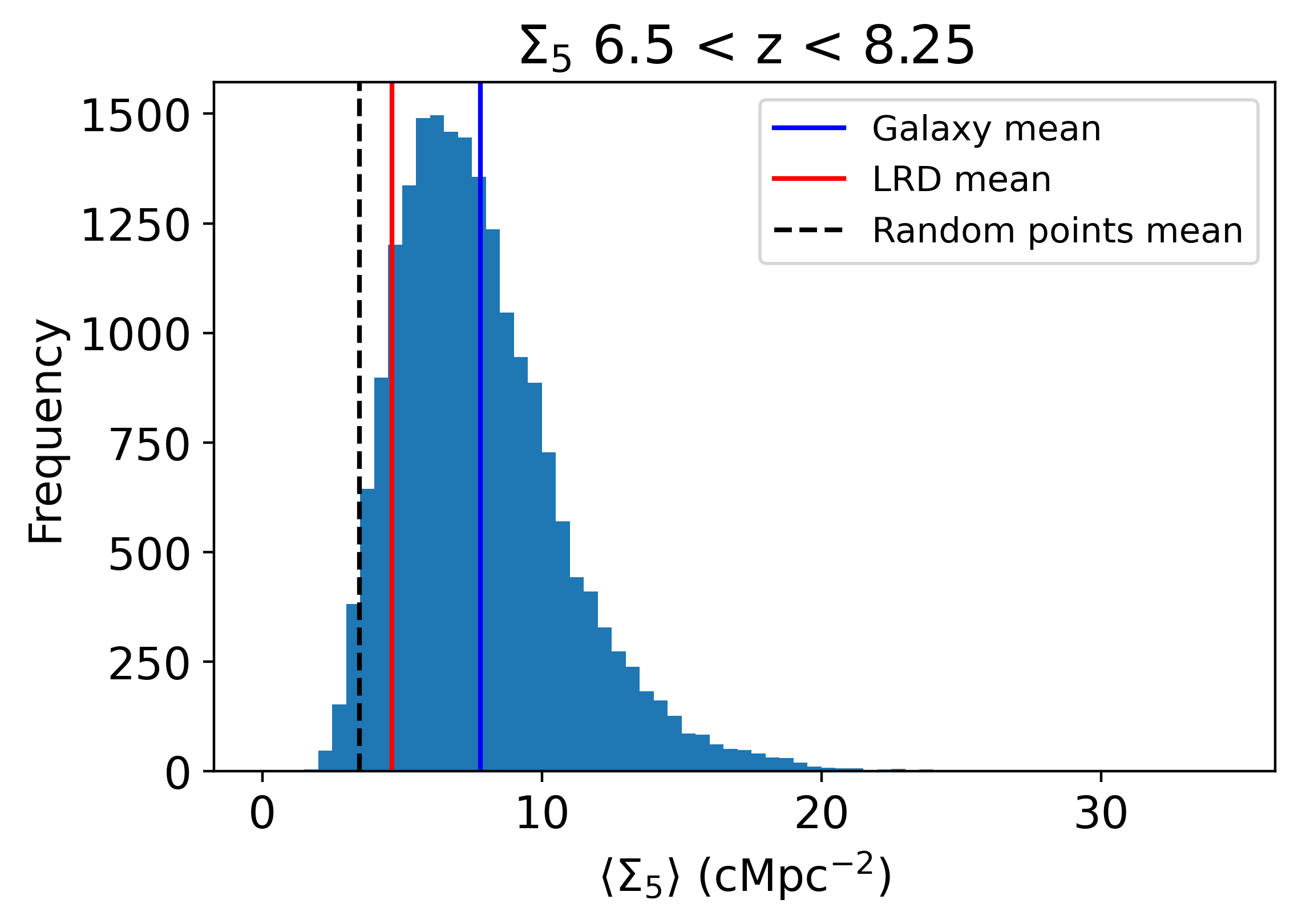}
\caption{The $\langle\Sigma_{5}\rangle$ of each run of a random galaxy sample for redshift bin 4.75 $ < z < $ 6.5 (top) and 6.5 $ < z < $ 8.25 (bottom). The $\langle\Sigma_{5}\rangle$ for our LRD and galaxy samples are shown in red and blue respectively, with the $\langle\Sigma_{5}\rangle$ of random points shown with a dashed black line for comparison.}
\label{fig:MCMC}
\end{figure}

To confirm that the above results are not due to the small size of our LRD sample or unknown effects, we randomly select a number of galaxies per redshift bin equal to the number of LRDs in each bin, so as to mimic any effects arising from the small size of the LRD sample. We then perform the same K-S test on the random galaxy sample and the parent galaxy sample, and calculate the $\langle\Sigma_{5}\rangle$ of the random galaxy sample. This is repeated 20,000 times, the results of which are shown in \autoref{fig:MCMC}. As the random galaxy sample originates from the parent galaxy sample, we expect significant results with low p-values to appear very infrequently. Ideally the frequency should be the same percentage as the corresponding threshold p-value, such that a p-value of 0.05 only appears in $\sim 5\%$ of runs. The K-S tests on the redshift bins 4.75 $ < z < $ 6.5 and 6.5 $ < z < $ 8.25 produce a p-value less than 0.05 only in $\lesssim4\%$ of the runs, suggesting that the K-S tests are behaving as expected. The distribution of the $\langle\Sigma_{5}\rangle$ of the random galaxy sample shows that the vast majority of random samples produce a $\langle\Sigma_{5}\rangle$ that is higher compared to the $\langle\Sigma_{5}\rangle$ of LRDs.  

\subsection{Halo masses}

Employing halo mass functions based on work by \cite{2013behroozi} and \cite{Tinker_2008}, we estimate the halo masses of LRDs.  We  use the abundance matching approach to measure these halo masses within a Planck cosmology \citep{PlanckCollab}.  The halo masses we calculate are upper limits of the halo masses of these systems.  We employ this approach as an experiment, under the hypothesis that the LRDs are galaxies and not AGN.  This allows us to assume that these systems are not variable, with a short life-time, but that they are galaxies of a similar semi-homogenous population.

\begin{figure}
\centering
\setkeys{Gin}{draft=False}
\includegraphics[width=0.5\textwidth]{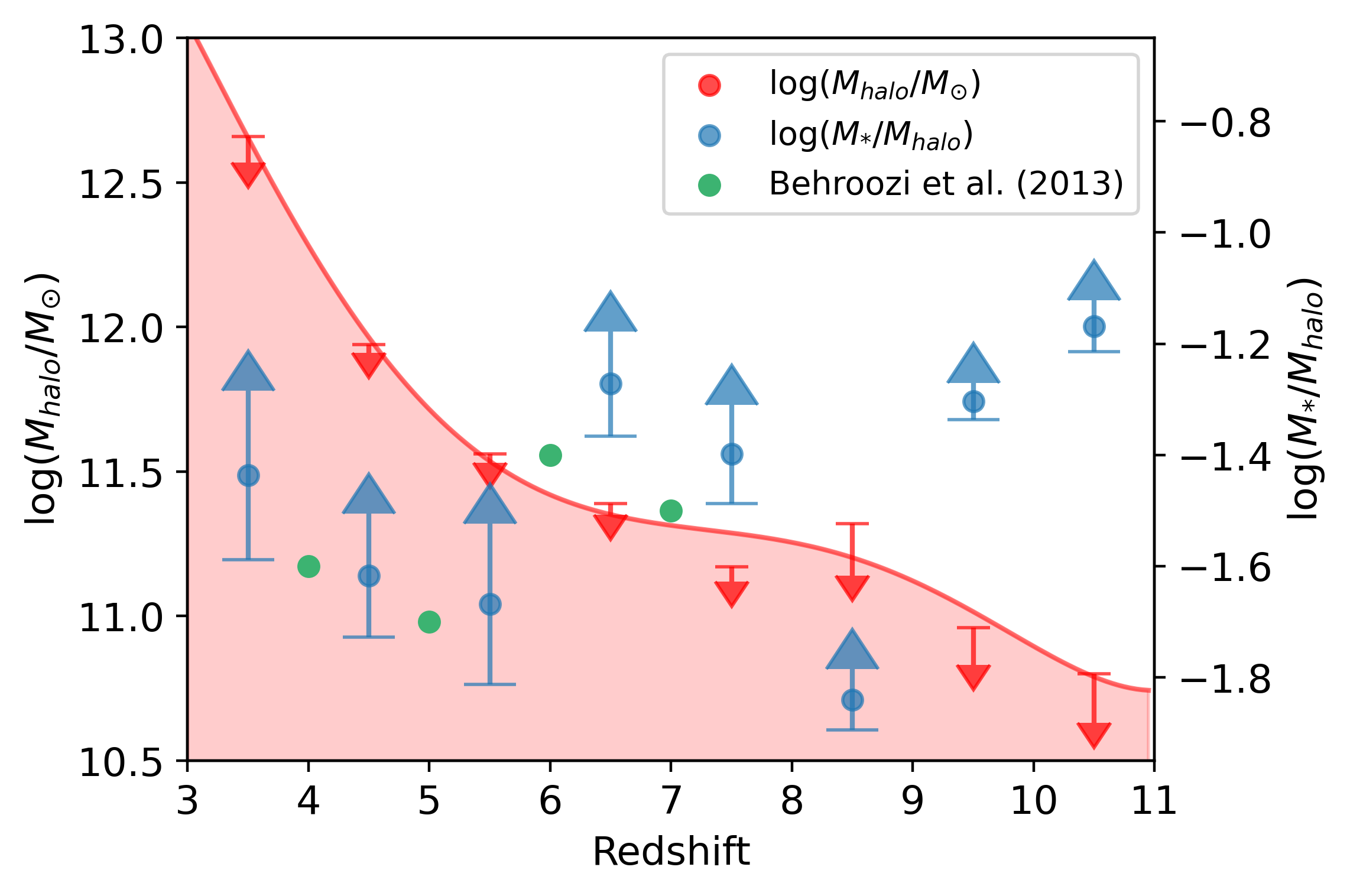}
\caption{Halo mass from abundance matching (red)  over the redshift range $3 < z < 11$ with redshift bin size $\Delta z = 1$. The SHMR (blue) is calculated from stellar masses extracted by $\cigale$ for non-AGN models.  The SHMR holds relatively constant throughout the bins.  The red shaded area is included for visualisation and is an estimated fit of our data.  Using this estimate fit, we include approximate points (green) for the corresponding SHMR of our halo mass at each redshift available in \cite{2013behroozi}.}
\label{fig:halo}
\end{figure}

The result of this is experiment is shown in Figure~\ref{fig:halo}.  What we can see is that the average halo mass grows considerably over time, such that under this hypothesis an LRD  grows its halo mass by almost a factor of $\sim 100$. This is a significant increase in the halo masses for these systems. When we examine the growth of stellar masses inferred for these systems in the non-AGN model fits, we see that they also grow with time for a LRD selected sample at a similar rate.  It is of course not at all clear or obvious that LRDs at high redshift are the same as those we see at lower redshift, or even for that matter how long the LRD phase lasts within these systems.

We compare the average stellar mass estimates of non-AGN models given by $\cigale$ against the halo masses given by the halo mass function, finding that the mean stellar to halo mass ratio (SHMR) varies around $\sim 10^{-1.4}$.  As the halo masses are upper limits, we expect the SHMRs to be lower limits. The value $10^{-1.4}$ is among the peak values for lower redshifts around $0 < z < 4$ found by simulations \citep{Girelli_2020, shmrcorrea}.  We create an approximate fit to our data using a polynomial of degree 5 to find estimates for halo mass at integer values of redshift.  Using these estimates of halo mass at integer redshifts, we extract corresponding SHMRs for each redshift from \cite{2013behroozi}. These corresponding SHMRs, shown in \autoref{fig:halo}, are slightly lower than ours, but seem to agree with our results. 

If LRDs are hosted in dark matter halos, their clustering can place constraints on their host halo masses and duty cycles.  For random points at $4.75 < z < 6.5$ and $6.5 < z < 8.25$, we find $\langle \Sigma_5 \rangle$ = 10.51 cMpc$^{-2}$ and 3.48 cMpc$^{-2}$, respectively.  Remarkably, random points have a slightly \textit{higher} average $\Sigma_5$ in the lower redshift bin than what we find for LRDs ($\langle \Sigma_5 \rangle \sim 9.6$ cMpc$^{-2}$).  The natural implication is that LRDs at lower redshifts cannot form in dense regions—i.e., there is a physical effect associated with dense regions (such as ionizing radiation or metal pollution) that inhibits LRDs from forming there.  Even for higher redshifts, the excess clustering above random implies a low bias relative to galaxies:
\begin{equation}
 \frac{b_\mathrm{LRDs}}{b_\mathrm{galaxies}} \approx \frac{\Sigma_\mathrm{5,LRDs}-\Sigma_\mathrm{5,Random}}{\Sigma_\mathrm{5,galaxies}-\Sigma_\mathrm{5,Random}} \sim 0.5, \end{equation} 
where the approximation comes because $\Sigma_5$ is an integral over slightly different physical scales for the different populations. From the cumulative number densities, the typical host halos of our galaxy sample (for $M_\ast > 10^{9.5}M_\odot$) in the higher redshift bin ($6.5 < z < 8.25$) would have $M_h > 10^{11.4}M_\odot$, with a mean bias of 10.8 \citep{Tinker10}, naively implying that the masses of the LRD hosts (with a mean bias half as much) would be $M_h \sim 10^{10.1} M_\odot$, which would typically host $10^7 M_\odot$ galaxies \citep{Behroozi19}.  Given the exceptionally low clustering in the low redshift bin, we view it as plausible that there is an isolation effect that also prevents LRDs from forming in the densest regions in the higher redshift bin, and so this halo mass estimate should be viewed as a lower limit.

\section{Discussion}

\subsection{SED modelling and BIC results}

Most LRDs in our and other samples appear to have broad H$\alpha$ lines (\S\ref{sec:broad}).  Most works interpret these broad lines as indicative of AGN \citep[][]{matthee2024, Kocevski_2025, Greene_2024}, which we consider in our SED fits. \cite{Baggen_2024} provide an alternative explanation to AGN, in which the broad lines are instead the result of a short-lived phase of galaxy evolution, and reflect the kinematics of extremely high densities of such compact galaxies. Alternatively, \cite{rusakov2025jwstslittlereddots} suggest that LRDs may be intrinsically narrow-line AGN in which the broadening of lines is partially due to electron scattering through a cocoon of ionized gas, potentially originating from mild outflows due to feedback from a burst of SF, or more simply accretion onto a SMBH. In the weak AGN or SF scenarios, it is possible that LRDs still contribute to reionization given that their main LyC escape and SF-driven outflow is directed away from the observer. LRDs would therefore not be a new class of objects, but rather compact dust star-formers with dust obscuration with relatively unobscured counterparts with visible rest-frame UV.  Given the mass measurements and limits based on clustering, it might be the case that these systems are forming star clusters which would not be expected to have a high clustering.   As our mass limits are 10$^{7}$ M$_{\odot}$, which is about 10 times higher than globular clusters today, it is likely that these are perhaps the more commonly observed super star clusters that likely dissolve at later times \citep[e.g.,][]{Guo2018}. 

Of the 31 LRDs with a better $\chi^{2}$ for the non-AGN models, the values of $\Delta\chi^{2} = \chi^{2}_{AGN} - \chi^2_{non-AGN}$ are usually small, with a typical of $\Delta\chi^{2} = $ 1, suggesting only minor improvements to fits are made when using an AGN component.  Model SEDs of JADES 32344 and CEERS 1919 are shown in \autoref{fig:SED_JADES} and \autoref{fig:SED_CEERS} respectively.  Of those that are not improved by including an AGN component, 12 are from JADES, 7 from CEERS, and 12 from NEP-TDF.  However, the BIC calculated for the models suggests that adding the $\skirtor$ SED module may lead to overfitting, as the BIC for AGN models is higher than that for the non-AGN models for 84 out of 124 LRDs in this sample (\autoref{fig:bic}), and more than half of LRDs showing substantial evidence ($\Delta BIC > 3.2$) pointing towards non-AGN models.  The $\Delta$BIC is especially high in the presence of MIRI data, where an AGN component is expected to dominate if present. LRDs with MIRI data are much more likely to show very strong evidence preferring the non-AGN model.  Before completely ruling out the possibility of LRDs hosting AGN, we note that it may be possible to model the SED of these objects with two AGN components, such as a reddened component and a scattered component with different normalisation and extinction. For example, \cite{labbe2024unambiguousagnbalmerbreak} use a composite AGN SED model to fit the SED of most optically-luminous LRD found to date and find a very strong fit. This implies that if AGN are found in LRDs, they cannot have a dusty torus as modelled by $\cigale$.

There is also a marked difference between the BICs calculated for LRDs in NEP-TDF and CEERS and the LRDs in JADES as seen in \autoref{fig:bic}.  For CEERS and NEP-TDF, which use HST and NIRCam data only, the BICs of AGN models are typically $\sim 30$, whilst for non-AGN models it is $\sim 25$.  For JADES, these values are $\sim 110$ and $\sim 100$ respectively.  All of the LRDs found in JADES have a lower $\chi^{2}$ for the non-AGN model, likely due to the presence of MIRI data. The BICs suggest that non-AGN models are more suitable, especially when MIRI data is used.  A further effect of MIRI data, as seen in \autoref{fig:stellarmass}, is that the difference in stellar mass between AGN and non-AGN models is much reduced compared to models with only NIRCam+HST data.

The SED models presented in this paper are not necessarily representative of the true composition of LRDs.  This is in part due to the poorly understood nature of LRDs.  Despite this, we explore the model compositions and attempt to draw some conclusions.  We find that AGN models without MIRI data tend to have high (f$_{AGN} \sim 0.7 - 0.99$) AGN to total IR luminosity fractions (\autoref{fig:bic}, \autoref{fig:chi}), whilst those with MIRI data tend to have a lower AGN fraction (f$_{AGN} = 0.1$).  There tends to be a more negative and greater range of $\Delta\chi^{2}$ associated with a larger fraction of AGN. Furthermore, a higher AGN fraction is also associated with a more negative and greater range of $\Delta$BIC. Both the BIC and $\chi^{2}$ of AGN versus non-AGN models with MIRI data indicate a better fit for the non-AGN models.

The presence of an AGN component sometimes leads to a significantly lower stellar mass, which is up to 2 dex lower, as seen in \autoref{fig:stellarmass}.  This somewhat alleviates the anomalously high central stellar mass densities implied for LRDs \citep{akins2024}.  Some model stellar masses ($\sim$ 50 $\%$) are largely unaffected by the presence of an AGN component.  In comparison, \cite{leung2024exploringnaturelittlered} model LRDs as galaxies using $\bagpipes$ \citep{carnall} and create AGN models using the $\qsogen$ code from \cite{temple}, and find that fitting as a galaxy results in stellar masses that are $\sim$ 2 dex higher than models with AGN.  We note that stellar masses for non-AGN models are typically lower and similar to AGN models for LRDs with HST+NIRCam+MIRI data.  This is of particular interest, as it suggests that the problem of overmassive stellar masses of LRDs \citep[e.g.][]{Kokorev_2023, Guia_2024} may be at least partially tackled when MIRI data are taken into consideration. 

There is no clear correlation visible in \autoref{fig:stellarmass_AGN} between the AGN fraction and the stellar mass.  Most fractions of AGN produce a similar range of stellar masses for LRDs, but the highest fraction (f$_{AGN} = $ 0.99) produces the largest range of stellar masses. 

\subsection{Environment results}

The relatively large size of redshift offset mask used to determine the nearest neighbours ($\Delta z < 0.2$, \S\ref{sec:loc_env}) reduces the impact on the $\langle\Sigma_{5}\rangle$ results of the somewhat poorer quality of photometric redshifts for LRDs compared to the rest of our sample  ($\S\ref{sec:redshift}$).  However, this will still have an impact on the strength of the results that is difficult to assess.

The result of both the K-S and D-A tests suggest that the lower $\langle\Sigma_{5}\rangle$ result of LRDs is not due to a smaller sample size, but due to a real difference between the distributions of LRDs and non-LRD galaxies.  LRDs have a tendency to be found in less dense environments.  Moreover, the range of $\Sigma_{5}$ of LRDs is also reduced compared to galaxies, as shown in \autoref{fig:cluster}, although this could be in part caused by the smaller sample size of LRDs. Only $\sim$ 4 $\%$ of the 20,000 randomly selected galaxy samples produce a p-value less than 0.05, suggesting that K-S tests are an effective method in determining that the difference in distributions of the $\Sigma_{5}$ of LRDs and galaxies is not coincidental or due to small sample size. The $\langle\Sigma_{5}\rangle$ of the random galaxy samples are also typically significantly higher than the $\langle\Sigma_{5}\rangle$ of our LRD sample, further suggesting that the lower $\Sigma_{5}$ of LRDs is not an accidental or random result.

We attempt to interpret the above results for LRDs as due to stellar mass by carrying out further K-S tests on various mass bins.  However, a lower $\Sigma_{5}$ is not necessarily a sign of a lower or higher stellar mass than the comparison galaxies, as we find that the stellar mass of galaxies has a very weak if any correlation with density, in agreement with \cite{li2024epochspaperxenvironmental}.  To make any estimates constraining halo mass, two-point clustering studies are required \citep{clustering}. 

An alternative explanation to the lower density environments in which LRDs are found could be that higher density environments speed up the LRD evolutionary phase, meaning that any LRDs that were in higher density environments have evolved past the LRD stage within the redshift range of this study. For example, \cite{morishita2025acceleratedemergenceevolvedgalaxies} find an increased fraction of galaxies with weak H$\alpha$+[\textrm{N}~\textsc{ii}] emission lines in two overdensities of galaxies at $z \sim$ 5.7 compared to field samples at similar redshifts. These weak emission line galaxies also exhibit a strong continuum break at 4000\AA\, which is associated with evolved stellar populations. This evidence is consistent with the suggestion that high-density environments accelerate the evolution of galaxies. Some works studying over-densities find that LRDs tend to be isolated. For example, \cite{fudamoto2025sapphiresgalaxyoverdensityheart} find an LRD candidate in a lower surface density area of an over-density. \cite{Champagne_2025} study a protocluster and find that objects meeting photometric LRD criteria \citep{Greene_2024} line the edge of the protocluster, and argue that galaxy evolution occurs 'inside-out' in dense environments. To extract further information, a more general study on the make up of populations found around the edges of over-densities may be necessary.

\subsection{Halo mass and SHMR results}

As our halo masses are upper limits, our SHMRs are expected to be lower limits.  To estimate a lower limit on halo masses, we assume that the SHMRs cannot exceed the maximum at $z = 0$, which is for the most massive systems.  This value is $\sim20\%$ of the baryon fraction ($17\%$) \citep{2013behroozi}, or approximately $10^{-1.5}$.  The inferred SHMR is greater than this in four redshift bins, suggesting one of a few scenarios.  This includes the idea that  all massive halos host LRDs, which might be the case if our LRDs are for example forming star clusters. Alternatively,  a large fraction of the light detected from LRDs must originate from AGN to increase their inferred measured stellar masses. However, we have shown that this is unlikely to be the case for the majority of LRDs given their SED fits.

The remarkably low clustering of LRDs compared to both galaxies and random points suggests a unique formation mechanism, which prevents LRDs from forming in the densest regions.  This is opposite to most other traditional formation mechanisms, for example, the formation of galaxies and black holes, which is expected to be easier in high-density regions.  This also poses a challenge to interpretation: all existing templates for galaxies and black holes are for objects that, at high redshifts, are easier to form in dense regions, and so may not be a good description of the physics occurring in LRDs.  At the least, however, we can be confident that LRDs do not look like traditional templates for AGN.

It is of course interesting to consider the objects that may be preferentially formed in lower-density regions.  One example is direct-collapse black holes \citep{Bromm03} and their potential precursors of supermassive stars \citep{Begelman10}, which need an ionizing radiation source to prevent gas fragmentation, but require extremely low metallicity for the same reason, and so may be anti-correlated with large-scale structure by $z\sim 5$.  A more exotic example is gas clouds that are supported by dark matter annihilation—and hence do not need a separate ionizing radiation source—but nonetheless require low metallicity to avoid fragmentation \citep{Banik19}.  Key observational signatures of both such object classes are (as is the case for LRDs), lower formation rates once the IGM has been metal enriched (especially at $z<3$; \citealt{DeCia18}) and lower clustering compared to galaxies.  However, we note that initial low metallicities during formation may be obscured by later gas accretion and star formation by the time the objects are bright enough to be observed as LRDs.  We remain agnostic as to whether either (or neither) of these two can explain some of the LRD population, but the clustering measured in this work strongly motivates spectroscopic follow-up studies to understand the nature of these mysterious objects.

\section{Conclusion}

The Little Red Dots (LRDs), discovered with the JWST, are some of the most mysterious objects yet found in the very distant universe.  These systems are defined by three major characteristics: very red SEDs and spectra, very compact to unresolved structures, and very bright.  In fact, when these were discovered it was thought that the large brightness of these systems made their inferred stellar masses larger than is acceptable within $\Lambda$CDM cosmology \citep[e.g.,][]{labbe2023uca}.  Thus, there is an opportunity and importance in uncovering the nature of these systems.  

In this paper we select a sample of a total of 124 LRDs from the CEERS, NEP-TDF and JADES field.  Of those with grating spectra, 14 of 16 LRDs show evidence of broad lines (FWHM $\>$ 1000 km s$^{-1}$), a similar fraction to \cite{Greene_2024}.  For the range 3 $ < z < $ 8, we find a number density of $\sim10^{-5}$ cMpc$^{-3}$, in line with \cite{pizzati2024littlereddotsreside}, which peaks at $\sim10^{-4}$ cMpc$^{-3}$ at 5 $ < z < $ 6.

We investigate and compare SED models for LRDs with and without AGN components.  To determine whether adding an AGN component to an SED model of an LRD results in overfitting we compare the $\chi^{2}$ and BIC of models with and without AGN.  We find that whilst the $\chi^{2}$ of $\sim75\%$ LRDs in our sample suggests that SED models containing AGN are more suitable, the BICs calculated for our sample suggests that an AGN component often results in overfitting the SEDs of LRDs.  This is particularly the case when using MIRI data, as we find that the presence of MIRI data results in a greater difference in the BICs of AGN and non-AGN models. AGN models with MIRI data also tend to have lower AGN fractions, whilst non-AGN models with MIRI data tend to have lower stellar masses.

Comparing the spatial distributions of LRDs to non-LRD galaxies, we find that the $\Sigma_{5}$ of LRDs tends to be lower than that of galaxies.  We employ a Kolmogorov-Smirnov test to determine whether this result is significant and find that the redshift bins 4.75 $< z <$ 6.5 and 6.5 $< z <$ 8.25 produce a p-value of 0.044 and 0.014 respectively. Both p-values are relatively low, giving a tentative indication that the $\Sigma_{5}$ distributions of general galaxies and LRDs are different.  Overall, this suggests that LRDs are typically found in lower density environments than galaxies.

We calculate upper limit estimates for the halo masses and stellar to halo mass ratio for our LRD sample using halo mass functions \citep{2013behroozi, Tinker_2008} and $\cigale$.  We find that the inferred SHMR is relatively constant ($\sim0.03$) over the range $3 < z < 11$.  Our halo mass and SHMR values agree with simulation data over the range $4 < z < 8$ and are comparable with the peak SHMR values found by simulations for lower redshifts. The SHMR is greater than the maximum SHMR at $z = 0$ for four redshift bins.  Because of this, we expect that either all massive halos host LRDs, in which case we would expect LRDs to be found in pairs, or AGN provide a large fraction of the light from LRDs. 

The low $\langle\Sigma_{5}\rangle$ of LRDs at redshift $4.75 < z < 6.5$ implies that LRDs are not formed in dense regions. In fact, even at higher redshifts ($6.5 < z < 8.25$), the low $\langle\Sigma_{5}\rangle$ values  suggest that LRD host halos have masses of $M_{h} \sim 10^{11.4} M_\odot$. These halos usually host galaxies with stellar masses $10^{7}M_\odot$. Alternatively, LRDs may only be able to form in lower density regions, in which case the halo mass calculated would be a lower limit.

The population of LRDs of the early universe will continue to surprise for the foreseeable future. This work already uncovers some curious properties, such as the low clustering of these objects. By further examining clustering, it may be possible to place further constraints on the characteristics of LRDs.

\section*{Acknowledgements}
We thank the JADES, PEARLS, and CEERS teams for their work in designing and preparing these public and GTO observations, and the STScI staff that carried them out. We acknowledge support from the ERC Advanced Investigator Grant EPOCHS
(788113), as well three studentships from the STFC.  RAW, SHC, and RAJ acknowledge support from NASA JWST Interdisciplinary Scientist grants NAG5-12460, NNX14AN10G and 80NSSC18K0200 from GSFC. LF acknowledges financial support from Coordenação de
Aperfeiçoamento de Pessoal de Nível Superior - Brazil (CAPES) in the form of a PhD studentship. This work is based on observations made with the NASA/ESA  \textit{Hubble Space Telescope} (HST) and NASA/ESA/CSA \textit{James Webb Space Telescope} (JWST) obtained from the $\MAST$ ($\mast$) at the \textit{Space Telescope Science Institute} (STScI), which is operated by the Association of Universities for Research in Astronomy, Inc., under NASA contract NAS 5-03127 for JWST, and NAS 5–26555 for HST. The observations used in this work are associated with JWST programme 1345, 1180, 1176, and 2738. The data described here may be obtained from the MAST archive at \dataset[doi:10.17909/5h64-g193]{https://dx.doi.org/10.17909/5h64-g193}.

Some of the data products presented herein were retrieved from the Dawn JWST Archive (DJA).  DJA is an initiative of the Cosmic Dawn Center (DAWN), which is funded by the Danish National Research Foundation under grant DNRF140.  

MMCE would like to thank Charlie Ellis for helping her with the presentation of the SED plots in this work.  

JAAT acknowledges support from the Simons Foundation and JWST program 3215. Support for program 3215 was provided by NASA through a grant from the Space Telescope Science Institute, which is operated by the Association of Universities for Research in Astronomy, Inc., under NASA contract NAS 5-03127.

This work made use of the following Python libraries: $\astropy$ \citep{AstropyCollaboration_2022}, $\scipy$ \citep{2020SciPy-NMeth}, and $\matplotlib$ \citep{matplot}.  
AZ acknowledges support by Grant No. 2020750 from the United States-Israel Binational Science Foundation (BSF) and Grant No. 2109066 from the United States National Science Foundation (NSF); and by the Israel Science Foundation Grant No. 864/23.



\bibliographystyle{aasjournal}
\bibliography{main} 

\begin{thebibliography}{}
\expandafter\ifx\csname natexlab\endcsname\relax\def\natexlab#1{#1}\fi
\providecommand{\url}[1]{\href{#1}{#1}}
\providecommand{\dodoi}[1]{doi:~\href{http://doi.org/#1}{\nolinkurl{#1}}}
\providecommand{\doeprint}[1]{\href{http://ascl.net/#1}{\nolinkurl{http://ascl.net/#1}}}
\providecommand{\doarXiv}[1]{\href{https://arxiv.org/abs/#1}{\nolinkurl{https://arxiv.org/abs/#1}}}

\bibitem[{Adams {et~al.}(2024)Adams, Conselice, Austin, Harvey, Ferreira, Trussler, Juodzbalis, Li, Windhorst, Cohen, Jansen, Summers, Tompkins, Driver, Robotham, D'Silva, Yan, Coe, Frye, Grogin, Koekemoer, Marshall, Pirzkal, Russell E.~Ryan, Maksym, Rutkowski, Willmer, Hammel, Nonino, Bhatawdekar, Wilkins, Bradley, Broadhurst, Cheng, Dole, Hathi, \& Zitrin}]{adams2024epochspaperiiultraviolet}
Adams, N.~J., Conselice, C.~J., Austin, D., {et~al.} 2024, EPOCHS Paper II: The Ultraviolet Luminosity Function from $7.5<z<13.5$ using 180 square arcminutes of deep, blank-fields from the PEARLS Survey and Public JWST data.
\newblock \doarXiv{2304.13721}

\bibitem[{Akins {et~al.}(2023)Akins, Casey, Allen, Bagley, Dickinson, Finkelstein, Franco, Harish, Haro, Ilbert, Kartaltepe, Koekemoer, Liu, Long, McCracken, Paquereau, Papovich, Pirzkal, Rhodes, Robertson, Shuntov, Toft, Yang, Barro, Bisigello, Buat, Champagne, Cooper, Costantin, de~la Vega, Drakos, Faisst, Fontana, Fujimoto, Gillman, Gómez-Guijarro, Gozaliasl, Hathi, Hayward, Hirschmann, Holwerda, Jin, Kocevski, Kokorev, Lambrides, Lucas, Magdis, Magnelli, McKinney, Mobasher, Pérez-González, Rich, Seillé, Talia, Urry, Valentino, Whitaker, Yung, \& Zavala}]{akins2023}
Akins, H.~B., Casey, C.~M., Allen, N., {et~al.} 2023, Two massive, compact, and dust-obscured candidate $z\sim 8$ galaxies discovered by JWST.
\newblock \doarXiv{2304.12347}

\bibitem[{Akins {et~al.}(2024)Akins, Casey, Lambrides, Allen, Andika, Brinch, Champagne, Cooper, Ding, Drakos, Faisst, Finkelstein, Franco, Fujimoto, Gentile, Gillman, Gozaliasl, Harish, Hayward, Hirschmann, Ilbert, Kartaltepe, Kocevski, Koekemoer, Kokorev, Liu, Long, McCracken, McKinney, Onoue, Paquereau, Renzini, Rhodes, Robertson, Shuntov, Silverman, Tanaka, Toft, Trakhtenbrot, Valentino, \& Zavala}]{akins2024}
Akins, H.~B., Casey, C.~M., Lambrides, E., {et~al.} 2024, COSMOS-Web: The over-abundance and physical nature of "little red dots"--Implications for early galaxy and SMBH assembly.
\newblock \doarXiv{2406.10341}

\bibitem[{{Alberts} {et~al.}(2024){Alberts}, {Lyu}, {Shivaei}, {Rieke}, {Perez-Gonzalez}, {Bonventura}, {Zhu}, {Helton}, {Ji}, {Morrison}, {Robertson}, {Stone}, {Sun}, {Williams}, \& {Willmer}}]{smiles}
{Alberts}, S., {Lyu}, J., {Shivaei}, I., {et~al.} 2024, arXiv e-prints, arXiv:2405.15972, \dodoi{10.48550/arXiv.2405.15972}

\bibitem[{Ananna {et~al.}(2024)Ananna, Ákos Bogdán, Kovács, Natarajan, \& Hickox}]{Ananna_2024}
Ananna, T.~T., Ákos Bogdán, Kovács, O.~E., Natarajan, P., \& Hickox, R.~C. 2024, The Astrophysical Journal Letters, 969, L18, \dodoi{10.3847/2041-8213/ad5669}

\bibitem[{Arrabal~Haro {et~al.}(2023)Arrabal~Haro, Dickinson, Finkelstein, Kartaltepe, Donnan, Burgarella, Carnall, Cullen, Dunlop, Fernández, Fujimoto, Jung, Krips, Larson, Papovich, Pérez-González, Amorín, Bagley, Buat, Casey, Chworowsky, Cohen, Ferguson, Giavalisco, Huertas-Company, Hutchison, Kocevski, Koekemoer, Lucas, McLeod, McLure, Pirzkal, Seillé, Trump, Weiner, Wilkins, \& Zavala}]{Arrabal_Haro_2023}
Arrabal~Haro, P., Dickinson, M., Finkelstein, S.~L., {et~al.} 2023, Nature, 622, 707–711, \dodoi{10.1038/s41586-023-06521-7}

\bibitem[{Baggen {et~al.}(2024)Baggen, van Dokkum, Brammer, de~Graaff, Franx, Greene, Labbé, Leja, Maseda, Nelson, Rix, Wang, \& Weibel}]{Baggen_2024}
Baggen, J. F.~W., van Dokkum, P., Brammer, G., {et~al.} 2024, The Astrophysical Journal Letters, 977, L13, \dodoi{10.3847/2041-8213/ad90b8}

\bibitem[{Bagley {et~al.}(2023)Bagley, Finkelstein, Koekemoer, Ferguson, Haro, Dickinson, Kartaltepe, Papovich, Pérez-González, Pirzkal, Somerville, Willmer, Yang, Yung, Fontana, Grazian, Grogin, Hirschmann, Kewley, Kirkpatrick, Kocevski, Lotz, Medrano, Morales, Pentericci, Ravindranath, Trump, Wilkins, Calabrò, Cooper, Costantin, de~la Vega, Hilbert, Hutchison, Larson, Lucas, McGrath, Ryan, Wang, \& Wuyts}]{Bagley_2023}
Bagley, M.~B., Finkelstein, S.~L., Koekemoer, A.~M., {et~al.} 2023, The Astrophysical Journal Letters, 946, L12, \dodoi{10.3847/2041-8213/acbb08}

\bibitem[{{Banik} {et~al.}(2019){Banik}, {Tan}, \& {Monaco}}]{Banik19}
{Banik}, N., {Tan}, J.~C., \& {Monaco}, P. 2019, \mnras, 483, 3592, \dodoi{10.1093/mnras/sty3298}

\bibitem[{{Begelman}(2010)}]{Begelman10}
{Begelman}, M.~C. 2010, \mnras, 402, 673, \dodoi{10.1111/j.1365-2966.2009.15916.x}

\bibitem[{{Behroozi} {et~al.}(2019){Behroozi}, {Wechsler}, {Hearin}, \& {Conroy}}]{Behroozi19}
{Behroozi}, P., {Wechsler}, R.~H., {Hearin}, A.~P., \& {Conroy}, C. 2019, \mnras, 488, 3143, \dodoi{10.1093/mnras/stz1182}

\bibitem[{{Behroozi} {et~al.}(2013){Behroozi}, {Wechsler}, \& {Conroy}}]{2013behroozi}
{Behroozi}, P.~S., {Wechsler}, R.~H., \& {Conroy}, C. 2013, \apj, 770, 57, \dodoi{10.1088/0004-637X/770/1/57}

\bibitem[{{Bertin} \& {Arnouts}(1996)}]{bertinarnouts1996}
{Bertin}, E., \& {Arnouts}, S. 1996, \aaps, 117, 393, \dodoi{10.1051/aas:1996164}

\bibitem[{{Boquien} {et~al.}(2019){Boquien}, {Burgarella}, {Roehlly}, {Buat}, {Ciesla}, {Corre}, {Inoue}, \& {Salas}}]{boquien2019}
{Boquien}, M., {Burgarella}, D., {Roehlly}, Y., {et~al.} 2019, \aap, 622, A103, \dodoi{10.1051/0004-6361/201834156}

\bibitem[{Brammer(2023)}]{brammer_2023_8319596}
Brammer, G. 2023, msaexp: NIRSpec analyis tools, 0.6.17,  Zenodo, \dodoi{10.5281/zenodo.8319596}

\bibitem[{{Brammer} {et~al.}(2008){Brammer}, {van Dokkum}, \& {Coppi}}]{brammer2008}
{Brammer}, G.~B., {van Dokkum}, P.~G., \& {Coppi}, P. 2008, \apj, 686, 1503, \dodoi{10.1086/591786}

\bibitem[{{Bromm} \& {Loeb}(2003)}]{Bromm03}
{Bromm}, V., \& {Loeb}, A. 2003, \apj, 596, 34, \dodoi{10.1086/377529}

\bibitem[{Brown {et~al.}(2021)Brown, Vallenari, Prusti, de~Bruijne, Babusiaux, Biermann, Creevey, Evans, Eyer, Hutton, Jansen, Jordi, Klioner, Lammers, Lindegren, Luri, Mignard, Panem, Pourbaix, Randich, Sartoretti, Soubiran, Walton, Arenou, Bailer-Jones, Bastian, Cropper, Drimmel, Katz, Lattanzi, van Leeuwen, Bakker, Cacciari, Castañeda, De~Angeli, Ducourant, Fabricius, Fouesneau, Frémat, Guerra, Guerrier, Guiraud, Jean-Antoine~Piccolo, Masana, Messineo, Mowlavi, Nicolas, Nienartowicz, Pailler, Panuzzo, Riclet, Roux, Seabroke, Sordo, Tanga, Thévenin, Gracia-Abril, Portell, Teyssier, Altmann, Andrae, Bellas-Velidis, Benson, Berthier, Blomme, Brugaletta, Burgess, Busso, Carry, Cellino, Cheek, Clementini, Damerdji, Davidson, Delchambre, Dell’Oro, Fernández-Hernández, Galluccio, García-Lario, Garcia-Reinaldos, González-Núñez, Gosset, Haigron, Halbwachs, Hambly, Harrison, Hatzidimitriou, Heiter, Hernández, Hestroffer, Hodgkin, Holl, Janßen, Jevardat~de Fombelle, Jordan, Krone-Martins, Lanzafame,
  Löffler, Lorca, Manteiga, Marchal, Marrese, Moitinho, Mora, Muinonen, Osborne, Pancino, Pauwels, Petit, Recio-Blanco, Richards, Riello, Rimoldini, Robin, Roegiers, Rybizki, Sarro, Siopis, Smith, Sozzetti, Ulla, Utrilla, van Leeuwen, van Reeven, Abbas, Abreu~Aramburu, Accart, Aerts, Aguado, Ajaj, Altavilla, Álvarez, Álvarez Cid-Fuentes, Alves, Anderson, Anglada~Varela, Antoja, Audard, Baines, Baker, Balaguer-Núñez, Balbinot, Balog, Barache, Barbato, Barros, Barstow, Bartolomé, Bassilana, Bauchet, Baudesson-Stella, Becciani, Bellazzini, Bernet, Bertone, Bianchi, Blanco-Cuaresma, Boch, Bombrun, Bossini, Bouquillon, Bragaglia, Bramante, Breedt, Bressan, Brouillet, Bucciarelli, Burlacu, Busonero, Butkevich, Buzzi, Caffau, Cancelliere, Cánovas, Cantat-Gaudin, Carballo, Carlucci, Carnerero, Carrasco, Casamiquela, Castellani, Castro-Ginard, Castro~Sampol, Chaoul, Charlot, Chemin, Chiavassa, Cioni, Comoretto, Cooper, Cornez, Cowell, Crifo, Crosta, Crowley, Dafonte, Dapergolas, David, David, de~Laverny,
  De~Luise, De~March, De~Ridder, de~Souza, de~Teodoro, de~Torres, del Peloso, del Pozo, Delbo, Delgado, Delgado, Delisle, Di~Matteo, Diakite, Diener, Distefano, Dolding, Eappachen, Edvardsson, Enke, Esquej, Fabre, Fabrizio, Faigler, Fedorets, Fernique, Fienga, Figueras, Fouron, Fragkoudi, Fraile, Franke, Gai, Garabato, Garcia-Gutierrez, García-Torres, Garofalo, Gavras, Gerlach, Geyer, Giacobbe, Gilmore, Girona, Giuffrida, Gomel, Gomez, Gonzalez-Santamaria, González-Vidal, Granvik, Gutiérrez-Sánchez, Guy, Hauser, Haywood, Helmi, Hidalgo, Hilger, Hładczuk, Hobbs, Holland, Huckle, Jasniewicz, Jonker, Juaristi~Campillo, Julbe, Karbevska, Kervella, Khanna, Kochoska, Kontizas, Kordopatis, Korn, Kostrzewa-Rutkowska, Kruszyńska, Lambert, Lanza, Lasne, Le~Campion, Le~Fustec, Lebreton, Lebzelter, Leccia, Leclerc, Lecoeur-Taibi, Liao, Licata, Lindstrøm, Lister, Livanou, Lobel, Madrero~Pardo, Managau, Mann, Marchant, Marconi, Marcos~Santos, Marinoni, Marocco, Marshall, Martin~Polo, Martín-Fleitas, Masip, Massari,
  Mastrobuono-Battisti, Mazeh, McMillan, Messina, Michalik, Millar, Mints, Molina, Molinaro, Molnár, Montegriffo, Mor, Morbidelli, Morel, Morris, Mulone, Munoz, Muraveva, Murphy, Musella, Noval, Ordénovic, Orrù, Osinde, Pagani, Pagano, Palaversa, Palicio, Panahi, Pawlak, Peñalosa~Esteller, Penttilä, Piersimoni, Pineau, Plachy, Plum, Poggio, Poretti, Poujoulet, Prša, Pulone, Racero, Ragaini, Rainer, Raiteri, Rambaux, Ramos, Ramos-Lerate, Re~Fiorentin, Regibo, Reylé, Ripepi, Riva, Rixon, Robichon, Robin, Roelens, Rohrbasser, Romero-Gómez, Rowell, Royer, Rybicki, Sadowski, Sagristà~Sellés, Sahlmann, Salgado, Salguero, Samaras, Sanchez~Gimenez, Sanna, Santoveña, Sarasso, Schultheis, Sciacca, Segol, Segovia, Ségransan, Semeux, Shahaf, Siddiqui, Siebert, Siltala, Slezak, Smart, Solano, Solitro, Souami, Souchay, Spagna, Spoto, Steele, Steidelmüller, Stephenson, Süveges, Szabados, Szegedi-Elek, Taris, Tauran, Taylor, Teixeira, Thuillot, Tonello, Torra, Torra, Turon, Unger, Vaillant, van Dillen, Vanel,
  Vecchiato, Viala, Vicente, Voutsinas, Weiler, Wevers, Wyrzykowski, Yoldas, Yvard, Zhao, Zorec, Zucker, Zurbach, \& Zwitter}]{brown2021}
Brown, A. G.~A., Vallenari, A., Prusti, T., {et~al.} 2021, Astronomy \&; Astrophysics, 650, C3, \dodoi{10.1051/0004-6361/202039657e}

\bibitem[{Bruzual \& Charlot(2003)}]{bruzualcharlot}
Bruzual, G., \& Charlot, S. 2003, Monthly Notices of the Royal Astronomical Society, 344, 1000, \dodoi{10.1046/j.1365-8711.2003.06897.x}

\bibitem[{Bunker {et~al.}(2024)Bunker, Cameron, Curtis-Lake, Jakobsen, Carniani, Curti, Witstok, Maiolino, D’Eugenio, Looser, Willott, Bonaventura, Hainline, Übler, Willmer, Saxena, Smit, Alberts, Arribas, Baker, Baum, Bhatawdekar, Bowler, Boyett, Charlot, Chen, Chevallard, Circosta, DeCoursey, de~Graaff, Egami, Eisenstein, Endsley, Ferruit, Giardino, Hausen, Helton, Hviding, Ji, Johnson, Jones, Kumari, Laseter, Lützgendorf, Maseda, Nelson, Parlanti, Perna, Rauscher, Rawle, Rix, Rieke, Robertson, Rodríguez Del~Pino, Sandles, Scholtz, Sharpe, Skarbinski, Stark, Sun, Tacchella, Topping, Villanueva, Wallace, Williams, \& Woodrum}]{Bunker_2024}
Bunker, A.~J., Cameron, A.~J., Curtis-Lake, E., {et~al.} 2024, Astronomy \& Astrophysics, 690, A288, \dodoi{10.1051/0004-6361/202347094}

\bibitem[{{Burgarella} {et~al.}(2005){Burgarella}, {Buat}, \& {Iglesias-P{\'a}ramo}}]{burgarella2005}
{Burgarella}, D., {Buat}, V., \& {Iglesias-P{\'a}ramo}, J. 2005, \mnras, 360, 1413, \dodoi{10.1111/j.1365-2966.2005.09131.x}

\bibitem[{Bushouse {et~al.}(2022)Bushouse, Eisenhamer, Dencheva, Davies, Greenfield, Morrison, Hodge, Simon, Grumm, Droettboom, Slavich, Sosey, Pauly, Miller, Jedrzejewski, Hack, Davis, Crawford, Law, Gordon, Regan, Cara, MacDonald, Bradley, Shanahan, Jamieson, Teodoro, \& Williams}]{bushouse_2022}
Bushouse, H., Eisenhamer, J., Dencheva, N., {et~al.} 2022, JWST Calibration Pipeline, 1.8.2,  Zenodo, \dodoi{10.5281/zenodo.7325378}

\bibitem[{{Calzetti} {et~al.}(2000){Calzetti}, {Armus}, {Bohlin}, {Kinney}, {Koornneef}, \& {Storchi-Bergmann}}]{calzetti}
{Calzetti}, D., {Armus}, L., {Bohlin}, R.~C., {et~al.} 2000, \apj, 533, 682, \dodoi{10.1086/308692}

\bibitem[{Cappellari {et~al.}(2006)Cappellari, Bacon, Bureau, Damen, Davies, De~Zeeuw, Emsellem, Falcón-Barroso, Krajnovic, Kuntschner, McDermid, Peletier, Sarzi, Van Den~Bosch, \& Van De~Ven}]{cappell}
Cappellari, M., Bacon, R., Bureau, M., {et~al.} 2006, Monthly Notices of the Royal Astronomical Society, 366, 1126, \dodoi{10.1111/j.1365-2966.2005.09981.x}

\bibitem[{Carnall {et~al.}(2018)Carnall, McLure, Dunlop, \& Davé}]{carnall}
Carnall, A.~C., McLure, R.~J., Dunlop, J.~S., \& Davé, R. 2018, Monthly Notices of the Royal Astronomical Society, 480, 4379, \dodoi{10.1093/mnras/sty2169}

\bibitem[{{Casey} {et~al.}(2025){Casey}, {Akins}, {Finkelstein}, {Franco}, {Fujimoto}, {Liu}, {Long}, {Magdis}, {Manning}, {McKinney}, {Shuntov}, \& {Tanaka}}]{2025Casey}
{Casey}, C.~M., {Akins}, H.~B., {Finkelstein}, S.~L., {et~al.} 2025, arXiv e-prints, arXiv:2505.18873, \dodoi{10.48550/arXiv.2505.18873}

\bibitem[{Castellano {et~al.}(2024)Castellano, Napolitano, Fontana, Roberts-Borsani, Treu, Vanzella, Zavala, Haro, Calabrò, Llerena, Mascia, Merlin, Paris, Pentericci, Santini, Bakx, Bergamini, Cupani, Dickinson, Filippenko, Glazebrook, Grillo, Kelly, Malkan, Mason, Morishita, Nanayakkara, Rosati, Sani, Wang, \& Yoon}]{castellano2024jwstnirspecspectroscopyremarkable}
Castellano, M., Napolitano, L., Fontana, A., {et~al.} 2024, JWST NIRSpec Spectroscopy of the Remarkable Bright Galaxy GHZ2/GLASS-z12 at Redshift 12.34.
\newblock \doarXiv{2403.10238}

\bibitem[{Chabrier(2003)}]{Chabrier_2003}
Chabrier, G. 2003, Publications of the Astronomical Society of the Pacific, 115, 763, \dodoi{10.1086/376392}

\bibitem[{Champagne {et~al.}(2025)Champagne, Wang, Yang, Fan, Hennawi, Sun, Bañados, Bosman, Costa, Habouzit, Jin, Jun, Li, Liu, Loiacono, Lupi, Mazzucchelli, Pudoka, Rojas-Ruiz, Tee, Trebitsch, Zhang, Zhuang, \& Zou}]{Champagne_2025}
Champagne, J.~B., Wang, F., Yang, J., {et~al.} 2025, The Astrophysical Journal, 981, 114, \dodoi{10.3847/1538-4357/adb1bc}

\bibitem[{Collaboration {et~al.}(2022)Collaboration, Price-Whelan, Lim, Earl, Starkman, Bradley, Shupe, Patil, Corrales, Brasseur, Nöthe, Donath, Tollerud, Morris, Ginsburg, Vaher, Weaver, Tocknell, Jamieson, van Kerkwijk, Robitaille, Merry, Bachetti, Günther, Authors, Aldcroft, Alvarado-Montes, Archibald, Bódi, Bapat, Barentsen, Bazán, Biswas, Boquien, Burke, Cara, Cara, Conroy, Conseil, Craig, Cross, Cruz, D’Eugenio, Dencheva, Devillepoix, Dietrich, Eigenbrot, Erben, Ferreira, Foreman-Mackey, Fox, Freij, Garg, Geda, Glattly, Gondhalekar, Gordon, Grant, Greenfield, Groener, Guest, Gurovich, Handberg, Hart, Hatfield-Dodds, Homeier, Hosseinzadeh, Jenness, Jones, Joseph, Kalmbach, Karamehmetoglu, Kałuszyński, Kelley, Kern, Kerzendorf, Koch, Kulumani, Lee, Ly, Ma, MacBride, Maljaars, Muna, Murphy, Norman, O’Steen, Oman, Pacifici, Pascual, Pascual-Granado, Patil, Perren, Pickering, Rastogi, Roulston, Ryan, Rykoff, Sabater, Sakurikar, Salgado, Sanghi, Saunders, Savchenko, Schwardt, Seifert-Eckert, Shih,
  Jain, Shukla, Sick, Simpson, Singanamalla, Singer, Singhal, Sinha, Sipőcz, Spitler, Stansby, Streicher, Šumak, Swinbank, Taranu, Tewary, Tremblay, de~Val-Borro, Kooten, Vasović, Verma, de~Miranda~Cardoso, Williams, Wilson, Winkel, Wood-Vasey, Xue, Yoachim, Zhang, Zonca, \& Contributors}]{AstropyCollaboration_2022}
Collaboration, T.~A., Price-Whelan, A.~M., Lim, P.~L., {et~al.} 2022, The Astrophysical Journal, 935, 167, \dodoi{10.3847/1538-4357/ac7c74}

\bibitem[{{Conroy} \& {Gunn}(2010)}]{conroyjames2010}
{Conroy}, C., \& {Gunn}, J.~E. 2010, {FSPS: Flexible Stellar Population Synthesis}, Astrophysics Source Code Library, record ascl:1010.043

\bibitem[{Conselice {et~al.}(2024)Conselice, Adams, Harvey, Austin, Ferreira, Ormerod, Duan, Trussler, Li, Juodzbalis, Westcott, Harris, Seeyave, Bluck, Windhorst, Bhatawdekar, Coe, Cohen, Cheng, Driver, Frye, Furtak, Grogin, Hathi, Holwerda, Jansen, Koekemoer, Marshall, Nonino, Robotham, Summers, Wilkins, Willmer, Yan, \& Zitrin}]{conselice2024epochsidiscoverystar}
Conselice, C.~J., Adams, N., Harvey, T., {et~al.} 2024, EPOCHS I. The Discovery and Star Forming Properties of Galaxies in the Epoch of Reionization at $6.5 < z < 18$ with PEARLS and Public JWST data.
\newblock \doarXiv{2407.14973}

\bibitem[{Correa \& Schaye(2020)}]{shmrcorrea}
Correa, C.~A., \& Schaye, J. 2020, Monthly Notices of the Royal Astronomical Society, 499, 3578, \dodoi{10.1093/mnras/staa3053}

\bibitem[{{Davis} {et~al.}(2007){Davis}, {Guhathakurta}, {Konidaris}, {Newman}, {Ashby}, {Biggs}, {Barmby}, {Bundy}, {Chapman}, {Coil}, {Conselice}, {Cooper}, {Croton}, {Eisenhardt}, {Ellis}, {Faber}, {Fang}, {Fazio}, {Georgakakis}, {Gerke}, {Goss}, {Gwyn}, {Harker}, {Hopkins}, {Huang}, {Ivison}, {Kassin}, {Kirby}, {Koekemoer}, {Koo}, {Laird}, {Le Floc'h}, {Lin}, {Lotz}, {Marshall}, {Martin}, {Metevier}, {Moustakas}, {Nandra}, {Noeske}, {Papovich}, {Phillips}, {Rich}, {Rieke}, {Rigopoulou}, {Salim}, {Schiminovich}, {Simard}, {Smail}, {Small}, {Weiner}, {Willmer}, {Willner}, {Wilson}, {Wright}, \& {Yan}}]{davis2007}
{Davis}, M., {Guhathakurta}, P., {Konidaris}, N.~P., {et~al.} 2007, \apjl, 660, L1, \dodoi{10.1086/517931}

\bibitem[{{De Cia} {et~al.}(2018){De Cia}, {Ledoux}, {Petitjean}, \& {Savaglio}}]{DeCia18}
{De Cia}, A., {Ledoux}, C., {Petitjean}, P., \& {Savaglio}, S. 2018, \aap, 611, A76, \dodoi{10.1051/0004-6361/201731970}

\bibitem[{de~Graaff {et~al.}(2024)de~Graaff, Brammer, Weibel, Lewis, Maseda, Oesch, Bezanson, Boogaard, Cleri, Cooper, Gottumukkala, Greene, Hirschmann, Hviding, Katz, Labbé, Leja, Matthee, McConachie, Miller, Naidu, Price, Rix, Setton, Suess, Wang, Whitaker, \& Williams}]{degraaff2024rubiescompletecensusbright}
de~Graaff, A., Brammer, G., Weibel, A., {et~al.} 2024, RUBIES: a complete census of the bright and red distant Universe with JWST/NIRSpec.
\newblock \doarXiv{2409.05948}

\bibitem[{{de Graaff} {et~al.}(2025){de Graaff}, {Rix}, {Naidu}, {Labbe}, {Wang}, {Leja}, {Matthee}, {Katz}, {Greene}, {Hviding}, {Baggen}, {Bezanson}, {Boogaard}, {Brammer}, {Dayal}, {van Dokkum}, {Goulding}, {Hirschmann}, {Maseda}, {McConachie}, {Miller}, {Nelson}, {Oesch}, {Setton}, {Shivaei}, {Weibel}, {Whitaker}, \& {Williams}}]{2025deGraaff}
{de Graaff}, A., {Rix}, H.-W., {Naidu}, R.~P., {et~al.} 2025, arXiv e-prints, arXiv:2503.16600, \dodoi{10.48550/arXiv.2503.16600}

\bibitem[{D'Eugenio {et~al.}(2024)D'Eugenio, Cameron, Scholtz, Carniani, Willott, Curtis-Lake, Bunker, Parlanti, Maiolino, Willmer, Jakobsen, Robertson, Johnson, Tacchella, Cargile, Rawle, Arribas, Chevallard, Curti, Egami, Eisenstein, Kumari, Looser, Rieke, Pino, Saxena, Übler, Venturi, Witstok, Baker, Bhatawdekar, Bonaventura, Boyett, Charlot, Danhaive, Hainline, Hausen, Helton, Ji, Ji, Jones, Joudžbalis, Maseda, Pérez-González, Perna, Puskás, Shivaei, Silcock, Simmonds, Smit, Sun, Villanueva, Williams, \& Zhu}]{deugenio2024jadesdatarelease3}
D'Eugenio, F., Cameron, A.~J., Scholtz, J., {et~al.} 2024, JADES Data Release 3 -- NIRSpec/MSA spectroscopy for 4,000 galaxies in the GOODS fields.
\newblock \doarXiv{2404.06531}

\bibitem[{{Diego} {et~al.}(2023){Diego}, {Meena}, {Adams}, {Broadhurst}, {Dai}, {Coe}, {Frye}, {Kelly}, {Koekemoer}, {Pascale}, {Willner}, {Zackrisson}, {Zitrin}, {Windhorst}, {Cohen}, {Jansen}, {Summers}, {Tompkins}, {Conselice}, {Driver}, {Yan}, {Grogin}, {Marshall}, {Pirzkal}, {Robotham}, {Ryan}, {Willmer}, {Bradley}, {Caminha}, {Caputi}, {Carleton}, \& {Kamieneski}}]{diego2023}
{Diego}, J.~M., {Meena}, A.~K., {Adams}, N.~J., {et~al.} 2023, \aap, 672, A3, \dodoi{10.1051/0004-6361/202245238}

\bibitem[{{Draine} \& {Li}(2007)}]{draine2007}
{Draine}, B.~T., \& {Li}, A. 2007, \apj, 657, 810, \dodoi{10.1086/511055}

\bibitem[{{Draine} {et~al.}(2014){Draine}, {Aniano}, {Krause}, {Groves}, {Sandstrom}, {Braun}, {Leroy}, {Klaas}, {Linz}, {Rix}, {Schinnerer}, {Schmiedeke}, \& {Walter}}]{draine2014}
{Draine}, B.~T., {Aniano}, G., {Krause}, O., {et~al.} 2014, \apj, 780, 172, \dodoi{10.1088/0004-637X/780/2/172}

\bibitem[{{Durkalec, A.} {et~al.}(2015){Durkalec, A.}, {Le Fèvre, O.}, {Pollo, A.}, {de la Torre, S.}, {Cassata, P.}, {Garilli, B.}, {Le Brun, V.}, {Lemaux, B. C.}, {Maccagni, D.}, {Pentericci, L.}, {Tasca, L. A. M.}, {Thomas, R.}, {Vanzella, E.}, {Zamorani, G.}, {Zucca, E.}, {Amorín, R.}, {Bardelli, S.}, {Cassarà, L. P.}, {Castellano, M.}, {Cimatti, A.}, {Cucciati, O.}, {Fontana, A.}, {Giavalisco, M.}, {Grazian, A.}, {Hathi, N. P.}, {Ilbert, O.}, {Paltani, S.}, {Ribeiro, B.}, {Schaerer, D.}, {Scodeggio, M.}, {Sommariva, V.}, {Talia, M.}, {Tresse, L.}, {Vergani, D.}, {Capak, P.}, {Charlot, S.}, {Contini, T.}, {Cuby, J. G.}, {Dunlop, J.}, {Fotopoulou, S.}, {Koekemoer, A.}, {López-Sanjuan, C.}, {Mellier, Y.}, {Pforr, J.}, {Salvato, M.}, {Scoville, N.}, {Taniguchi, Y.}, \& {Wang, P. W.}}]{clustering}
{Durkalec, A.}, {Le Fèvre, O.}, {Pollo, A.}, {et~al.} 2015, A\&A, 583, A128, \dodoi{10.1051/0004-6361/201425343}

\bibitem[{Durodola {et~al.}(2024)Durodola, Pacucci, \& Hickox}]{durodola2024}
Durodola, E., Pacucci, F., \& Hickox, R.~C. 2024, Exploring the AGN Fraction of a Sample of JWST's Little Red Dots at $5 < z < 8$: Overmassive Black Holes Are Strongly Favored.
\newblock \doarXiv{2406.10329}

\bibitem[{Eisenstein {et~al.}(2023)Eisenstein, Willott, Alberts, Arribas, Bonaventura, Bunker, Cameron, Carniani, Charlot, Curtis-Lake, D'Eugenio, Endsley, Ferruit, Giardino, Hainline, Hausen, Jakobsen, Johnson, Maiolino, Rieke, Rieke, Rix, Robertson, Stark, Tacchella, Williams, Willmer, Baker, Baum, Bhatawdekar, Boyett, Chen, Chevallard, Circosta, Curti, Danhaive, DeCoursey, de~Graaff, Dressler, Egami, Helton, Hviding, Ji, Jones, Kumari, Lützgendorf, Laseter, Looser, Lyu, Maseda, Nelson, Parlanti, Perna, Puskás, Rawle, Pino, Sandles, Saxena, Scholtz, Sharpe, Shivaei, Silcock, Simmonds, Skarbinski, Smit, Stone, Suess, Sun, Tang, Topping, Übler, Villanueva, Wallace, Whitler, Witstok, \& Woodrum}]{eisenstein2023overviewjwstadvanceddeep}
Eisenstein, D.~J., Willott, C., Alberts, S., {et~al.} 2023, Overview of the JWST Advanced Deep Extragalactic Survey (JADES).
\newblock \doarXiv{2306.02465}

\bibitem[{Frye {et~al.}(2023)Frye, Pascale, Foo, Leimbach, Garuda, Robles, Summers, Diaz, Kamieneski, Furtak, Cohen, Diego, Beauchesne, Windhorst, Willner, Koekemoer, Zitrin, Caminha, Caputi, Coe, Conselice, Dai, Dole, Driver, Grogin, Harrington, Jansen, Kneib, Lehnert, Lowenthal, Marshall, Menanteau, Pampliega, Pirzkal, Polletta, Richard, Robotham, Ryan, Rutkowski, Sifón, Tompkins, Wang, Yan, \& Yun}]{Frye_2023}
Frye, B.~L., Pascale, M., Foo, N., {et~al.} 2023, The Astrophysical Journal, 952, 81, \dodoi{10.3847/1538-4357/acd929}

\bibitem[{Fudamoto {et~al.}(2025)Fudamoto, Helton, Lin, Sun, Behroozi, Hsiao, Egami, Bunker, Harikane, Ouchi, Liu, Liu, Maiolino, Ji, Jin, Tee, Wang, Willmer, Xu, \& Zhu}]{fudamoto2025sapphiresgalaxyoverdensityheart}
Fudamoto, Y., Helton, J.~M., Lin, X., {et~al.} 2025, SAPPHIRES: A Galaxy Over-Density in the Heart of Cosmic Reionization at $z=8.47$.
\newblock \doarXiv{2503.15597}

\bibitem[{Furtak {et~al.}(2023)Furtak, Zitrin, Plat, Fujimoto, Wang, Nelson, Labbé, Bezanson, Brammer, van Dokkum, Endsley, Glazebrook, Greene, Leja, Price, Smit, Stark, Weaver, Whitaker, Atek, Chevallard, Curtis-Lake, Dayal, Feltre, Franx, Fudamoto, Marchesini, Mowla, Pan, Suess, Vidal-García, \& Williams}]{Furtak_2023}
Furtak, L.~J., Zitrin, A., Plat, A., {et~al.} 2023, The Astrophysical Journal, 952, 142, \dodoi{10.3847/1538-4357/acdc9d}

\bibitem[{Furtak {et~al.}(2024)Furtak, Labb{\'e}, Zitrin, Greene, Dayal, Chemerynska, Kokorev, Miller, Goulding, de~Graaff, Bezanson, Brammer, Cutler, Leja, Pan, Price, Wang, Weaver, Whitaker, Atek, Bogd{\'a}n, Charlot, Curtis-Lake, van Dokkum, Endsley, Feldmann, Fudamoto, Fujimoto, Glazebrook, Juneau, Marchesini, Maseda, Nelson, Oesch, Plat, Setton, Stark, \& Williams}]{Furtak2024}
Furtak, L.~J., Labb{\'e}, I., Zitrin, A., {et~al.} 2024, Nature, 628, 57, \dodoi{10.1038/s41586-024-07184-8}

\bibitem[{Girelli {et~al.}(2020)Girelli, Pozzetti, Bolzonella, Giocoli, Marulli, \& Baldi}]{Girelli_2020}
Girelli, G., Pozzetti, L., Bolzonella, M., {et~al.} 2020, Astronomy \& Astrophysics, 634, A135, \dodoi{10.1051/0004-6361/201936329}

\bibitem[{Glazebrook {et~al.}(2024)Glazebrook, Nanayakkara, Schreiber, Lagos, Kawinwanichakij, Jacobs, Chittenden, Brammer, Kacprzak, Labbe, Marchesini, Marsan, Oesch, Papovich, Remus, Tran, Esdaile, \& Chandro-Gomez}]{glazebrook}
Glazebrook, K., Nanayakkara, T., Schreiber, C., {et~al.} 2024, Nature, 628, \dodoi{10.1038/s41586-024-07191-9}

\bibitem[{Grazian {et~al.}(2024)Grazian, Giallongo, Boutsia, Cristiani, Fontanot, Bischetti, Bisigello, Bongiorno, Calderone, Tegli, Cupani, Lucia, D’Odorico, Feruglio, Fiore, Gandolfi, Girardi, Guarneri, Hirschmann, Porru, Rodighiero, Saccheo, Simioni, Trost, \& Viitanen}]{Grazian_2024}
Grazian, A., Giallongo, E., Boutsia, K., {et~al.} 2024, The Astrophysical Journal, 974, 84, \dodoi{10.3847/1538-4357/ad6980}

\bibitem[{Greene {et~al.}(2024)Greene, Labbe, Goulding, Furtak, Chemerynska, Kokorev, Dayal, Volonteri, Williams, Wang, Setton, Burgasser, Bezanson, Atek, Brammer, Cutler, Feldmann, Fujimoto, Glazebrook, de~Graaff, Khullar, Leja, Marchesini, Maseda, Matthee, Miller, Naidu, Nanayakkara, Oesch, Pan, Papovich, Price, van Dokkum, Weaver, Whitaker, \& Zitrin}]{Greene_2024}
Greene, J.~E., Labbe, I., Goulding, A.~D., {et~al.} 2024, The Astrophysical Journal, 964, 39, \dodoi{10.3847/1538-4357/ad1e5f}

\bibitem[{Grogin {et~al.}(2011)Grogin, Kocevski, Faber, Ferguson, Koekemoer, Riess, Acquaviva, Alexander, Almaini, Ashby, Barden, Bell, Bournaud, Brown, Caputi, Casertano, Cassata, Castellano, Challis, Chary, Cheung, Cirasuolo, Conselice, Cooray, Croton, Daddi, Dahlen, Davé, de~Mello, Dekel, Dickinson, Dolch, Donley, Dunlop, Dutton, Elbaz, Fazio, Filippenko, Finkelstein, Fontana, Gardner, Garnavich, Gawiser, Giavalisco, Grazian, Guo, Hathi, Häussler, Hopkins, Huang, Huang, Jha, Kartaltepe, Kirshner, Koo, Lai, Lee, Li, Lotz, Lucas, Madau, McCarthy, McGrath, McIntosh, McLure, Mobasher, Moustakas, Mozena, Nandra, Newman, Niemi, Noeske, Papovich, Pentericci, Pope, Primack, Rajan, Ravindranath, Reddy, Renzini, Rix, Robaina, Rodney, Rosario, Rosati, Salimbeni, Scarlata, Siana, Simard, Smidt, Somerville, Spinrad, Straughn, Strolger, Telford, Teplitz, Trump, van~der Wel, Villforth, Wechsler, Weiner, Wiklind, Wild, Wilson, Wuyts, Yan, \& Yun}]{Grogin_2011}
Grogin, N.~A., Kocevski, D.~D., Faber, S.~M., {et~al.} 2011, The Astrophysical Journal Supplement Series, 197, 35, \dodoi{10.1088/0067-0049/197/2/35}

\bibitem[{Guia {et~al.}(2024)Guia, Pacucci, \& Kocevski}]{Guia_2024}
Guia, C.~A., Pacucci, F., \& Kocevski, D.~D. 2024, Research Notes of the AAS, 8, 207, \dodoi{10.3847/2515-5172/ad7262}

\bibitem[{{G{\"u}ltekin} {et~al.}(2009){G{\"u}ltekin}, {Richstone}, {Gebhardt}, {Lauer}, {Tremaine}, {Aller}, {Bender}, {Dressler}, {Faber}, {Filippenko}, {Green}, {Ho}, {Kormendy}, {Magorrian}, {Pinkney}, \& {Siopis}}]{gultekin2009}
{G{\"u}ltekin}, K., {Richstone}, D.~O., {Gebhardt}, K., {et~al.} 2009, \apj, 698, 198, \dodoi{10.1088/0004-637X/698/1/198}

\bibitem[{{Guo} {et~al.}(2018){Guo}, {Rafelski}, {Bell}, {Conselice}, {Dekel}, {Faber}, {Giavalisco}, {Koekemoer}, {Koo}, {Lu}, {Mandelker}, {Primack}, {Ceverino}, {de Mello}, {Ferguson}, {Hathi}, {Kocevski}, {Lucas}, {P{\'e}rez-Gonz{\'a}lez}, {Ravindranath}, {Soto}, {Straughn}, \& {Wang}}]{Guo2018}
{Guo}, Y., {Rafelski}, M., {Bell}, E.~F., {et~al.} 2018, \apj, 853, 108, \dodoi{10.3847/1538-4357/aaa018}

\bibitem[{{Habouzit} \& {Department of Astronomy}(2025)}]{2025habouzit}
{Habouzit}, M., \& {Department of Astronomy}, Chemin~d'Ecogia, U. o.~G. 2025, \mnras, 537, 2323, \dodoi{10.1093/mnras/staf167}

\bibitem[{{Hainline} {et~al.}(2024{\natexlab{a}}){Hainline}, {Johnson}, {Robertson}, {Tacchella}, {Helton}, {Sun}, {Eisenstein}, {Simmonds}, {Topping}, {Whitler}, {Willmer}, {Rieke}, {Suess}, {Hviding}, {Cameron}, {Alberts}, {Baker}, {Baum}, {Bhatawdekar}, {Bonaventura}, {Boyett}, {Bunker}, {Carniani}, {Charlot}, {Chevallard}, {Chen}, {Curti}, {Curtis-Lake}, {D'Eugenio}, {Egami}, {Endsley}, {Hausen}, {Ji}, {Looser}, {Lyu}, {Maiolino}, {Nelson}, {Pusk{\'a}s}, {Rawle}, {Sandles}, {Saxena}, {Smit}, {Stark}, {Williams}, {Willott}, \& {Witstok}}]{hainline2024}
{Hainline}, K.~N., {Johnson}, B.~D., {Robertson}, B., {et~al.} 2024{\natexlab{a}}, \apj, 964, 71, \dodoi{10.3847/1538-4357/ad1ee4}

\bibitem[{{Hainline} {et~al.}(2024{\natexlab{b}}){Hainline}, {Helton}, {Johnson}, {Sun}, {Topping}, {Leisenring}, {Baker}, {Eisenstein}, {Hausen}, {Hviding}, {Lyu}, {Robertson}, {Tacchella}, {Williams}, {Willmer}, \& {Roellig}}]{hainline2024bd}
{Hainline}, K.~N., {Helton}, J.~M., {Johnson}, B.~D., {et~al.} 2024{\natexlab{b}}, \apj, 964, 66, \dodoi{10.3847/1538-4357/ad20d1}

\bibitem[{Haro {et~al.}(2023)Haro, Dickinson, Finkelstein, Fujimoto, Fernández, Kartaltepe, Jung, Cole, Burgarella, Chworowsky, Hutchison, Morales, Papovich, Simons, Amorín, Backhaus, Bagley, Bisigello, Calabrò, Castellano, Cleri, Davé, Dekel, Ferguson, Fontana, Gawiser, Giavalisco, Harish, Hathi, Hirschmann, Holwerda, Huertas-Company, Koekemoer, Larson, Lucas, Mobasher, Pérez-González, Pirzkal, Rose, Santini, Trump, de~la Vega, Wang, Weiner, Wilkins, Yang, Yung, \& Zavala}]{ArrabalHaro_2023}
Haro, P.~A., Dickinson, M., Finkelstein, S.~L., {et~al.} 2023, The Astrophysical Journal Letters, 951, L22, \dodoi{10.3847/2041-8213/acdd54}

\bibitem[{He {et~al.}(2023)He, Akiyama, Enoki, Ichikawa, Inayoshi, Kashikawa, Kawaguchi, Matsuoka, Nagao, Onoue, Oogi, Schulze, Toba, \& Ueda}]{he2023}
He, W., Akiyama, M., Enoki, M., {et~al.} 2023, Black hole mass and Eddington ratio distributions of less-luminous quasars at $z\sim4$ in the Subaru Hyper Suprime-Cam Wide field.
\newblock \doarXiv{2311.08922}

\bibitem[{{Heintz} {et~al.}(2024){Heintz}, {Watson}, {Brammer}, {Vejlgaard}, {Hutter}, {Strait}, {Matthee}, {Oesch}, {Jakobsson}, {Tanvir}, {Laursen}, {Naidu}, {Mason}, {Killi}, {Jung}, {Hsiao}, {Abdurro'uf}, {Coe}, {Arrabal Haro}, {Finkelstein}, \& {Toft}}]{2024heintz}
{Heintz}, K.~E., {Watson}, D., {Brammer}, G., {et~al.} 2024, Science, 384, 890, \dodoi{10.1126/science.adj0343}

\bibitem[{Hoaglin {et~al.}(1983)Hoaglin, Mosteller, \& Tukey}]{hoaglin2000understanding}
Hoaglin, D.~C., Mosteller, F., \& Tukey, J.~W. 1983, Understanding Robust and Exploratory Data Analysis (New York: Wiley) (Wiley)

\bibitem[{{Holwerda} {et~al.}(2024){Holwerda}, {Hsu}, {Hathi}, {Bisigello}, {de la Vega}, {Haro}, {Bagley}, {Dickinson}, {Finkelstein}, {Kartaltepe}, {Koekemoer}, {Papovich}, {Pirzkal}, {Cook}, {Robertson}, {Casey}, {Aganze}, {P{\'e}rez-Gonz{\'a}lez}, {Lucas}, {Jogee}, {Wilkins}, {Burgarella}, \& {Kirkpatrick}}]{2024Holwerda}
{Holwerda}, B.~W., {Hsu}, C.-C., {Hathi}, N., {et~al.} 2024, \mnras, 529, 1067, \dodoi{10.1093/mnras/stae316}

\bibitem[{Hu(2008)}]{hu2008}
Hu, J. 2008, Monthly Notices of the Royal Astronomical Society, 386, 2242, \dodoi{10.1111/j.1365-2966.2008.13195.x}

\bibitem[{Hunter(2007)}]{matplot}
Hunter, J.~D. 2007, Computing in Science \& Engineering, 9, 90, \dodoi{10.1109/MCSE.2007.55}

\bibitem[{Illingworth {et~al.}(2017)Illingworth, Magee, Bouwens, Oesch, Labbe, van Dokkum, Whitaker, Holden, Franx, \& Gonzalez}]{illingworth2017hubblelegacyfieldshlfgoodss}
Illingworth, G., Magee, D., Bouwens, R., {et~al.} 2017, The Hubble Legacy Fields (HLF-GOODS-S) v1.5 Data Products: Combining 2442 Orbits of GOODS-S/CDF-S Region ACS and WFC3/IR Images.
\newblock \doarXiv{1606.00841}

\bibitem[{{Inayoshi} \& {Maiolino}(2025)}]{2025ApJ...980L..27I}
{Inayoshi}, K., \& {Maiolino}, R. 2025, \apjl, 980, L27, \dodoi{10.3847/2041-8213/adaebd}

\bibitem[{{Inoue}(2011)}]{inoue_nebular}
{Inoue}, A.~K. 2011, \mnras, 415, 2920, \dodoi{10.1111/j.1365-2966.2011.18906.x}

\bibitem[{{Ji} {et~al.}(2025){Ji}, {Maiolino}, {{\"U}bler}, {Scholtz}, {D'Eugenio}, {Sun}, {Perna}, {Turner}, {Arribas}, {Bennett}, {Bunker}, {Carniani}, {Charlot}, {Cresci}, {Curti}, {Egami}, {Fabian}, {Inayoshi}, {Isobe}, {Jones}, {Juod{\v{z}}balis}, {Kumari}, {Lyu}, {Mazzolari}, {Parlanti}, {Robertson}, {Rodr{\'\i}guez Del Pino}, {Schneider}, {Sijacki}, {Tacchella}, {Trinca}, {Valiante}, {Venturi}, {Volonteri}, {Willott}, {Witten}, \& {Witstok}}]{2025Ji}
{Ji}, X., {Maiolino}, R., {{\"U}bler}, H., {et~al.} 2025, arXiv e-prints, arXiv:2501.13082, \dodoi{10.48550/arXiv.2501.13082}

\bibitem[{Kass \& Raftery(1995)}]{Raftery1995}
Kass, R.~E., \& Raftery, A.~E. 1995, Journal of the American Statistical Association, 90, 773.
\newblock \url{http://www.jstor.org/stable/2291091}

\bibitem[{Kocevski {et~al.}(2023)Kocevski, Onoue, Inayoshi, Trump, Haro, Grazian, Dickinson, Finkelstein, Kartaltepe, Hirschmann, Fujimoto, Juneau, Amorin, Bagley, Barro, Bell, Bisigello, Calabro, Cleri, Cooper, Ding, Grogin, Ho, Inoue, Jiang, Jones, Koekemoer, Li, Li, McGrath, Molina, Papovich, Perez-Gonzalez, Pirzkal, Wilkins, Yang, \& Yung}]{kocevski2023}
Kocevski, D.~D., Onoue, M., Inayoshi, K., {et~al.} 2023, Hidden Little Monsters: Spectroscopic Identification of Low-Mass, Broad-Line AGN at $z>5$ with CEERS.
\newblock \doarXiv{2302.00012}

\bibitem[{Kocevski {et~al.}(2025)Kocevski, Finkelstein, Barro, Taylor, Calabrò, Laloux, Buchner, Trump, Leung, Yang, Dickinson, Pérez-González, Pacucci, Inayoshi, Somerville, McGrath, Akins, Bagley, Bowler, Bisigello, Carnall, Casey, Cheng, Cleri, Costantin, Cullen, Davis, Donnan, Dunlop, Ellis, Ferguson, Fujimoto, Fontana, Giavalisco, Grazian, Grogin, Hathi, Hirschmann, Huertas-Company, Holwerda, Illingworth, Juneau, Kartaltepe, Koekemoer, Li, Lucas, Magee, Mason, McLeod, McLure, Napolitano, Papovich, Pirzkal, Rodighiero, Santini, Wilkins, \& Yung}]{Kocevski_2025}
Kocevski, D.~D., Finkelstein, S.~L., Barro, G., {et~al.} 2025, The Astrophysical Journal, 986, 126, \dodoi{10.3847/1538-4357/adbc7d}

\bibitem[{Koekemoer {et~al.}(2011)Koekemoer, Faber, Ferguson, Grogin, Kocevski, Koo, Lai, Lotz, Lucas, McGrath, Ogaz, Rajan, Riess, Rodney, Strolger, Casertano, Castellano, Dahlen, Dickinson, Dolch, Fontana, Giavalisco, Grazian, Guo, Hathi, Huang, van~der Wel, Yan, Acquaviva, Alexander, Almaini, Ashby, Barden, Bell, Bournaud, Brown, Caputi, Cassata, Challis, Chary, Cheung, Cirasuolo, Conselice, Cooray, Croton, Daddi, Davé, de~Mello, de~Ravel, Dekel, Donley, Dunlop, Dutton, Elbaz, Fazio, Filippenko, Finkelstein, Frazer, Gardner, Garnavich, Gawiser, Gruetzbauch, Hartley, Häussler, Herrington, Hopkins, Huang, Jha, Johnson, Kartaltepe, Khostovan, Kirshner, Lani, Lee, Li, Madau, McCarthy, McIntosh, McLure, McPartland, Mobasher, Moreira, Mortlock, Moustakas, Mozena, Nandra, Newman, Nielsen, Niemi, Noeske, Papovich, Pentericci, Pope, Primack, Ravindranath, Reddy, Renzini, Rix, Robaina, Rosario, Rosati, Salimbeni, Scarlata, Siana, Simard, Smidt, Snyder, Somerville, Spinrad, Straughn, Telford, Teplitz, Trump,
  Vargas, Villforth, Wagner, Wandro, Wechsler, Weiner, Wiklind, Wild, Wilson, Wuyts, \& Yun}]{Koekemoer_2011}
Koekemoer, A.~M., Faber, S.~M., Ferguson, H.~C., {et~al.} 2011, The Astrophysical Journal Supplement Series, 197, 36, \dodoi{10.1088/0067-0049/197/2/36}

\bibitem[{Kokorev {et~al.}(2023)Kokorev, Fujimoto, Labbe, Greene, Bezanson, Dayal, Nelson, Atek, Brammer, Caputi, Chemerynska, Cutler, Feldmann, Fudamoto, Furtak, Goulding, de~Graaff, Leja, Marchesini, Miller, Nanayakkara, Oesch, Pan, Price, Setton, Smit, Stefanon, Wang, Weaver, Whitaker, Williams, \& Zitrin}]{Kokorev_2023}
Kokorev, V., Fujimoto, S., Labbe, I., {et~al.} 2023, The Astrophysical Journal Letters, 957, L7, \dodoi{10.3847/2041-8213/ad037a}

\bibitem[{Kokorev {et~al.}(2024)Kokorev, Caputi, Greene, Dayal, Trebitsch, Cutler, Fujimoto, Labbé, Miller, Iani, Navarro-Carrera, \& Rinaldi}]{kokorev2024}
Kokorev, V., Caputi, K.~I., Greene, J.~E., {et~al.} 2024, arXiv e-prints.
\newblock \doarXiv{2401.09981}

\bibitem[{{Kokubo} \& {Harikane}(2024)}]{kokubo2024}
{Kokubo}, M., \& {Harikane}, Y. 2024, arXiv e-prints, arXiv:2407.04777, \dodoi{10.48550/arXiv.2407.04777}

\bibitem[{Kormendy \& Ho(2013)}]{kormendyho2013}
Kormendy, J., \& Ho, L.~C. 2013, Annual Review of Astronomy and Astrophysics, 51, 511, \dodoi{https://doi.org/10.1146/annurev-astro-082708-101811}

\bibitem[{Labbe {et~al.}(2023)Labbe, Greene, Bezanson, Fujimoto, Furtak, Goulding, Matthee, Naidu, Oesch, Atek, Brammer, Chemerynska, Coe, Cutler, Dayal, Feldmann, Franx, Glazebrook, Leja, Marchesini, Maseda, Nanayakkara, Nelson, Pan, Papovich, Price, Suess, Wang, Whitaker, Williams, \& Zitrin}]{labbe2023uca}
Labbe, I., Greene, J.~E., Bezanson, R., {et~al.} 2023, UNCOVER: Candidate Red Active Galactic Nuclei at 3 $<z<$ 7 with JWST and ALMA.
\newblock \doarXiv{2306.07320}

\bibitem[{Labbe {et~al.}(2024)Labbe, Greene, Matthee, Treiber, Kokorev, Miller, Kramarenko, Setton, Ma, Goulding, Bezanson, Naidu, Williams, Atek, Brammer, Cutler, Chemerynska, Cloonan, Dayal, de~Graaff, Fudamoto, Fujimoto, Furtak, Glazebrook, Heintz, Leja, Marchesini, Nanayakkara, Nelson, Oesch, Pan, Price, Shivaei, Sobral, Suess, van Dokkum, Wang, Weaver, Whitaker, \& Zitrin}]{labbe2024unambiguousagnbalmerbreak}
Labbe, I., Greene, J.~E., Matthee, J., {et~al.} 2024, An unambiguous AGN and a Balmer break in an Ultraluminous Little Red Dot at z=4.47 from Ultradeep UNCOVER and All the Little Things Spectroscopy.
\newblock \doarXiv{2412.04557}

\bibitem[{Langeroodi \& Hjorth(2023)}]{langeroodi}
Langeroodi, D., \& Hjorth, J. 2023, The Astrophysical Journal Letters, 957, L27, \dodoi{10.3847/2041-8213/acfeec}

\bibitem[{{Larson} {et~al.}(2023){Larson}, {Hutchison}, {Bagley}, {Finkelstein}, {Yung}, {Somerville}, {Hirschmann}, {Brammer}, {Holwerda}, {Papovich}, {Morales}, \& {Wilkins}}]{larson2023}
{Larson}, R.~L., {Hutchison}, T.~A., {Bagley}, M., {et~al.} 2023, \apj, 958, 141, \dodoi{10.3847/1538-4357/acfed4}

\bibitem[{Leung {et~al.}(2024)Leung, Finkelstein, Pérez-González, Morales, Taylor, Barro, Kocevski, Akins, Carnall, Óscar A.~Chávez~Ortiz, Cleri, Cullen, Donnan, Dunlop, Ellis, Grogin, Hirschmann, Koekemoer, Kokorev, Lucas, McLeod, Papovich, \& Yung}]{leung2024exploringnaturelittlered}
Leung, G. C.~K., Finkelstein, S.~L., Pérez-González, P.~G., {et~al.} 2024, Exploring the Nature of Little Red Dots: Constraints on AGN and Stellar Contributions from PRIMER MIRI Imaging.
\newblock \doarXiv{2411.12005}

\bibitem[{Li {et~al.}(2025)Li, Conselice, Sarron, Harvey, Austin, Adams, Trussler, Duan, Ferreira, Westcott, Harris, Dole, Grogin, Frye, Koekemoer, Robertson, Windhorst, Polletta, Hathi, \& Jansen}]{li2024epochspaperxenvironmental}
Li, Q., Conselice, C.~J., Sarron, F., {et~al.} 2025, Monthly Notices of the Royal Astronomical Society, 539, 1796, \dodoi{10.1093/mnras/staf543}

\bibitem[{Li {et~al.}(2024)Li, Inayoshi, Chen, Ichikawa, \& Ho}]{li2024}
Li, Z., Inayoshi, K., Chen, K., Ichikawa, K., \& Ho, L.~C. 2024, Little Red Dots: Rapidly Growing Black Holes Reddened by Extended Dusty Flows.
\newblock \doarXiv{2407.10760}

\bibitem[{{Lopes} {et~al.}(2016){Lopes}, {Rembold}, {Ribeiro}, {Nascimento}, \& {Vajgel}}]{lopes2016}
{Lopes}, P.~A.~A., {Rembold}, S.~B., {Ribeiro}, A.~L.~B., {Nascimento}, R.~S., \& {Vajgel}, B. 2016, \mnras, 461, 2559, \dodoi{10.1093/mnras/stw1497}

\bibitem[{Ma {et~al.}(2024)Ma, Sun, Cheng, Yan, Sun, Foo, Egami, Diego, Cohen, Jansen, Summers, Windhorst, D'Silva, Koekemoer, Coe, Conselice, Driver, Frye, Grogin, Marshall, Nonino, au2, Pirzkal, Robotham, Russell E.~Ryan, Willmer, Adams, Hathi, Dole, Willner, Espada, Furtak, Hsiao, Li, Chen, Jolly, \& Chen}]{ma2024jwstviewinfantgalaxies}
Ma, Z., Sun, B., Cheng, C., {et~al.} 2024, JWST view of four infant galaxies at z=8.31-8.49 in the MACS0416 field and implications for reionization.
\newblock \doarXiv{2406.04617}

\bibitem[{Madau {et~al.}(2024)Madau, Giallongo, Grazian, \& Haardt}]{madau2024cosmicreionizationjwstera}
Madau, P., Giallongo, E., Grazian, A., \& Haardt, F. 2024, Cosmic Reionization in the JWST Era: Back to AGNs?
\newblock \doarXiv{2406.18697}

\bibitem[{{Maiolino} {et~al.}(2024){Maiolino}, {Risaliti}, {Signorini}, {Trefoloni}, {Juodzbalis}, {Scholtz}, {Uebler}, {D'Eugenio}, {Carniani}, {Fabian}, {Ji}, {Mazzolari}, {Bertola}, {Brusa}, {Bunker}, {Charlot}, {Comastri}, {Cresci}, {DeCoursey}, {Egami}, {Fiore}, {Gilli}, {Perna}, {Tacchella}, \& {Venturi}}]{maiolino2024}
{Maiolino}, R., {Risaliti}, G., {Signorini}, M., {et~al.} 2024, arXiv e-prints, arXiv:2405.00504, \dodoi{10.48550/arXiv.2405.00504}

\bibitem[{{Maraston}(2005)}]{2005Maraston}
{Maraston}, C. 2005, \mnras, 362, 799, \dodoi{10.1111/j.1365-2966.2005.09270.x}

\bibitem[{{Marley} {et~al.}(2021){Marley}, {Saumon}, {Visscher}, {Lupu}, {Freedman}, {Morley}, {Fortney}, {Seay}, {Smith}, {Teal}, \& {Wang}}]{marley2021}
{Marley}, M.~S., {Saumon}, D., {Visscher}, C., {et~al.} 2021, \apj, 920, 85, \dodoi{10.3847/1538-4357/ac141d}

\bibitem[{Maseda {et~al.}(2024)Maseda, de~Graaff, Franx, Rix, Carniani, Laseter, Dudzevičiūtė, Rawle, Parlanti, Arribas, Bunker, Cameron, Charlot, Curti, D’Eugenio, Jones, Kumari, Maiolino, Übler, Saxena, Smit, Willott, \& Witstok}]{Maseda_2024}
Maseda, M.~V., de~Graaff, A., Franx, M., {et~al.} 2024, Astronomy \& Astrophysics, 689, A73, \dodoi{10.1051/0004-6361/202449914}

\bibitem[{Matsuoka {et~al.}(2018)Matsuoka, Strauss, Kashikawa, Onoue, Iwasawa, Tang, Lee, Imanishi, Nagao, Akiyama, Asami, Bosch, Furusawa, Goto, Gunn, Harikane, Ikeda, Izumi, Kawaguchi, Kato, Kikuta, Kohno, Komiyama, Lupton, Minezaki, Miyazaki, Murayama, Niida, Nishizawa, Noboriguchi, Oguri, Ono, Ouchi, Price, Sameshima, Schulze, Shirakata, Silverman, Sugiyama, Tait, Takada, Takata, Tanaka, Toba, Utsumi, Wang, \& Yamashita}]{Matsuoka_2018}
Matsuoka, Y., Strauss, M.~A., Kashikawa, N., {et~al.} 2018, The Astrophysical Journal, 869, 150, \dodoi{10.3847/1538-4357/aaee7a}

\bibitem[{Matthee {et~al.}(2024{\natexlab{a}})Matthee, Naidu, Brammer, Chisholm, Eilers, Goulding, Greene, Kashino, Labbe, Lilly, Mackenzie, Oesch, Weibel, Wuyts, Xiao, Bordoloi, Bouwens, van Dokkum, Illingworth, Kramarenko, Maseda, Mason, Meyer, Nelson, Reddy, Shivaei, Simcoe, \& Yue}]{matthee2024}
Matthee, J., Naidu, R.~P., Brammer, G., {et~al.} 2024{\natexlab{a}}, Little Red Dots: an abundant population of faint AGN at z~5 revealed by the EIGER and FRESCO JWST surveys.
\newblock \doarXiv{2306.05448}

\bibitem[{Matthee {et~al.}(2024{\natexlab{b}})Matthee, Naidu, Kotiwale, Furtak, Kramarenko, Mackenzie, Greene, Adamo, Bouwens, Cesare, Eilers, de~Graaff, Heintz, Kashino, Maseda, Tacchella, \& Torralba}]{matthee2024environmentalevidenceoverlymassive}
Matthee, J., Naidu, R.~P., Kotiwale, G., {et~al.} 2024{\natexlab{b}}, Environmental Evidence for Overly Massive Black Holes in Low Mass Galaxies and a Black Hole - Halo Mass Relation at $z \sim 5$.
\newblock \doarXiv{2412.02846}

\bibitem[{Morishita {et~al.}(2025)Morishita, Liu, Stiavelli, Treu, Trenti, Chartab, Roberts-Borsani, Vulcani, Bergamini, Castellano, \& Grillo}]{morishita2025acceleratedemergenceevolvedgalaxies}
Morishita, T., Liu, Z., Stiavelli, M., {et~al.} 2025, Accelerated Emergence of Evolved Galaxies in Early Overdensities at $z\sim5.7$.
\newblock \doarXiv{2408.10980}

\bibitem[{{Naidu} {et~al.}(2025){Naidu}, {Matthee}, {Katz}, {de Graaff}, {Oesch}, {Smith}, {Greene}, {Brammer}, {Weibel}, {Hviding}, {Chisholm}, {Labb\textbackslash'e}, {Simcoe}, {Witten}, {Atek}, {Baggen}, {Belli}, {Bezanson}, {Boogaard}, {Bose}, {Covelo-Paz}, {Dayal}, {Fudamoto}, {Furtak}, {Giovinazzo}, {Goulding}, {Gronke}, {Heintz}, {Hirschmann}, {Illingworth}, {Inoue}, {Johnson}, {Leja}, {Leonova}, {McConachie}, {Maseda}, {Natarajan}, {Nelson}, {Setton}, {Shivaei}, {Sobral}, {Stefanon}, {Tacchella}, {Toft}, {Torralba}, {van Dokkum}, {van der Wel}, {Volonteri}, {Walter}, {Wang}, \& {Watson}}]{2025Naidu}
{Naidu}, R.~P., {Matthee}, J., {Katz}, H., {et~al.} 2025, arXiv e-prints, arXiv:2503.16596, \dodoi{10.48550/arXiv.2503.16596}

\bibitem[{Niida {et~al.}(2020)Niida, Nagao, Ikeda, Akiyama, Matsuoka, He, Matsuoka, Toba, Onoue, Kobayashi, Taniguchi, Furusawa, Harikane, Imanishi, Kashikawa, Kawaguchi, Komiyama, Shirakata, Terashima, \& Ueda}]{Niida_2020}
Niida, M., Nagao, T., Ikeda, H., {et~al.} 2020, The Astrophysical Journal, 904, 89, \dodoi{10.3847/1538-4357/abbe11}

\bibitem[{{Noll} {et~al.}(2009){Noll}, {Burgarella}, {Giovannoli}, {Buat}, {Marcillac}, \& {Mu{\~n}oz-Mateos}}]{noll2009}
{Noll}, S., {Burgarella}, D., {Giovannoli}, E., {et~al.} 2009, \aap, 507, 1793, \dodoi{10.1051/0004-6361/200912497}

\bibitem[{O'Brien {et~al.}(2024)O'Brien, Jansen, Grogin, Cohen, Smith, Silver, au2, Windhorst, Carleton, Koekemoer, Hathi, Willmer, Frye, Alpaslan, Ashby, Ashcraft, Bonoli, Brisken, Cappelluti, Civano, Conselice, Dhillon, Driver, Duncan, Dupke, Elvis, Fazio, Finkelstein, Gim, Griffiths, Hammel, Hyun, Im, Jones, Kim, Ladjelate, Larson, Malhotra, Marshall, Milam, Pierel, Rhoads, Rodney, Röttgering, Rutkowski, R.~E.~Ryan, Ward, White, van Weeren, Zhao, Summers, D'Silva, au2, Robotham, Coe, Nonino, Pirzkal, Yan, \& Acharya}]{obrien2024treasurehunttransientsvariabilitydiscovered}
O'Brien, R., Jansen, R.~A., Grogin, N.~A., {et~al.} 2024, TREASUREHUNT: Transients and Variability Discovered with HST in the JWST North Ecliptic Pole Time Domain Field.
\newblock \doarXiv{2401.04944}

\bibitem[{{Oke}(1974)}]{oke1974}
{Oke}, J.~B. 1974, \apjs, 27, 21, \dodoi{10.1086/190287}

\bibitem[{{Oke} \& {Gunn}(1983)}]{Oke1983}
{Oke}, J.~B., \& {Gunn}, J.~E. 1983, \apj, 266, 713, \dodoi{10.1086/160817}

\bibitem[{Perrin {et~al.}(2014)Perrin, Sivaramakrishnan, Lajoie, Elliott, Pueyo, Ravindranath, \& Albert}]{Perrin2014}
Perrin, M.~D., Sivaramakrishnan, A., Lajoie, C.-P., {et~al.} 2014, in Space Telescopes and Instrumentation 2014: Optical, Infrared, and Millimeter Wave, Vol. 9143, SPIE, 1174--1184

\bibitem[{Perrin {et~al.}(2012)Perrin, Soummer, Elliott, Lallo, \& Sivaramakrishnan}]{perrin2012}
Perrin, M.~D., Soummer, R., Elliott, E.~M., Lallo, M.~D., \& Sivaramakrishnan, A. 2012, in Space Telescopes and Instrumentation 2012: Optical, Infrared, and Millimeter Wave, ed. M.~C. Clampin, G.~G. Fazio, H.~A. MacEwen, \& J.~M.~O. Jr., Vol. 8442, International Society for Optics and Photonics (SPIE), 84423D, \dodoi{10.1117/12.925230}

\bibitem[{Pizzati {et~al.}(2024)Pizzati, Hennawi, Schaye, Eilers, Huang, Schindler, \& Wang}]{pizzati2024littlereddotsreside}
Pizzati, E., Hennawi, J.~F., Schaye, J., {et~al.} 2024, "Little Red Dots" cannot reside in the same dark matter halos as comparably luminous unobscured quasars.
\newblock \doarXiv{2409.18208}

\bibitem[{{Planck Collaboration} {et~al.}(2016){Planck Collaboration}, {Ade, P. A. R.}, {Aghanim, N.}, {Arnaud, M.}, {Ashdown, M.}, {Aumont, J.}, {Baccigalupi, C.}, {Banday, A. J.}, {Barreiro, R. B.}, {Bartlett, J. G.}, {Bartolo, N.}, {Battaner, E.}, {Battye, R.}, {Benabed, K.}, {Benoît, A.}, {Benoit-Lévy, A.}, {Bernard, J.-P.}, {Bersanelli, M.}, {Bielewicz, P.}, {Bock, J. J.}, {Bonaldi, A.}, {Bonavera, L.}, {Bond, J. R.}, {Borrill, J.}, {Bouchet, F. R.}, {Boulanger, F.}, {Bucher, M.}, {Burigana, C.}, {Butler, R. C.}, {Calabrese, E.}, {Cardoso, J.-F.}, {Catalano, A.}, {Challinor, A.}, {Chamballu, A.}, {Chary, R.-R.}, {Chiang, H. C.}, {Chluba, J.}, {Christensen, P. R.}, {Church, S.}, {Clements, D. L.}, {Colombi, S.}, {Colombo, L. P. L.}, {Combet, C.}, {Coulais, A.}, {Crill, B. P.}, {Curto, A.}, {Cuttaia, F.}, {Danese, L.}, {Davies, R. D.}, {Davis, R. J.}, {de Bernardis, P.}, {de Rosa, A.}, {de Zotti, G.}, {Delabrouille, J.}, {Désert, F.-X.}, {Di Valentino, E.}, {Dickinson, C.}, {Diego, J. M.}, {Dolag, K.},
  {Dole, H.}, {Donzelli, S.}, {Doré, O.}, {Douspis, M.}, {Ducout, A.}, {Dunkley, J.}, {Dupac, X.}, {Efstathiou, G.}, {Elsner, F.}, {Enßlin, T. A.}, {Eriksen, H. K.}, {Farhang, M.}, {Fergusson, J.}, {Finelli, F.}, {Forni, O.}, {Frailis, M.}, {Fraisse, A. A.}, {Franceschi, E.}, {Frejsel, A.}, {Galeotta, S.}, {Galli, S.}, {Ganga, K.}, {Gauthier, C.}, {Gerbino, M.}, {Ghosh, T.}, {Giard, M.}, {Giraud-Héraud, Y.}, {Giusarma, E.}, {Gjerløw, E.}, {González-Nuevo, J.}, {Górski, K. M.}, {Gratton, S.}, {Gregorio, A.}, {Gruppuso, A.}, {Gudmundsson, J. E.}, {Hamann, J.}, {Hansen, F. K.}, {Hanson, D.}, {Harrison, D. L.}, {Helou, G.}, {Henrot-Versillé, S.}, {Hernández-Monteagudo, C.}, {Herranz, D.}, {Hildebrandt, S. R.}, {Hivon, E.}, {Hobson, M.}, {Holmes, W. A.}, {Hornstrup, A.}, {Hovest, W.}, {Huang, Z.}, {Huffenberger, K. M.}, {Hurier, G.}, {Jaffe, A. H.}, {Jaffe, T. R.}, {Jones, W. C.}, {Juvela, M.}, {Keihänen, E.}, {Keskitalo, R.}, {Kisner, T. S.}, {Kneissl, R.}, {Knoche, J.}, {Knox, L.}, {Kunz, M.},
  {Kurki-Suonio, H.}, {Lagache, G.}, {Lähteenmäki, A.}, {Lamarre, J.-M.}, {Lasenby, A.}, {Lattanzi, M.}, {Lawrence, C. R.}, {Leahy, J. P.}, {Leonardi, R.}, {Lesgourgues, J.}, {Levrier, F.}, {Lewis, A.}, {Liguori, M.}, {Lilje, P. B.}, {Linden-Vørnle, M.}, {López-Caniego, M.}, {Lubin, P. M.}, {Macías-Pérez, J. F.}, {Maggio, G.}, {Maino, D.}, {Mandolesi, N.}, {Mangilli, A.}, {Marchini, A.}, {Maris, M.}, {Martin, P. G.}, {Martinelli, M.}, {Martínez-González, E.}, {Masi, S.}, {Matarrese, S.}, {McGehee, P.}, {Meinhold, P. R.}, {Melchiorri, A.}, {Melin, J.-B.}, {Mendes, L.}, {Mennella, A.}, {Migliaccio, M.}, {Millea, M.}, {Mitra, S.}, {Miville-Deschênes, M.-A.}, {Moneti, A.}, {Montier, L.}, {Morgante, G.}, {Mortlock, D.}, {Moss, A.}, {Munshi, D.}, {Murphy, J. A.}, {Naselsky, P.}, {Nati, F.}, {Natoli, P.}, {Netterfield, C. B.}, {Nørgaard-Nielsen, H. U.}, {Noviello, F.}, {Novikov, D.}, {Novikov, I.}, {Oxborrow, C. A.}, {Paci, F.}, {Pagano, L.}, {Pajot, F.}, {Paladini, R.}, {Paoletti, D.}, {Partridge, B.},
  {Pasian, F.}, {Patanchon, G.}, {Pearson, T. J.}, {Perdereau, O.}, {Perotto, L.}, {Perrotta, F.}, {Pettorino, V.}, {Piacentini, F.}, {Piat, M.}, {Pierpaoli, E.}, {Pietrobon, D.}, {Plaszczynski, S.}, {Pointecouteau, E.}, {Polenta, G.}, {Popa, L.}, {Pratt, G. W.}, {Prézeau, G.}, {Prunet, S.}, {Puget, J.-L.}, {Rachen, J. P.}, {Reach, W. T.}, {Rebolo, R.}, {Reinecke, M.}, {Remazeilles, M.}, {Renault, C.}, {Renzi, A.}, {Ristorcelli, I.}, {Rocha, G.}, {Rosset, C.}, {Rossetti, M.}, {Roudier, G.}, {Rouillé d’Orfeuil, B.}, {Rowan-Robinson, M.}, {Rubiño-Martín, J. A.}, {Rusholme, B.}, {Said, N.}, {Salvatelli, V.}, {Salvati, L.}, {Sandri, M.}, {Santos, D.}, {Savelainen, M.}, {Savini, G.}, {Scott, D.}, {Seiffert, M. D.}, {Serra, P.}, {Shellard, E. P. S.}, {Spencer, L. D.}, {Spinelli, M.}, {Stolyarov, V.}, {Stompor, R.}, {Sudiwala, R.}, {Sunyaev, R.}, {Sutton, D.}, {Suur-Uski, A.-S.}, {Sygnet, J.-F.}, {Tauber, J. A.}, {Terenzi, L.}, {Toffolatti, L.}, {Tomasi, M.}, {Tristram, M.}, {Trombetti, T.}, {Tucci, M.},
  {Tuovinen, J.}, {Türler, M.}, {Umana, G.}, {Valenziano, L.}, {Valiviita, J.}, {Van Tent, F.}, {Vielva, P.}, {Villa, F.}, {Wade, L. A.}, {Wandelt, B. D.}, {Wehus, I. K.}, {White, M.}, {White, S. D. M.}, {Wilkinson, A.}, {Yvon, D.}, {Zacchei, A.}, \& {Zonca, A.}}]{PlanckCollab}
{Planck Collaboration}, {Ade, P. A. R.}, {Aghanim, N.}, {et~al.} 2016, A\&A, 594, A13, \dodoi{10.1051/0004-6361/201525830}

\bibitem[{Pontoppidan {et~al.}(2022)Pontoppidan, Barrientes, Blome, Braun, Brown, Carruthers, Coe, DePasquale, Espinoza, Marin, Gordon, Henry, Hustak, James, Jenkins, Koekemoer, LaMassa, Law, Lockwood, Moro-Martin, Mullally, Pagan, Player, Proffitt, Pulliam, Ramsay, Ravindranath, Reid, Robberto, Sabbi, Ubeda, Balogh, Flanagan, Gardner, Hasan, Meinke, \& Nota}]{Pontoppidan_2022}
Pontoppidan, K.~M., Barrientes, J., Blome, C., {et~al.} 2022, The Astrophysical Journal Letters, 936, L14, \dodoi{10.3847/2041-8213/ac8a4e}

\bibitem[{Pérez-González {et~al.}(2024)Pérez-González, Barro, Rieke, Lyu, Rieke, Alberts, Williams, Hainline, Sun, Puskas, Annunziatella, Baker, Bunker, Egami, Ji, Johnson, Robertson, Pino, Rujopakarn, Shivaei, Tacchella, Willmer, \& Willott}]{pérezgonzález2024}
Pérez-González, P.~G., Barro, G., Rieke, G.~H., {et~al.} 2024, What is the nature of Little Red Dots and what is not, MIRI SMILES edition.
\newblock \doarXiv{2401.08782}

\bibitem[{{Rieke} {et~al.}(2023){Rieke}, {Robertson}, {Tacchella}, {Hainline}, {Johnson}, {Hausen}, {Ji}, {Willmer}, {Eisenstein}, {Pusk{\'a}s}, {Alberts}, {Arribas}, {Baker}, {Baum}, {Bhatawdekar}, {Bonaventura}, {Boyett}, {Bunker}, {Cameron}, {Carniani}, {Charlot}, {Chevallard}, {Chen}, {Curti}, {Curtis-Lake}, {Danhaive}, {DeCoursey}, {Dressler}, {Egami}, {Endsley}, {Helton}, {Hviding}, {Kumari}, {Looser}, {Lyu}, {Maiolino}, {Maseda}, {Nelson}, {Rieke}, {Rix}, {Sandles}, {Saxena}, {Sharpe}, {Shivaei}, {Skarbinski}, {Smit}, {Stark}, {Stone}, {Suess}, {Sun}, {Topping}, {{\"U}bler}, {Villanueva}, {Wallace}, {Williams}, {Willott}, {Whitler}, {Witstok}, \& {Woodrum}}]{2023ApJS..269...16R}
{Rieke}, M.~J., {Robertson}, B., {Tacchella}, S., {et~al.} 2023, \apjs, 269, 16, \dodoi{10.3847/1538-4365/acf44d}

\bibitem[{Rusakov {et~al.}(2025)Rusakov, Watson, Nikopoulos, Brammer, Gottumukkala, Harvey, Heintz, Nielsen, Sim, Sneppen, Vijayan, Adams, Austin, Conselice, Goolsby, Toft, \& Witstok}]{rusakov2025jwstslittlereddots}
Rusakov, V., Watson, D., Nikopoulos, G.~P., {et~al.} 2025, JWST's little red dots: an emerging population of young, low-mass AGN cocooned in dense ionized gas.
\newblock \doarXiv{2503.16595}

\bibitem[{{Setton} {et~al.}(2024){Setton}, {Greene}, {de Graaff}, {Ma}, {Leja}, {Matthee}, {Bezanson}, {Boogaard}, {Cleri}, {Katz}, {Labbe}, {Maseda}, {McConachie}, {Miller}, {Price}, {Suess}, {van Dokkum}, {Wang}, {Weibel}, {Whitaker}, \& {Williams}}]{2024setton}
{Setton}, D.~J., {Greene}, J.~E., {de Graaff}, A., {et~al.} 2024, arXiv e-prints, arXiv:2411.03424, \dodoi{10.48550/arXiv.2411.03424}

\bibitem[{{Stalevski} {et~al.}(2012){Stalevski}, {Fritz}, {Baes}, {Nakos}, \& {Popovi{\'c}}}]{2012skirtor}
{Stalevski}, M., {Fritz}, J., {Baes}, M., {Nakos}, T., \& {Popovi{\'c}}, L.~{\v{C}}. 2012, \mnras, 420, 2756, \dodoi{10.1111/j.1365-2966.2011.19775.x}

\bibitem[{{Stalevski} {et~al.}(2016){Stalevski}, {Ricci}, {Ueda}, {Lira}, {Fritz}, \& {Baes}}]{2016skirtor}
{Stalevski}, M., {Ricci}, C., {Ueda}, Y., {et~al.} 2016, \mnras, 458, 2288, \dodoi{10.1093/mnras/stw444}

\bibitem[{Temple {et~al.}(2021)Temple, Hewett, \& Banerji}]{temple}
Temple, M.~J., Hewett, P.~C., \& Banerji, M. 2021, Monthly Notices of the Royal Astronomical Society, 508, 737, \dodoi{10.1093/mnras/stab2586}

\bibitem[{Tinker {et~al.}(2008)Tinker, Kravtsov, Klypin, Abazajian, Warren, Yepes, Gottlöber, \& Holz}]{Tinker_2008}
Tinker, J., Kravtsov, A.~V., Klypin, A., {et~al.} 2008, The Astrophysical Journal, 688, 709, \dodoi{10.1086/591439}

\bibitem[{{Tinker} {et~al.}(2010){Tinker}, {Robertson}, {Kravtsov}, {Klypin}, {Warren}, {Yepes}, \& {Gottl{\"o}ber}}]{Tinker10}
{Tinker}, J.~L., {Robertson}, B.~E., {Kravtsov}, A.~V., {et~al.} 2010, \apj, 724, 878, \dodoi{10.1088/0004-637X/724/2/878}

\bibitem[{van Dokkum(2008)}]{van_Dokkum_2008}
van Dokkum, P.~G. 2008, The Astrophysical Journal, 674, 29–50, \dodoi{10.1086/525014}

\bibitem[{Virtanen {et~al.}(2020)Virtanen, Gommers, Oliphant, Haberland, Reddy, Cournapeau, Burovski, Peterson, Weckesser, Bright, {van der Walt}, Brett, Wilson, Millman, Mayorov, Nelson, Jones, Kern, Larson, Carey, Polat, Feng, Moore, {VanderPlas}, Laxalde, Perktold, Cimrman, Henriksen, Quintero, Harris, Archibald, Ribeiro, Pedregosa, {van Mulbregt}, \& {SciPy 1.0 Contributors}}]{2020SciPy-NMeth}
Virtanen, P., Gommers, R., Oliphant, T.~E., {et~al.} 2020, Nature Methods, 17, 261, \dodoi{10.1038/s41592-019-0686-2}

\bibitem[{Werner \& Schermelleh-Engel(2010)}]{werner}
Werner, C., \& Schermelleh-Engel, K. 2010

\bibitem[{Whitaker {et~al.}(2019)Whitaker, Ashas, Illingworth, Magee, Leja, Oesch, van Dokkum, Mowla, Bouwens, Franx, Holden, Labbé, Rafelski, Teplitz, \& Gonzalez}]{Whitaker_2019}
Whitaker, K.~E., Ashas, M., Illingworth, G., {et~al.} 2019, The Astrophysical Journal Supplement Series, 244, 16, \dodoi{10.3847/1538-4365/ab3853}

\bibitem[{Williams {et~al.}(2023)Williams, Alberts, Ji, Hainline, Lyu, Rieke, Endsley, Suess, Johnson, Florian, Shivaei, Rujopakarn, Baker, Bhatawdekar, Boyett, Bunker, Carniani, Charlot, Curtis-Lake, DeCoursey, de~Graaff, Egami, Eisenstein, Gibson, Hausen, Helton, Maiolino, Maseda, Nelson, Perez-Gonzalez, Rieke, Robertson, Sun, Tacchella, Willmer, \& Willott}]{williams2023}
Williams, C.~C., Alberts, S., Ji, Z., {et~al.} 2023, The galaxies missed by Hubble and ALMA: the contribution of extremely red galaxies to the cosmic census at 3<z<8.
\newblock \doarXiv{2311.07483}

\bibitem[{Windhorst {et~al.}(2022)Windhorst, Cohen, Jansen, Summers, Tompkins, Conselice, Driver, Yan, Coe, Frye, Grogin, Koekemoer, Marshall, O’Brien, Pirzkal, Robotham, Ryan, Willmer, Carleton, Diego, Keel, Porto, Redshaw, Scheller, Wilkins, Willner, Zitrin, Adams, Austin, Arendt, Beacom, Bhatawdekar, Bradley, Broadhurst, Cheng, Civano, Dai, Dole, D’Silva, Duncan, Fazio, Ferrami, Ferreira, Finkelstein, Furtak, Gim, Griffiths, Hammel, Harrington, Hathi, Holwerda, Honor, Huang, Hyun, Im, Joshi, Kamieneski, Kelly, Larson, Li, Lim, Ma, Maksym, Manzoni, Meena, Milam, Nonino, Pascale, Petric, Pierel, del Carmen~Polletta, Röttgering, Rutkowski, Smail, Straughn, Strolger, Swirbul, Trussler, Wang, Welch, Wyithe, Yun, Zackrisson, Zhang, \& Zhao}]{Windhorst_2023}
Windhorst, R.~A., Cohen, S.~H., Jansen, R.~A., {et~al.} 2022, The Astronomical Journal, 165, 13, \dodoi{10.3847/1538-3881/aca163}

\bibitem[{{Yue} {et~al.}(2024){Yue}, {Eilers}, {Ananna}, {Panagiotou}, {Kara}, \& {Miyaji}}]{yue2024}
{Yue}, M., {Eilers}, A.-C., {Ananna}, T.~T., {et~al.} 2024, arXiv e-prints, arXiv:2404.13290, \dodoi{10.48550/arXiv.2404.13290}

\bibitem[{Zhang {et~al.}(2024)Zhang, Jiang, Liu, \& Ho}]{zhang2024analysismultiepochjwstimages}
Zhang, Z., Jiang, L., Liu, W., \& Ho, L.~C. 2024, Analysis of Multi-epoch JWST Images of $\sim 300$ Little Red Dots: Tentative Detection of Variability in a Minority of Sources.
\newblock \doarXiv{2411.02729}

\end{thebibliography}





\end{document}